# Introducing the Single-Atom Real-Space Global Minimization Method for Solving Small Structures in Single Crystal X-ray Crystallography

Authors

**Xiaodong Zhang[a]\***

[a]Chemistry Department, Tulane University, 6400 Freret Street, New Orleans, Louisiana, 70118, United States

Correspondence email: xzhang2@tulane.edu

**Synopsis**    A novel method dubbed as the single-atom real-space global minimization for solving small structures in single crystal X-ray crystallography is presented.

**Abstract**    A new method for solving small X-ray structures with up to couple of hundreds of atoms in the unit cell has been developed. The method works by locating atoms one-by-one via global minimization of a newly defined single-atom R1 factor in real-space. In total forty test cases (of twenty samples), every resulting model has at least 52% atoms located correctly, and in more than half of the cases this percentage is 90% and higher.  For most heavier-atoms-containing (heavier than Si) small structures it is sufficient to use a single-solution strategy (using the as-collected reflection intensities). But for most lighter-atoms-only (C, N, O and F) small structures it is necessary to try multi-solution strategy: try different B values for modifying the reflection intensities by multiplying them with $\exp(B s^2)$ before feeding to the method, where $s = \sin\theta/\lambda$, and pick the best solution. This method has been demonstrated to be very effective for locating heavier atoms, so, it has potential applications in macromolecular crystallography by assisting determination of heavy-atom substructures.

## 1. Introduction

In crystallography structure determination is usually divided into the structure solution step and the structure refinement step. In the structure solution step most methods do Fourier synthesis of electron density map, a calculation in which both the amplitudes and phase angles of the structure factors are required. Though the amplitudes are proportional to the square root of the observed reflection intensities, the phase angles are not directly measured and need to be recovered from the observed intensities. This is the so-called phase problem. Therefore, the central theme of structure solution in X-ray crystallography has been solving this phase problem, and many methods have been developed. Traditional direct methods (Hauptman & Karle, 1953; Cochran, 1955; Hauptman, 1975; Karle & Karle, 1966; Main *et al.*, 1971; Giacovazzo, 1977, 1980) utilize triplet and/or quartet invariants to estimate all phase angles from a small set of starting phase angles along with the observed reflection intensities, and usually employ multi-solution strategy by trying many sets of possible starting phase angles. But nowadays the phase problem is almost always solved in two-step processes: the first step is to get an approximate or partial model of the structure, which is used to establish starting phasing of the whole target structure, while the second step iteratively improves this phasing via some dual-space recycling, which may involve reciprocal space phase refinement and/or real space electron density modifications. In the most popular program for solving small-structures, SHELXT (Sheldrick, 2015a), the starting model is established by the Patterson superposition minimum function (Buerger, 1959), and the dual-space recycling involves real-space peak-picking and reciprocal-space expansion of phases from about 40% most reliable phases (Sheldrick *et al.*, 2001; Schneider & Sheldrick, 2002). In macromolecular crystallography, molecular-replacement methods (MR; Hoppe, 1957; Rossmann & Blow, 1962; Rossmann, 1972; Rossmann & Arnold, 1993) establish the starting model by optimally placing known models into the target unit cell; in isomorphous replacement methods (Perutz, 1956; Kendrew *et al.*, 1958) the heavy-atom substructure of the derivative is determined from the measured intensity change between the derivative sample and the native (before adding heavy atoms) sample; and in single- or multi-wavelength anomalous dispersion/diffraction methods (SAD or MAD; Hendrickson & Teeter, 1981; Wang, 1985; Dauter *et al.*, 1999, 2002; Dodson, 2003; Hendrickson, 1991; Smith, 1998) the anomalous scatterer substructure is derived from the anomalous differences in the observed reflection intensities. The dual-space recycling step of macromolecular crystallography involves various electron-density-modification techniques: solvent flattening (Wang, 1985; Leslie 1988) or solvent flipping (Abrahams & Leslie, 1996), and histogram matching and noncrystallographic symmetry averaging (Cowtan & Zhang, 1999; Terwilliger, 2002; Brunger *et al.*, 1998), etc.

Though the common belief is that the phase problem has been solved for small structure molecules (Burla *et al.*, 2007), one may still wonder if the phase problem is unavoidable, or, if a route towards a structure solution must go through calculation of the electron density map? The answer to this question is NO. First, the calculation of the Patterson function does not involve the phase angles. Therefore, like the molecular replacement methods, when a structure model is built by placing model parts into the unit cell of the unknown structure via methods involving Patterson functions (Crowther, 1972; Egert, 1983) there is no phase problem. Or, when models are set up via various Patterson deconvolution techniques (Pavelcik, 1988, 1994; Burla *et al.*, 2004, 2006, 2007; Caliandro *et al.*, 2008) there is no phase problem, either. Unfortunately, the models developed by these methods are usually not complete enough to be directly used in final structure refinement; instead, these models are usually used to establish the starting phasing of the target structures. Besides the Patterson function, there is another route to avoid the phase problem. The reflection intensities can be calculated from a trial model by using the atomic scattering factors and the atomic locations, without dealing with the phase angles. Therefore, when a structure is solved via optimization techniques by optimally matching the calculated and the observed reflection intensities, the phase problem does not exist on the first place, either.

Optimization (also called minimization or maximization) has been applied in crystallography for many purposes. One prominent example is the application of the least squares technique (one of the minimization methods) in structure refinement (Sheldrick, 2015b). Optimization has also been applied in locating partial structures by matching Patterson peaks (Egert, 1983), and in improving phasing by annealing (Sheldrick, 1990). Maximization of an origin-free modulus sum function expressed in terms of the absolute electron density is applied for phasing (Rius, 2020). Iterative peaklist optimization has been used in structure solution (Sheldrick & Gould, 1995). A recent paper (Burla *et al.*, 2018) reports that a diagonal least-squares technique (a simplified version of the least-squares method) has been successfully applied to produce *ab initio* phasing for small crystal structures. In fact, in this example the calculation goes all the way through the structure refinement by only using the least-squares techniques; therefore, this example demonstrates that optimization techniques can indeed solve small crystal structures.

This paper reports a new optimization method that can also solve small structures without dealing with the phase angles. The method is dubbed as the single-atom real-space global minimization method. Section 2 explains the theory of this method. Section 3 presents the results of applying this method to twenty example data sets. Section 4 ends with some concluding discussions.

## 2. The single-atom real-space global minimization method

A reflection is indexed by hkl, and its intensity is proportional to the square of the amplitude of the structure factor. This paper directly uses the square of the amplitude of the structure factor $F_o^2(hkl)$ to represent the observed reflection intensity, and its corresponding calculated reflection intensity is $F_c^2(hkl)$, which can be calculated from a structure model in following way:

$$F_c^2(hkl) = [\Sigma_j \, f_j(hkl) \, \cos 2\pi(hx_j+ky_j+lz_j)]^2 + [\Sigma_j \, f_j(hkl) \, \sin 2\pi(hx_j+ky_j+lz_j)]^2 \qquad (1)$$

here atom j is located at $(x_j,y_j,z_j)$ and has atomic scattering factor $f_j(hkl)$, which is calculated with isotropic displacement parameter U=0.0025 Å². Note that this value of U is selected by accident, but it turns out to be a good choice. Test calculations indicate that if, for example U=0.05 Å$^2$ is used instead, the locations of the atoms are determined poorly.

Equation (1) can also be expressed in following form:

$$F_c^2(hkl) = \Sigma_j \, f_j^2(hkl) + 2\Sigma_{j<j'} \, f_j(hkl) \, f_j(hkl) \cos 2\pi[h(x_j-x_{j'})+k(y_j-y_{j'})+l(z_j-z_{j'})] \qquad (2)$$

In this expression the first sum $\Sigma_j \, f_j^2(hkl)$ is positive and only depends on the number and types of the atoms and does not depend on the location of the atoms. The second sum $2\Sigma_{j<j'} \, f_j(hkl) \, f_j(hkl)$ $\cos 2\pi[h(x_j-x_{j'})+k(y_j-y_{j'})+l(z_j-z_{j'})]$ depends on the relative locations of the pairs of atoms and are contributed by positive and negative values because the cosine function has value from -1 to +1. These positive and negative contributions are largely cancelling each other when the intensities of all reflections are added together, and so, following approximation can be reached:

$$\Sigma_{hkl} \, F_c^2(hkl) \approx \Sigma_{hkl} \, \Sigma_j \, f_j^2(hkl) \qquad (3)$$

This suggests that the observed intensities $F_o^2(hkl)$ should be scaled such that their sum equals $\Sigma_{hkl} \, \Sigma_j$ $f_j^2(hkl)$.

This paper uses following traditional definition of $R_1$ factor to evaluate the mismatch between the calculated and the observed intensities:

$$R_1 = \Sigma_{hkl} \, |F_c(hkl) - F_o(hkl)| \, / \, \Sigma_{hkl} \, F_o(hkl) \qquad (4)$$

The atoms are located one-by-one in the order from heavier atoms to lighter atoms. For N atoms in a unit cell, at a particular step, atoms 1 to j-1 have already been located, and atom j is under investigation in this step, while atoms j+1 to N have not been located yet. Consulting equations (1) and (2), contribution of atoms 1 to j to reflection intensity $F_c^2(hkl)$ can be handled like equation (1) while contribution of atoms j+1 to N can be handled partially like equation (2) by only retaining their positive contribution to the first sum there. That is:

$$F_c^2(hkl) \approx [f_1(hkl) \cos 2\pi(hx_1+ky_1+lz_1) + \cdots + f_j(hkl) \cos 2\pi(hx_j+ky_j+lz_j)]^2$$
$$+ \, [f_1(hkl) \sin 2\pi(hx_1+ky_1+lz_1) + \cdots + f_j(hkl) \sin 2\pi(hx_j+ky_j+lz_j)]^2$$
$$+ \, f_{j+1}^2(hkl) + \cdots + f_N^2(hkl) \qquad (5)$$

Note that coordinates $(x_1,y_1,z_1)$ to $(x_{j-1},y_{j-1},z_{j-1})$ have already been determined, and in current step the algorithm is determining $(x_j,y_j,z_j)$. Therefore, using the approximation of equation (5) the residue $R_1$ factor only depends on $(x_j,y_j,z_j)$, and so, $(x_j,y_j,z_j)$ can be determined by globally minimizing $R_1$. The form of $R_1$ factor with approximation of equation (5) can be called the single-atom $R_1$ factor.

In current implementation, the calculation is performed in the minimal P1 space group, and any centering is also ignored. This strategy has been proved to be robust in some other programs (Sheldrick & Gould, 1995; Burla et al., 2000; Caliandro et al., 2007; Sheldrick, 2015a). The reflection data are expanded to P1 by complementing any missing symmetry equivalents, and Friedel pairs are also fully expanded. Duplicate reflections are merged by taking simple (non-weighted) average. The calculation result naturally takes care of any existing crystallographic symmetry and centering, which can later be recovered by examining the resulting cell content. However, this extra step is not fully performed for the examples in this report: only the space group is extracted by using PLATON (Spek, 2009) program to analyze the final model, but shifting cell origin to match the convention and extracting asymmetric part of the structure are not performed.

When the calculation starts, the first atom can be placed in an arbitrary location, for which the default choice is (0.3,0.3,0.3). For locating one of any other atoms, a coarse grid with step size 0.4 Å is set within the unit cell and the single-atom $R_1$ is evaluated at the grid points to locate where $R_1$ reaches the lowest value. Then a local smaller grid is set with half of the previous step size, centered at the best point previously found, with one step up and one step down in each dimension. This local grid consists of 3 by 3 by 3 total 27 points, and the smallest $R_1$ within this local grid is located. This step is repeated with further halved step size until the step size is smaller than 0.001 Å.

After many tests it is decided necessary to prevent ghost atoms around the atoms that have already been located. To prevent ghost atoms the grid points that are near each already determined atoms within certain radius are excluded. This exclusion radius is chosen at 2.2 Å for heavy atoms like I, Mo, Pd and Se, etc., and at 1.2 Å for all other atoms.

## 3. Example Applications

For each example data set, the single-atom real-space global minimization method builds an initial model. In most cases, the initial model is chemically recognizable and therefore, is directly used in the refinement program SHELXL (Sheldrick, 2015b) to be manually improved into the final model. When an initial model cannot be directly used in final refinement, it is first improved with the aid of a custom 2Fo-Fc Fourier recycling algorithm (using a custom density modification scheme; see detailed implementation of this algorithm in the supporting information) before final manual improvement in SHELXL. To assess the quality of an initial model, it is compared with the final model. The details of the algorithm for handling this comparison are deferred in the supporting information. When an atom is located within 0.5 Å of its correct location, it is said to be located within ballpark. The comparison result shows how many of all atoms are located within ballpark. To assess how well the method works for determining heavy-atom substructures, similar comparison of initial and final models after deleting all the light atoms from them is also carried out to yield how many heavy-atoms are located within ballpark. In this paper, atoms of Si and heavier are considered as heavier atoms, and C, N, O, and F are light atoms.

Usually a single-solution strategy (using the as-collected reflection intensities) is carried out. When desirable, better results can be achieved by carrying out following multi-solution strategy: the reflection intensities can be modified by multiplying $\exp(B s^2)$, where $s = \sin\theta/\lambda$, and B is a constant whose value user can set at will (typically in the range -2 to +6). Obviously, B=0 corresponds to the as-collected data set. After trying multiple B values, pick the best solution (the one with most atoms located within ballpark).

Table 1 lists the crystallographic information of twenty samples. The samples will be referred to by their corresponding sample numbers. Information about data collection is deferred in the supporting information. Table 2 lists the results of applying the single-atom real-space global minimization method to the sample data sets with resolution as collected as well as truncated to 1 Å. Visual comparison between the initial models (both of data as-collected and as truncated) and the final model of each sample is available in the supporting information.

**Table 1**   Crystallographic information of the samples

| Sample | Formula (excluding H) | Z | atoms in cell | a(Å) | b(Å) | c(Å) | $\alpha$(°) | $\beta$(°) | $\gamma$(°) | space group |
|---|---|---|---|---|---|---|---|---|---|---|
| 1 | $S_8$ | 16 | 128 | 10.38 | 12.75 | 24.43 | 90 | 90 | 90 | Fddd |
| 2 | $S_6C_{24}$ | 1 | 30 | 9.14 | 9.61 | 9.90 | 72.06 | 71.95 | 63.88 | P-1 |
| 3 | $S_2O_2C_{12}$ | 2 | 32 | 5.86 | 10.34 | 10.74 | 90 | 104.50 | 90 | P2(1) |
| 4 | $SO_3NC_{13}$ | 2 | 36 | 7.44 | 7.42 | 10.60 | 90 | 75.82 | 90 | P2(1) |
| 5 | $Cl_4O_2C_{14}$ | 2 | 40 | 12.24 | 14.04 | 3.83 | 90 | 90 | 90 | Pba2 |
| 6 | $SN_2C_{18}$ | 2 | 42 | 6.96 | 9.00 | 12.25 | 92.74 | 95.60 | 106.98 | P-1 |
| 7 | $Pd_2Cl_4P_4C_{34}$ | 1 | 44 | 7.28 | 10.62 | 12.94 | 75.02 | 80.91 | 88.73 | P-1 |
| 8 | $ION_2C_7$ | 4 | 44 | 18.34 | 7.17 | 8.26 | 90 | 90 | 90 | Pnma |
| 9 | $Cl_2O_2C_{14}$ | 4 | 76 | 17.76 | 3.86 | 17.99 | 90 | 102.76 | 90 | P2/n |
| 10 | $CuS_2P_2F_{12}N_8C_{18}$ | 2 | 86 | 10.93 | 14.67 | 8.14 | 90 | 90 | 90 | P2(1)2(1)2 |
| 11 | $NiS_2N_2C_{18}$ | 4 | 92 | 11.45 | 10.91 | 16.37 | 90 | 104.16 | 90 | P2(1)/n |
| 12 | $Pt_2S_8Cl_4O_2N_4C_{76}$ | 1 | 96 | 11.60 | 14.16 | 14.99 | 68.04 | 68.03 | 74.14 | P-1 |
| 13 | $NiCl_2S_2P_2C_{44}$ | 2 | 102 | 12.11 | 12.11 | 14.63 | 67.10 | 79.65 | 73.17 | P-1 |
| 14 | $SO_2N_2C_{26}$ | 4 | 120 | 12.17 | 13.18 | 14.58 | 91.76 | 107.65 | 98.31 | P-1 |
| 15 | $NiCl_2S_4C_{28}$ | 4 | 164 | 8.25 | 23.86 | 15.58 | 90 | 94.69 | 90 | P2(1)/c |
| 16 | $IMo_3S_{13}N_3C_{27}$ | 4 | 188 | 11.87 | 37.77 | 13.14 | 90 | 92.34 | 90 | P2(1)/c |
| 17 | $O_2C_4$ | 8 | 48 | 15.12 | 3.9 | 15.84 | 90 | 106.49 | 90 | C2/c |
| 18 | $C_{46}$ | 1 | 46 | 5.95 | 10.80 | 12.97 | 103.77 | 99.95 | 90.46 | P-1 |
| 19 | $C_{32}$ | 4 | 128 | 11.27 | 14.80 | 15.78 | 90 | 97.94 | 90 | P2(1)/n |
| 20 | $C_{78}$ | 2 | 156 | 12.34 | 15.98 | 16.57 | 114.10 | 90.70 | 103.20 | P-1 |

**Table 2**   Results of applying the single-atom real-space global minimization method to the sample data sets with resolution as collected as well as truncated to 1 Å (when the optimal model is obtained with B other than 0, the results with B=0 are shown in parentheses)

| sample | full resolution as collected | | | | resolution truncated to 1 A | | | |
|---|---|---|---|---|---|---|---|---|
| | Resolution (Å): # reflections | B value used | # heavy atoms in ballpark | # all atoms in ballpark | % reflections used | B value used | # heavy atoms in ballpark | # all atoms in ballpark |
| 1 | 7.65 - 0.55 : 16578 | 0 | 126/128=98% | 126/128=98% | 15% | -0.2 (0) | 86/128=67% (74/128=58%) | 86/128=67% (74/128=58%) |
| 2 | 9.22 - 0.79 : 5373 | 1.726 (0) | 5/6=83% (4/6=67%) | 18/30=60% (13/30=43%) | 44% | 0 | 5/6=83% | 22/30=73% |
| 3 | 10.4 - 0.84 : 4219 | 0 | 4/4=100% | 32/32=100% | 60% | 0 | 4/4=100% | 26/32=81% |
| 4 | 7.42 - 0.65 : 7633 | 0 | 2/2=100% | 34/36=94% | 20% | 0.2 (0) | 2/2=100% (2/2=100%) | 28/36=78% (25/36=65%) |
| 5 | 14.04 - 0.84 : 4110 | 0 | 8/8=100% | 38/40=95% | 59% | 0 | 8/8=100% | 38/40=95% |
| 6 | 12.15 - 0.65 : 10807 | 0 | 2/2=100% | 39/42=93% | 25% | 0 | 2/2=100% | 38/42=90% |
| 7 | 10.26 - 0.8 : 6712 | 0 | 10/10=100% | 43/44=98% | 58% | 0 | 10/10=100% | 43/44=98% |
| 8 | 9.17 - 0.59 : 21572 | 0 | 4/4=100% | 44/44=100% | 21% | 0 | 4/4=100% | 40/44=91% |
| 9 | 11.16 - 0.85 : 8133 | 0 | 8/8=100% | 67/76=88% | 59% | 0 | 8/8=100% | 57/76=75% |
| 10 | 10.93 - 0.77 : 5450 | 0 | 6/8=75% | 76/86=88% | 61% | 5.6644 (0) | 6/8=75% (6/8=75%) | 45/86=52% (28/86=33%) |
| 11 | 11.1 - 0.8 : 14355 | 0 | 8/8=100% | 90/92=98% | 50% | 0 | 8/8=100% | 88/92=96% |
| 12 | 9.89 - 0.77 : 52407 | 0 | 14/14=100% | 75/96=78% | 75% | 0 | 14/14=100% | 72/96=75% |
| 13 | 13.43 - 0.75 : 18194 | 0 | 12/12=100% | 102/102=100% | 40% | 0 | 12/12=100% | 101/102=99% |
| 14 | 13.85 - 0.65 : 33417 | 0 | 3/4=75% | 113/124=91% | 26% | 0 | 4/4=100% | 106/124=85% |
| 15 | 23.86 - 0.83 : 19760 | 0 | 47/48=98% | 153/164=93% | 59% | 0 | 48/48=100% | 138/164=84% |
| **16** | 11.31 - 0.69 : 67158 | 0 | 65/68=96% | 131/188=70% | 30% | 0 | 68/68=100% | 164/188=87% |
| **17** | 15.19 - 0.67 : 9478 | 1.4 (0) | 0/0 (0/0) | 48/48=100% (39/48=81%) | 49% | 0 | 0/0 | 39/48=81% |
| **18** | 12.39 - 0.73 : 8412 | 0.856 178074 (0) | 0/0 (0/0) | 46/46=100% (29/46=63%) | 38% | 0 | 0/0 | 44/46=96% |
| **19** | 14.8 - 0.81 : 18698 | -0.4 (0) | 0/0 (0/0) | 122/128=95% (57/128=45%) | 55% | -0.2 (0) | 0/0 (0/0) | 122/128=95% (31/128=24%) |
| **20** | 15.02 - 0.77 : 26654 | 1.6 (0) | 0/0 (0/0) | 116/156=74% (39/156=25%) | 45% | 2.4234 (0) | 0/0 (0/0) | 96/156=62% (32/156=21%) |

## 4. Discussions

One thing must be pointed out first, that is, the calculation of the single atom real-space global minimization is very time consuming. For a typical size unit cell of about 15 Å on each side, with 0.4 Å coarse step size, there are about 50000 grid points over which the calculation must be performed. With the current Python implementation of this algorithm (of which, source codes are available in the supporting information), running on Microsoft Surface Pro 7, to locate one atom it takes from a few seconds up to about half hour, depending on the size of the unit cell. Because of this time limitation only preliminary explorations have been carried out.

Each atom (except the first atom) is located by globally minimizing the single-atom $R_1$ factor in real space. Usually when a new atom is added, the $R_1$ factor value drops. One should refrain from inferring meaning from this trend. The single-atom $R_1$ factor is approximate, and when new atom is added, its definition changes (because the formula now contains one more atom). A comparison between single-atom $R_1$ factor value from one step to next step carries not much useful information. The single-atom $R_1$ factor serves as a tool at each step to locate an atom; it does not accurately measure the overall fitting quality.

One mathematical hunch can be pointed out: the heavier atoms have larger impact on $R_1$ factor, so, at start even though the single-atom $R_1$ factor is not so accurate, it is still possible to pick out the heavier atoms. When coming to late stage, the approximation of the single-atom $R_1$ factor becomes more accurate, and this makes the determination of the lighter atoms better than otherwise. This mathematical hunch partially explains why the algorithm works as it is designed.

Examining the results in table 2, in general, the single-atom real-space global minimization method works well: structures are determined to completeness (measured as percent of all atoms located within ballpark) ranging from 52% up to 100%.

There are twenty samples. For each sample there are two cases: one case with full data resolution (as collected) and one case with resolution truncated to 1 Å. So, total 40 cases. Out of total 40 cases, only 3 cases have model completeness below 70%, and only 10 cases have model completeness below 80%. Especially, there are 22 cases (more than half) have model completeness 90% and higher.

Model completeness (52% to 100% complete) of the heavier-atoms-containing cases is about the same as that (62% to 100% complete) of the light-atoms-only cases. However, 75% of the light-atoms-only cases (6/8) need multi-solution strategy, while only 12% of the heavier-atoms-containing cases (4/32) need multi-solution strategy. For most heavier-atoms-containing cases the single-solution strategy is sufficient.

Comparing the truncated-resolution cases against the full-resolution cases, out of total 20 samples: 3 samples have no change in model completeness after truncating resolution; 6 samples see slight decrease in model completeness (decrease by 1 to 6 percent points); 9 samples see more decrease in model completeness (9 to 36 percent points decrease). As exceptions, samples 2 and 16 see model completeness increase after truncating data resolution (increasing by 13 and 17 percent points, respectively).

Out of total 40 cases, 10 cases have tried multi-solution strategy. In these tried cases, multi-solution strategy brings about 9 to 71 percent points increase in model completeness, where the most dramatic improvement happens in sample 19 with truncated resolution: model completeness has improved from 24% in the single-solution result to 95% in the multi-solution result – an increase of 71 percent points.

Examining the data in table 2 to see how well the method can determine heavy-atom substructures when only single-solution strategy is used: out of 32 total cases, 26 cases have heavy-atom substructure completeness 96% and higher, while the remaining 6 cases have heavy-atom substructure completeness ranging from 58% to 83%. Therefore, the method is proved to be very effective for locating heavier atoms.

It is interesting to compare this paper's method (the single-atom real-space global minimization method) with the diagonal least-squares technique that is reported in a recent paper (Burla *et al.*, 2018). The two approaches apply different optimization methods: this paper applies global minimization to the single-atom $R_1$ factor, while the other paper uses diagonal least-squares technique. Mathematically, least-squares method requires the starting solution to be close to the correct solution to converge, because it uses Taylor expansion to linearize the model. Global minimization does not have this limitation. To overcome this limitation, Burla *et al.* use multi-solution strategy by gambling on randomly set atomic locations. In this paper, for most cases the single-solution strategy is sufficient (as the results are already very good and there is not much room for further improvement); but when there is room for further improvement, using multi-solution strategy by gambling on the B value may greatly improve the results. Least-squares method can adjust all atoms together, so, supposedly it has an edge in efficiency. Unfortunately, the multi-solution strategy makes the whole algorithm time consuming. This paper's global minimization approach is also very time consuming, possibly even more time consuming, because, as analyzed in above, to locate one atom computation must be performed over tens of thousands of coarse grid points. In this regard, both methods are future oriented. Considering super computers with intense parallel processing, or even the future of quantum computing, when computing power is not of a concern, then, if these methods are proved to have reliability (need much more work for assessment of this), they may turn into future working force. Finally, the two approaches may combine to form a better algorithm: one may start with single-atom real-space global minimization to locate a few heavier atoms and continue with the diagonal least-squares method to complete the model.

Because the single-atom real-space global minimization method is very effective for locating heavier atoms, while the first step of experimental phasing in macromolecular crystallography is to determine heavy-atom substructures, the new method may find application in macromolecular crystallography. However, considering large unit cells in macromolecular crystals, the algorithm should be fully optimized and computers with intense parallel processing should probably be used.

**Acknowledgements**    The author thanks Professor James P. Donahue for the opportunity of collecting data for his research group. The author is grateful of Professor Robert A. Pascal for providing three light-atoms-only data sets. The author also thanks Professor Joel T. Mague, Assistant Professor Alex McSkimming and Dr. Michael Johnson for assistance in operating the XRD lab.

# Supporting information

### S1. Information on data collection of the samples

All crystals were coated with paratone oil and mounted on the end of a nylon loop attached to the end of the goniometer. Data were collected at 150 K under a dry $N_2$ stream supplied under the control of an Oxford Cryostream 800 attachment. The data collection instrument was a Bruker D8 Quest Photon 3 diffractometer equipped with a Mo fine-focus sealed tube providing radiation at $\lambda = 0.71073$ nm or a Bruker D8 Venture diffractometer operating with a Photon 100 CMOS detector and a Cu Incoatec I microfocus source generating X-rays at $\lambda = 1.54178$ nm.

### S2. Algorithm for comparing the initial model with the final correct model

To compare the initial model with the final model, origin of the unit cell of the final model should be shifted until best match (judged by maximum number of atoms located within 0.5 Å) is realized. Realistically one may set 0.2 Å grid points for the origin to be shifted over. However, if calculation is carried out in this way, it is very time consuming. Instead following simplified version of calculation is adopted: take one atom in the initial model as the reference point, and shift cell origin of the final model such that one of its atoms overlaps with the reference point. Try each atom in initial model as reference point and try each atom in the final model to overlap the reference point to find the best match between the two models. Because when the structure is non-central symmetric the final model could be inverted, its inverted version is also checked for locating the best match.

### S3. Implementation of 2Fo-Fc Fourier recycling algorithm

When the single-atom real-space global minimization method does not yield a satisfactory solution, a 2Fo-Fc Fourier recycling algorithm is used to improve the solution to an acceptable result. This algorithm includes following steps:

(1) The solution of the single-atom real-space global minimization method is used as the initial trial model.

(2) Structure factor amplitudes $F_c$ and phase angles $\varphi$ are calculated from the trial model.

(3) Use $2F_o$-$F_c$ and the phase angles $\varphi$ to establish the electron density function.

(4) Calculate the electron density values on the grid points which can be later converted into complex structure factors via inverse FFT (fast Fourier transform). If the Patterson superposition minimum function is used as input, it can replace the electron density function in this step.

(5) Do Fourier recycling batch: a) electron density grid values are masked (see details of how this is done below) and then converted to complex structure factors via inverse FFT, b) $F_c$ and φ are calculated from complex structure factors, c) scale $F_c$, note that $F_o$ was scaled at very beginning of calculation, here it is necessary to scale $F_c$ because the electron density grid values are masked and/or electron density was derived from Patterson superposition minimum function, d) use $2F_o$-$F_c$ and φ to re-establish the complex structure factors, e) the complex structure factors are converted to electron density grid values via FFT, f) repeat a to e until 20 cycles, g) final complex structure factors are used to re-calculate the $F_c$ and φ, and after re-scaling $F_c$, use $2F_o$-$F_c$ and the phase angles φ to re-establish the electron density function.

(6) Peaks of the electron density are located and assigned to atoms in the order of taller peaks to shorter peaks, and ghost peaks are excluded in the similar way as done in the single-atom real-space global minimization method, in addition, peaks that are potentially forming triangular bonding with the already accepted peaks with bond length less than 1.5 Å are also excluded, thus reach an improved model.

(7) Turn the improved model into a new trial model and the cycle repeats until a satisfactory model is reached as judged by visual inspection.

**How the electron density grid values are masked**: Let E be original unmasked electron density grid values. F will contain final masked electron density grid values but starts with all zero values, and mask M also starts with all zero values. The electron density will be added to F one peak at a time, until all N peaks are added, where N is the number of atoms in the unit cell. To add the first peak, the algorithm locates the highest value in E, takes its location as (xyz), turns all the grid values of M within 1.2 Å (2.2 Å if heavy atom)of (xyz) to 1, adds E multiply by M and multiply by a Gaussian peak of unit height centered at (xyz) with 0.7 Å full width at half height to F. Replace E with E multiply 1-M, and set M all zero again, and then move on to locate the next peak.

**S4. Visual comparison between the initial models and the final model of each sample**

Sample 1: S8, Z=16

as-collected             truncated resolution             final model

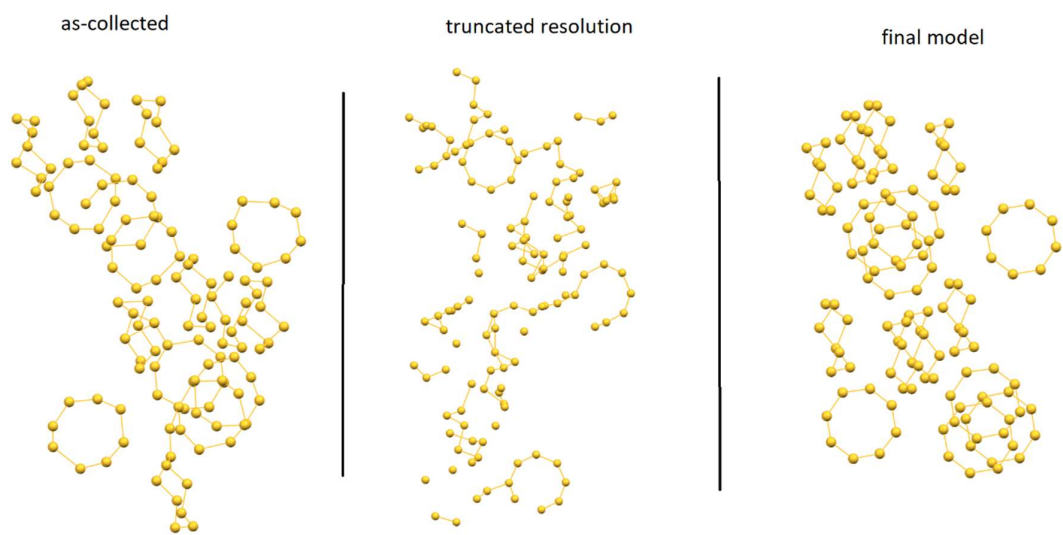

Sample 2: S6C24, Z=1

as-collected             truncated resolution             final model

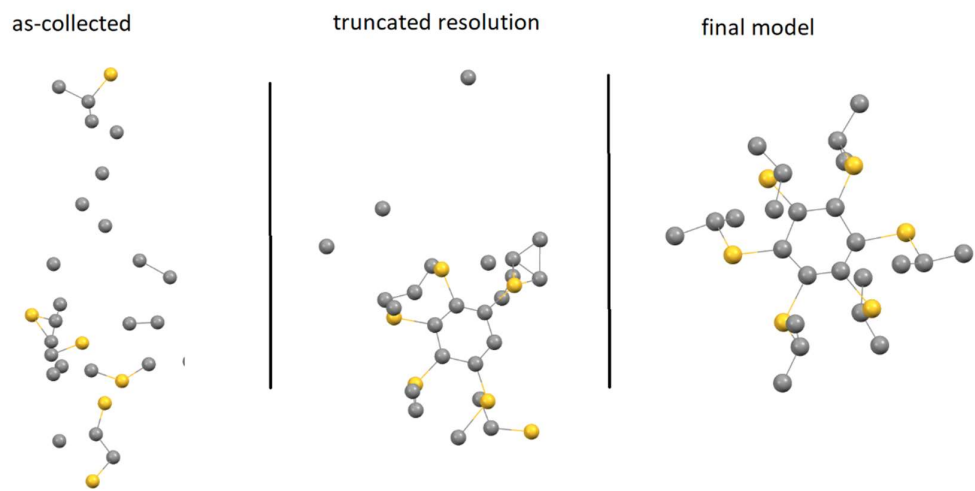

Sample 3: S2O2C12, Z=2

as-collected                    truncated resolution            final model

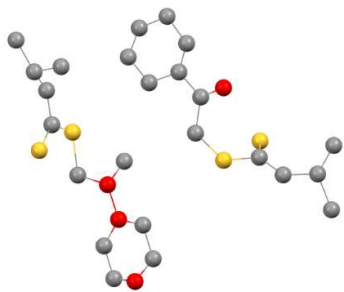
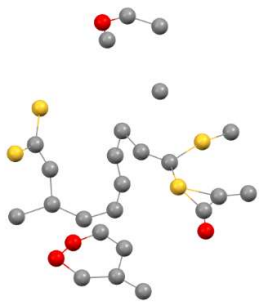
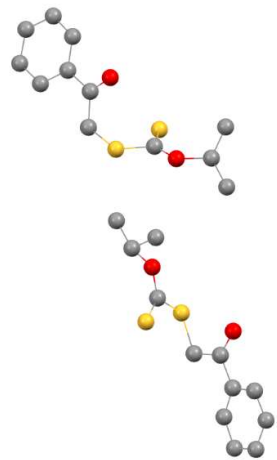

sample 4: SO3NC13, Z=2

as-collected                    truncated resolution            final model

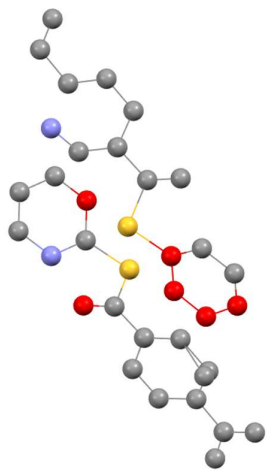
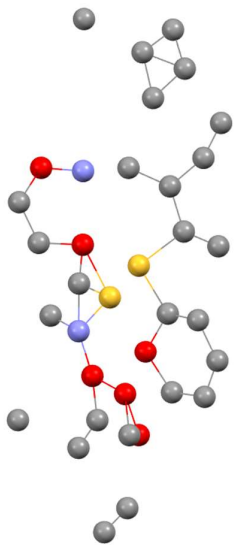
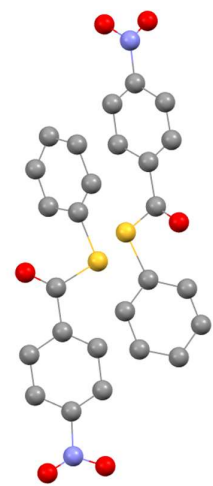

sample 5: Cl4O2C14, Z=2

as-collected                    truncated resolution                    final model

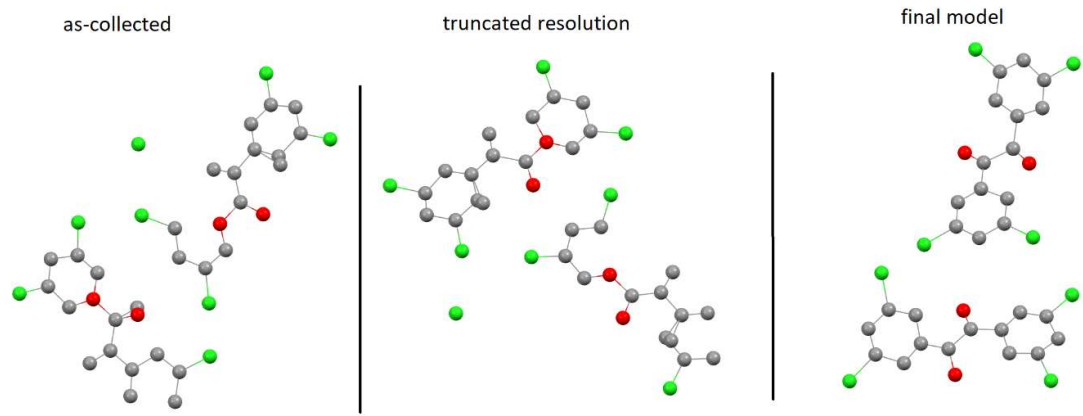

Sample 6: SN2C18, Z=2

as-collected                    truncated resolution                    final model

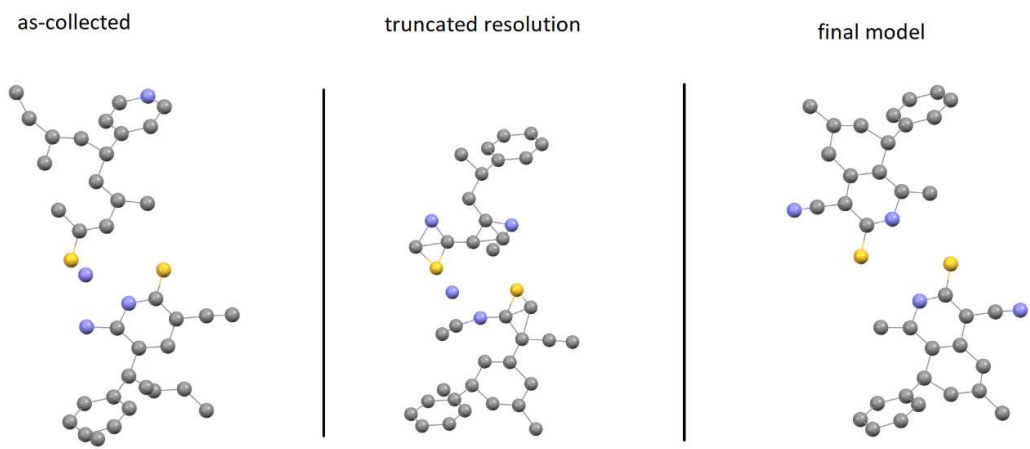

Sample 7: Pd2Cl4P4C34, Z=1

as-collected                              truncated resolution                              final model

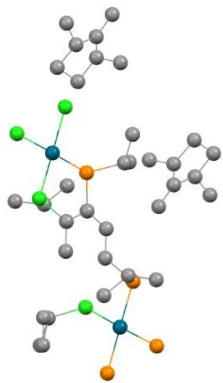
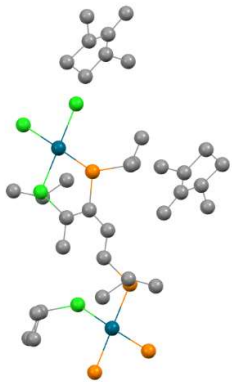
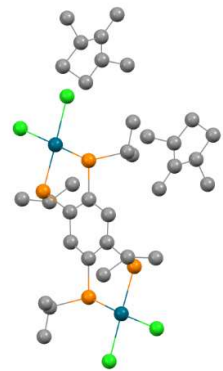

Sample 8: ION2C7, Z=4

as-collected                              truncated resolution                              final model

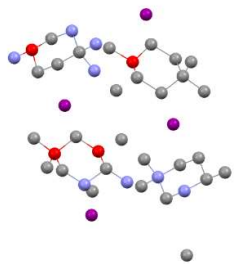
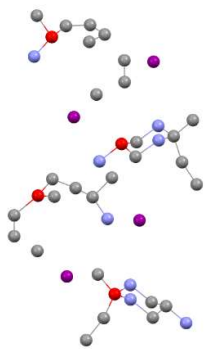
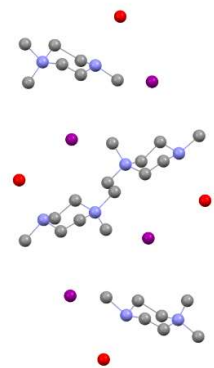

Sample 9: Cl2O2C14, Z=4

as-collected    truncated resolution    final model

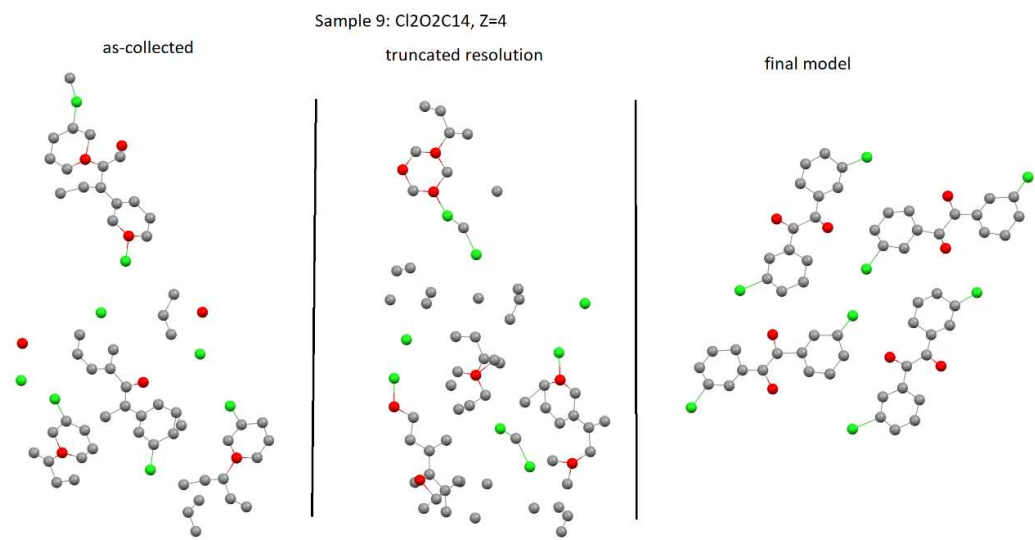

Sample 10: CuS2P2F12N8C18, Z=2

as-collected    truncated resolution    final model

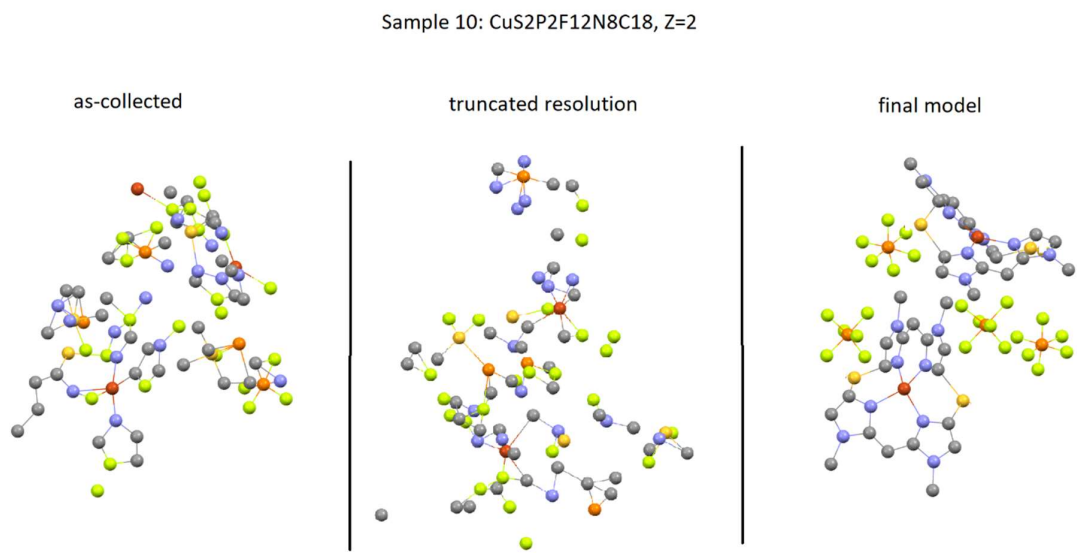

Sample 11: NiS2N2C18, Z=4

as-collected    truncated resolution    final model

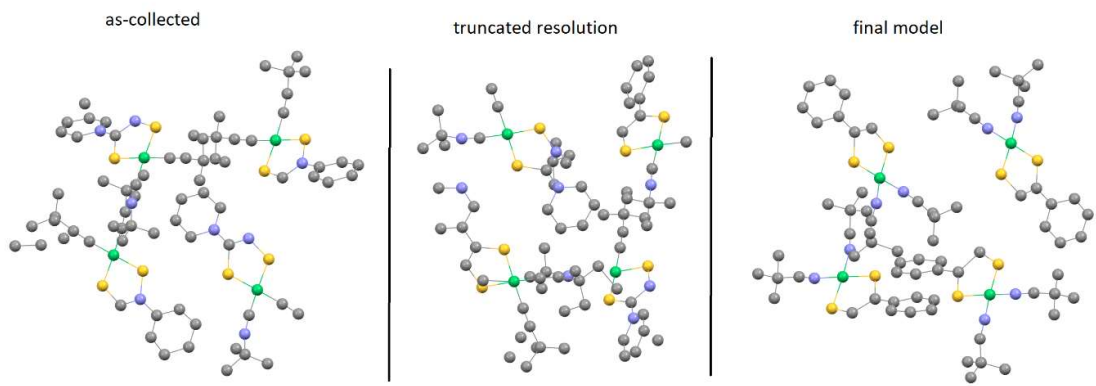

Sample 12: Pt2S8P4O2N4C76, Z=1

as-collected    truncated resolution    final model

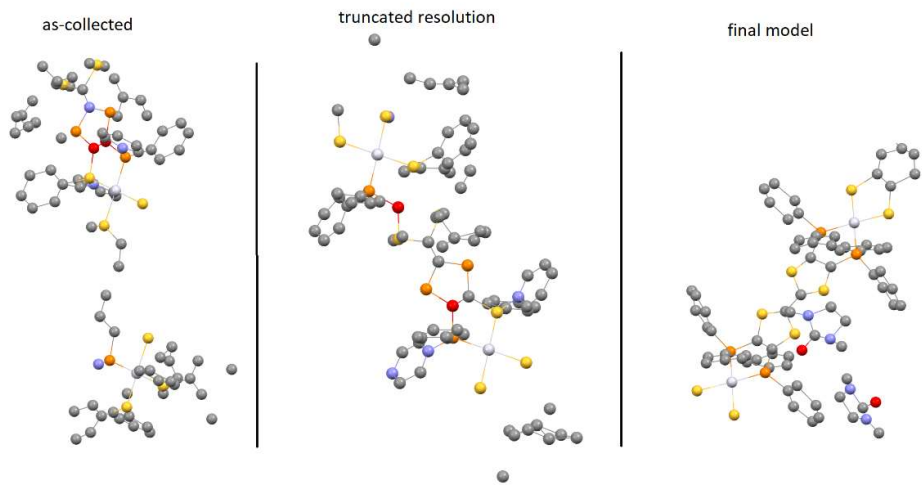

Sample 13: NiCl2S2P2C44, Z=2

as-collected                    truncated resolution                    final model

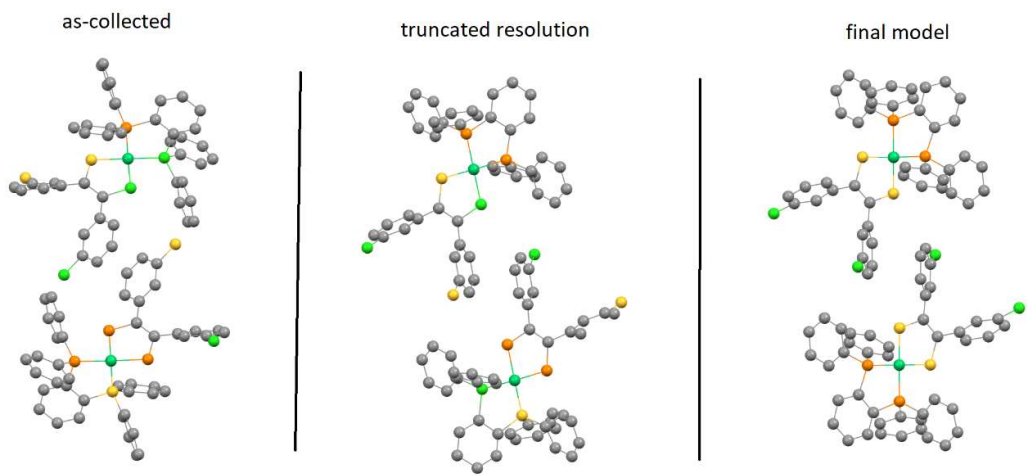

Sample 14: SO2N2C26, Z=4

as-collected                    truncated resolution                    final model

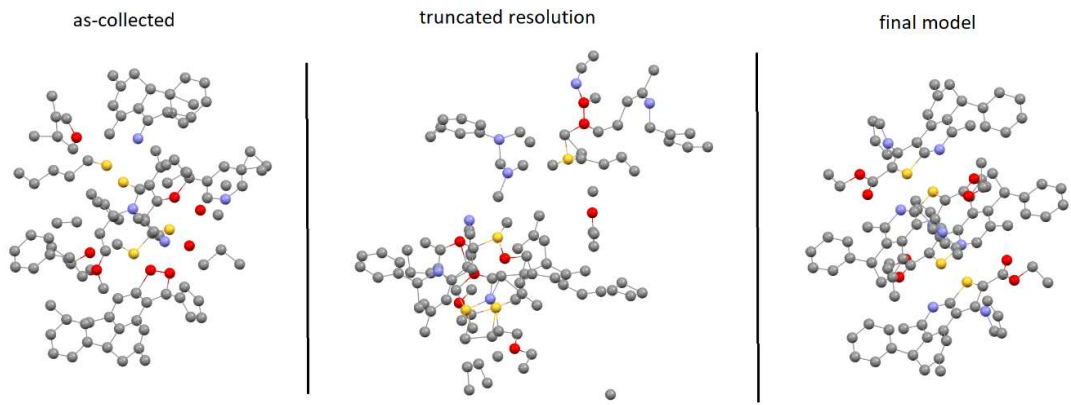

Sample 15: NiCl8S4C28, Z=4

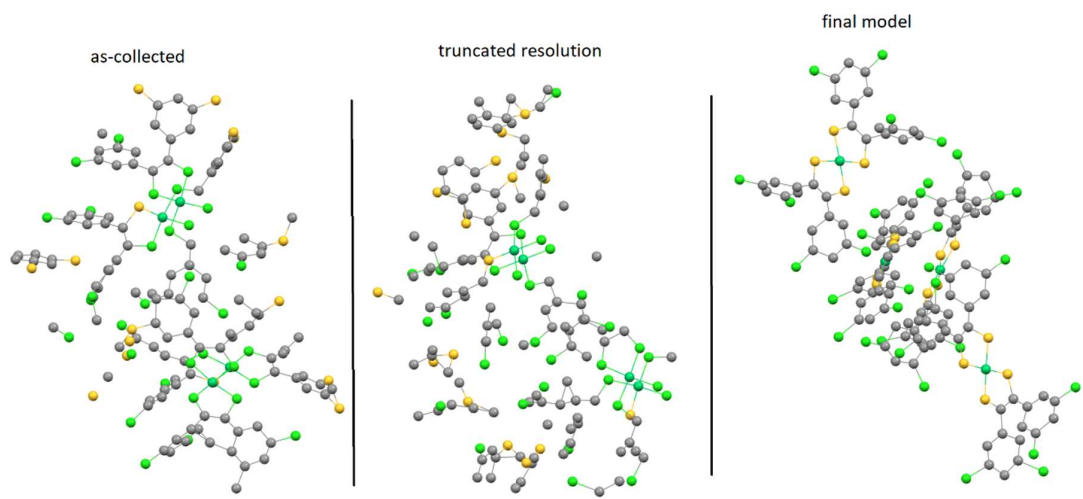

Sample 16: IMo3S13N3C27, Z=4

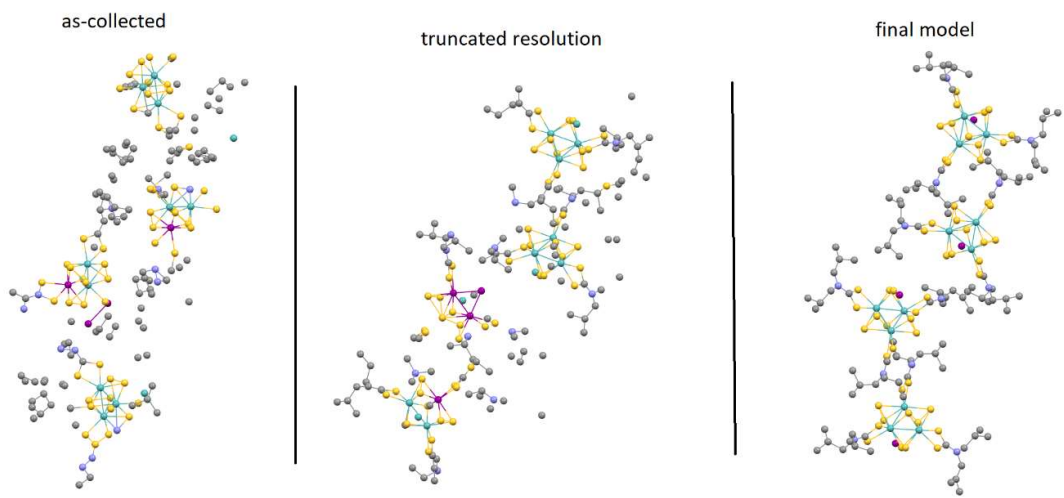

Sample 17: O2C4, Z=8

as-collected                    truncated resolution                    final model

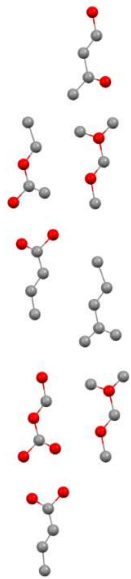 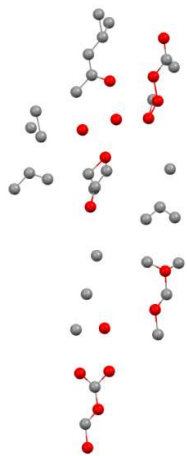 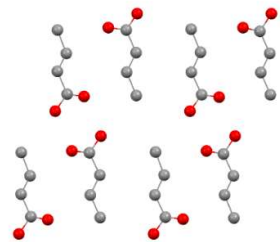

Sample 18: C46, Z=1

as-collected                    truncated resolution                    final model

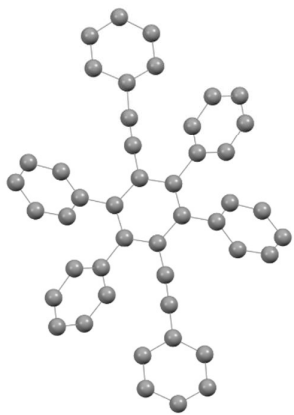 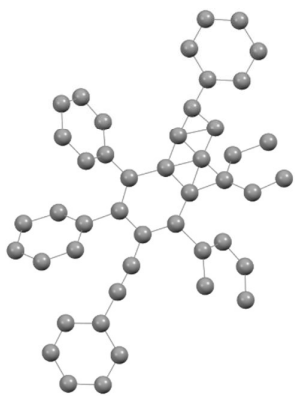 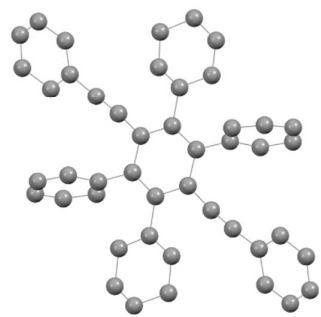

Sample 19: C32, Z=4

as-collected                          truncated resolution                          final modle

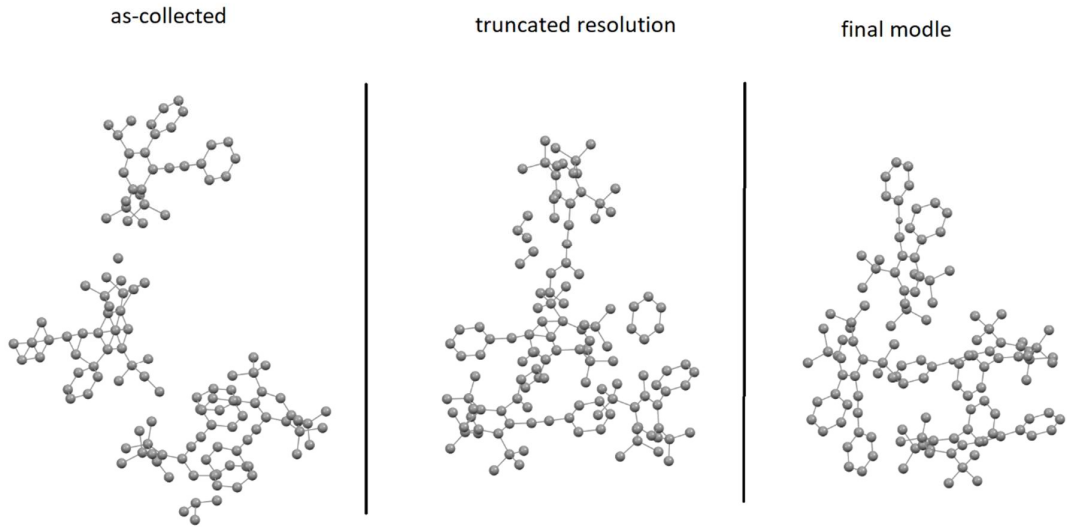

Sample 20: C78, Z=2

as-collected                          truncated resolution                          final model

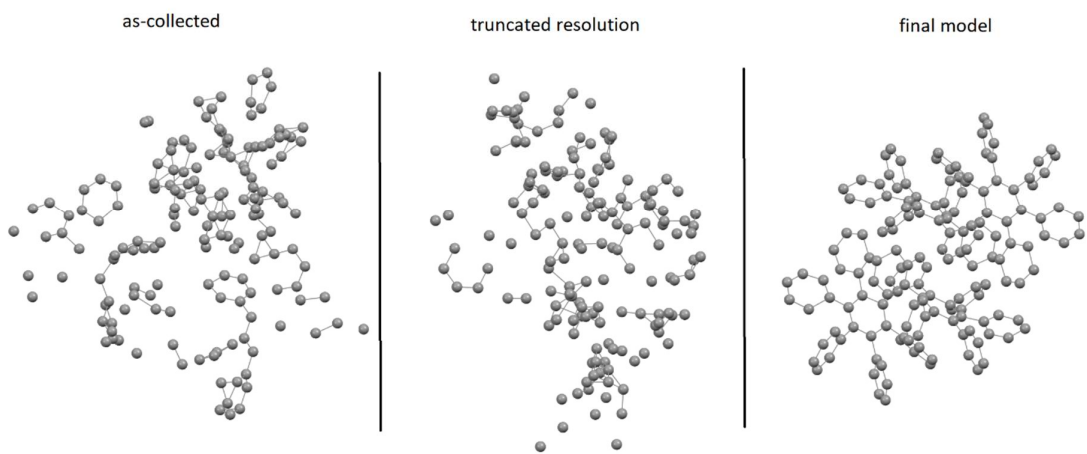

## S5. List of Python source codes

Python source codes are listed below. Readers may use these codes to test the method with their own data sets. Anyone needs assistance to set up the Python programs of this method may contact the author: Dr Xiaodong Zhang, xzhang2@tulane.edu. Brief description of the codes is given here. The modules tools2, read_data, expand_p1bar and elements3 should be placed in a particular folder, in the author's example, that folder is "D:\xray_data\extra_data\", and Windows 11 operating system is used. The main user interface is s_JPD1253.py, in which the line

sys.path.insert(0,'D:\\xray_data\\extra_data') should reflect the actual folder where the above four modules are located. Module s_JPD1253.py should be located where res and hkl files are located and rename res and hkl files to a.res and a.hkl.  Change the line molecule = 'CuS2P2F12N8C18' to reflect your molecular formula. If you need to study what Z value should be, change Z = 2 to negative value like Z = -2. Run s_JPD1253.py and test with various Z values to see which Z value gives meaningful density. Set Z to what you believe its value should be. Set fast=1 to 1 to run single-atom real-space global minimization, to 5 to run 2Fo-Fc Fourier recycling. When running 2Fo-Fc Fourier recycling, use max_runs=1000 to control maximum batches of 20-cycle recycling. The initial model for 2Fo-Fc Fourier recycling is read from the res file. Typically, there is no need to adjust other parameters. All results are saved in the history.txt file. When running 2Fo-Fc Fourier recycling, intermediate results are periodically copied to clipboard, which can be examined by ShelXle program.

It is necessary to expand the hkl file into P1 space group by complementing any missing symmetry equivalents, as well as fully expand Friedel pairs. For those data sets with no missing symmetry equivalents following module expand_p1bar.py can be used. If some symmetry equivalents are missing, it should be handled case by case.

### S5.1. Module expand_p1bar.py for expanding hkl file into P1 space group

```
# expand to Triclinic P1

from read_data import read_hkl3

def st1(x):

        r = str(x)

        n = 4-len(r)

        if n>0:

                r = ' '*n + r

        return r

def st2(x):

        r = str(x)

        n = 8-len(r)

        if n>0:
```

```
        r = ' '*n + r

    return r

# read in

h,k,l,F2,sigF2 = read_hkl3()

# expand Friedel opposites

reflections = []

for i in range(len(h)):

        reflections.append((h[i],k[i],l[i],F2[i],sigF2[i]))

        reflections.append((-h[i],-k[i],-l[i],F2[i],sigF2[i]))

reflections.sort(key=lambda r:r[2])

reflections.sort(key=lambda r:r[1])

reflections.sort(key=lambda r:r[0])

# merge

n = len(reflections)

print(n)

reflects = []

i = 0

while i < n:

        h,k,l,F2,sigF2 = reflections[i]

        i+=1

        count = 0

        for j in range(i, n):

                hh,kk,ll,F22,sigF22 = reflections[i]

                if h==hh and k==kk and l==ll:
```

```
                count += 1

                i += 1

                F2 += F22

                sigF2 += sigF22

        else:

                break

    F2 /= count+1

    sigF2 /= count+1

    reflects.append((h,k,l,round(F2,3),round(sigF2,3)))

print(len(reflects))

print('writing new reflection file a.hkl: ')

with open('a.hkl', 'w') as f:

        for (h,k,l,F2,sigF2) in reflects:

                f.write(st1(h)+st1(k)+st1(l)+st2(F2)+st2(sigF2)+'\n')

        f.write(st1(0)+st1(0)+st1(0)+st2(0)+st2(0)+'\n')

print('done')
```

## S5.2. Main user interface s_JPD1253.py

```
import sys

sys.path.insert(0,'D:\\xray_data\\extra_data')

import time

from tools2 import *

starttime=time()

res_file='a.res'

hkl_file='a.hkl'
```

```
A = matrix_A(res_file)

molecule = 'CuS2P2F12N8C18'

Z = 2

fast=1

        #1: use min-cycle to build model from scratch

        #2: use min-cycle to continue building model

        #3: use r1 hole to continue building model

        #4: build model by patterson technique

        #5: dual space cycling using FFT

startfrom=1 #r1 hole,min-cycles, gambling: atom to start from

max_runs=1000  #max dual-space or min cycles

cutlimit = 3.0  #dual-space: 1.5 probably enough, may choose 2.5

nextra = 0

mB = 1 # defaul 1, how much to sharpen F2, 0 no sharpen, 2 fully sharpened, 1 halfway sharpened

patterson = False

ntotal_initial=None # num atoms in min, if none will decide automatically

s_dual=0.4

n_refine=3 # num of cycles of refinement

n_improve = 1 # 1 for 2*Fo-1*Fc

improve_only = 1 # this flag no longer being used

extension=False   # recommend False

double_first=-1  # -1 not to double, 1 to double first run for catching weaker peaks

if Z<0:

    Z = 1

    print('When Z=1, density = ',round(density(molecule,Z,A),2))
```

```
    while True:

        Z = int(input('Please enter Z: '))

        print('When Z=',Z,', density = ',round(density(molecule,Z,A),2))

        done = input('Done? ')

        if done: sys.exit()

print('When Z=',Z,', density = ',round(density(molecule,Z,A),2))

get_structure_solution(res_file,hkl_file,molecule,Z,ntotal_initial,

    starttime,s_dual,n_refine,max_runs,improve_only,fast=fast,

    n_improve=n_improve,cutlimit=cutlimit,extension=extension,

    double_first=double_first,startfrom=startfrom,nextra=nextra,

    mB=mB,patterson=patterson)

print('done')
```

### S5.3. Module tools2.py

```
from math import radians, degrees

from read_data import get_cell,  read_hkl3, read_atoms, read_fcf

import numpy as np

import matplotlib.pyplot as plt

from scipy.optimize import minimize

import sys

from random import sample, choice

from time import time

from multiprocessing import Pool

from pyperclip import copy as cp, paste as pst

from elements3 import elements

from random import random

from scipy.fft import fftn, ifftn
```

```python
sin = np.sin

cos = np.cos

exp = np.exp

sqrt = np.sqrt

atan = np.arctan

acos = np.arccos

pi = np.pi

tpi = 2*pi

the_heavy = ['Mo','Se','I','Pd','Re','Pt',] # dd=2.2 for the heavy, otherwise 1.2

def put_in_cell(x):

    while x < 0:

        x += 1

    while x >= 1:

        x -= 1

    return x

def st1(x):

    r = str(x)

    n = 4-len(r)

    if n>0:

        r = ' '*n + r

    return r

def st2(x):
```

```python
    r = str(x)

    n = 8-len(r)

    if n>0:

        r = ' '*n + r

    return r

def st3(x):

    r = str(x)

    n = 6-len(r)

    if n>0:

        r += ' '*n

    return r

def matrix_A(choice=None):

    # choice = int 0,1,2... or str filename

    lam,a,b,c,alfa,beta,gamma = get_cell(choice)

    #print(lam,a,b,c,alfa,beta,gamma)

    alfa = radians(alfa)

    beta = radians(beta)

    gamma = radians(gamma)

    A = np.array([

    [a*a,a*b*cos(gamma),a*c*cos(beta)],

    [a*b*cos(gamma),b*b,b*c*cos(alfa)],

    [a*c*cos(beta),b*c*cos(alfa),c*c]

    ])

    return A
```

```python
def abc(A):
    a,b,c = sqrt(A[0,0]),sqrt(A[1,1]),sqrt(A[2,2])
    return (a,b,c)

def alfabetagamma(A):
    a,b,c = abc(A)
    alfa,beta,gamma = A[1,2]/b/c,A[0,2]/a/c,A[0,1]/a/b
    alfa,beta,gamma =  acos(alfa),acos(beta),acos(gamma)
    alfa,beta,gamma = degrees(alfa),degrees(beta),degrees(gamma)
    return (alfa,beta,gamma)

def matrix_inv(A):
    # inverse of A
    return np.linalg.inv(A)

def dis_exact(p,A): # exact
    # returns |p|^2
    x,y,z = p
    p1 = np.array([x,y,z])
    return np.matmul(p1,np.matmul(A,p1[:,np.newaxis]))[0]

def dis(p,A): # shortest
    # returns |p|^2
    x,y,z = p
```

```python
    x,y,z = put_in_cell(x),put_in_cell(y),put_in_cell(z)

    x,y,z = min(x,1-x),min(y,1-y),min(z,1-z)

    p1 = np.array([x,y,z])

    return np.matmul(p1,np.matmul(A,p1[:,np.newaxis]))[0]

def shortest(p1,p2,A): # shortest

    # returns shortest |p1-p2|^2 and where p2 should be to get this shortest

    x1,y1,z1 = p1

    x2,y2,z2 = p2

    x,y,z = put_in_cell(x2-x1),put_in_cell(y2-y1),put_in_cell(z2-z1)

    if x>0.5:x-=1

    if y>0.5:y-=1

    if z>0.5:z-=1

    p1 = np.array([x,y,z])

    return (np.matmul(p1,np.matmul(A,p1[:,np.newaxis]))[0],(x1+x,y1+y,z1+z))

def bond_length(p1,p2,A): # shortest length

    x1,y1,z1 = p1

    x2,y2,z2 = p2

    x,y,z = x1-x2,y1-y2,z1-z2

    p = (x,y,z)

    r = dis(p,A)

    return sqrt(r)

def dot(p1,p2,A):

    p1 = np.array(p1)

    p2 = np.array(p2)

    return np.matmul(p1,np.matmul(A,p2[:,np.newaxis]))[0]
```

```python
def cross(p1,p2,A):
    x1,y1,z1 = p1
    x2,y2,z2 = p2
    T = matrix_inv(A)
    cell_V = sqrt(np.linalg.det(A))

    Z =np.array( [y1*z2-z1*y2,z1*x2-x1*z2,x1*y2-y1*x2] )*cell_V #base a*,b*,c*
    Z = np.matmul(T,Z[:,np.newaxis])  # switch to base a, b, c
    Z = Z.reshape(3)
    return Z

def cell_volume(A):
    return sqrt(np.linalg.det(A))

def length(p1,p2,A): # return actual length, not shortest length
    p1 = np.array(p1)
    p2 = np.array(p2)
    p = p1-p2
    return sqrt(np.matmul(p,np.matmul(A,p[:,np.newaxis]))[0])

def length2(p,A): # actual length
    return length(p,(0,0,0),A)

def dist(pt,pts,A): # shortest
    # returns min of |pt-p|^2 for p in pts
    ds = np.array([dis((pt[0]-p[0],pt[1]-p[1],pt[2]-p[2]),A) for p in pts])
    return np.min(ds)
```

```python
def dist_arg(pt,pts,A): # shortest
    # returns i of min of |pt-p[i]|^2 for p[i] in pts
    ds = np.array([dis((pt[0]-p[0],pt[1]-p[1],pt[2]-p[2]),A) for p in pts])
    return np.argmin(ds)

def notnear(pt,pts,s,A): # shortest
    ss = s*s
    for p in pts:
        x = (pt[0]-p[0],pt[1]-p[1],pt[2]-p[2])
        if dis(x,A)<ss:
            return False
    return True

def expand(pts):
    points = []
    for p in pts:
        x,y,z = p[0],p[1],p[2]
        points.append((x,y,z))
        points.append((x+1,y,z))
        points.append((x-1,y,z))
        points.append((x,y+1,z))
        points.append((x,y-1,z))
        points.append((x,y,z+1))
        points.append((x,y,z-1))

        points.append((x+1,y+1,z+1))
        points.append((x-1,y-1,z-1))
```

```
        points.append((x-1,y+1,z+1))

        points.append((x+1,y-1,z-1))

        points.append((x+1,y-1,z+1))

        points.append((x-1,y+1,z-1))

        points.append((x+1,y+1,z-1))

        points.append((x-1,y-1,z+1))

        points.append((x+1,y+1,z))

        points.append((x+1,y-1,z))

        points.append((x-1,y+1,z))

        points.append((x-1,y-1,z))

        points.append((x+1,y,z+1))

        points.append((x+1,y,z-1))

        points.append((x-1,y,z+1))

        points.append((x-1,y,z-1))

        points.append((x,y+1,z+1))

        points.append((x,y-1,z+1))

        points.append((x,y+1,z-1))

        points.append((x,y-1,z-1))

    return points

def dis_old(p,A):

    # returns |p|^2

    p = np.array(p)

    return np.matmul(p,np.matmul(A,p[:,np.newaxis]))[0]
```

```python
def notnear2_old(pt,pts,s,A):

    pts2 = expand(pts)

    ss = s*s

    for p in pts:

        x = (pt[0]-p[0],pt[1]-p[1],pt[2]-p[2])

        if dis_old(x,A)<ss:

            return False

    return True

def notnear3(pt,solution,atomj,A): # shortest

    for i,p in enumerate(solution):

        atom = atomj[i]

        if atom in the_heavy:

            dd = 2.2

        else:

            dd = 1.2

        if not notnear2_old(pt,[p],dd,A):

            return False

    return True

def notnear2(pt,pts,s,A):

    pts2 = expand(pts)

    ss = s*s

    for p in pts:

        x = (pt[0]-p[0],pt[1]-p[1],pt[2]-p[2])
```

```
        if dis(x,A)<ss:

            return False

    return True

def notnear3_new(pt,solution,atomj,A): # shortest

    for i,p in enumerate(solution):

        atom = atomj[i]

        if atom in the_heavy:

            dd = 2.2

        else:

            dd = 1.2

        if not notnear2(pt,[p],dd,A):

            return False

    return True

def trianglebonding(p,solution,A):

    if len(solution)<2:

        return False

    x,y,z = p

    n = len(solution)

    for i in range(n-1):

        xi,yi,zi = solution[i]

        ri = dis_old([x-xi,y-yi,z-zi],A)

        if ri>=2.25: continue

        for j in range(i+1,n):

            xj,yj,zj = solution[j]

            rj = dis_old([x-xj,y-yj,z-zj],A)
```

```
        if rj>=2.25: continue

        rij = dis_old([xi-xj,yi-yj,zi-zj],A)

        if rij>=2.25: continue

        if ri<2.25 and rj<2.25 and rij<2.25: return True

    return False

def trianglebonding1(p,solution,A):

    #return False

    if len(solution)<2:

        return False

    rlim =2.25 # 4

    x,y,z = p

    n = len(solution)

    for i in range(n-1):

        xi,yi,zi = solution[i]

        ri = dis_old([x-xi,y-yi,z-zi],A)

        if ri>=rlim: continue

        for j in range(i+1,n):

            xj,yj,zj = solution[j]

            rj = dis_old([x-xj,y-yj,z-zj],A)

            if rj>=rlim: continue

            rij = dis_old([xi-xj,yi-yj,zi-zj],A)

            if rij>=rlim: continue

            if ri<2.25 and rj<rlim and rij<rlim: return True

            #if ri<rlim and rj<2.25 and rij<rlim: return True

            #if ri<rlim and rj<rlim and rij<rlim: return True

    return False
```

```python
def analyze_simple_molecule(molecule): # no parentheses
    # CH3OH --> {'C':1, 'H':4, 'O':1}
    atoms = {}
    n = len(molecule)
    i = 0
    while i < n:
        symbol = molecule[i]
        i += 1
        if i < n:
            c = molecule[i]
            if c.islower():
                symbol += c
                i += 1
        num = ''
        while True:
            if i < n:
                c = molecule[i]
                if c.isdigit():
                    num += c
                    i += 1
                    if i == n:
                        break
                else:
                    break
            else:
                break
```

```python
        if num == '':
            num = 1
        else:
            num = int(num)
        atoms[symbol] = atoms.setdefault(symbol, 0) + num
    return atoms

def remove_parentheses(molecule):
    # (CH3O)2CO-->C2H6O2C1O1

    # replace {} and [] with ()
    new_molecule = ''
    for i in range(len(molecule)):
        c = molecule[i]
        if c in ['{','[']:
            new_molecule += '('
        elif c in ['}',']']:
            new_molecule += ')'
        else:
            new_molecule += c

    j = new_molecule.find(')')
    if j==-1:
        return molecule
    i = new_molecule[:j].rfind('(')
    if j==len(new_molecule)-1:
        return new_molecule[:i]+new_molecule[i+1:j]
```

```python
        num = ''
        k = j+1
        while k < len(new_molecule):
            c = new_molecule[k]
            if c.isdigit():
                num += c
                k += 1
            else:
                break
        if num=='':
            num = 1
        else:
            num = int(num)
        part1 = new_molecule[:i]
        part2 = new_molecule[i+1:j]
        part3 = new_molecule[k:]
        atoms = analyze_simple_molecule(part2)
        for a in atoms:
            atoms[a] *= num
        part2 = ''
        for a in atoms:
            part2 += a + str(atoms[a])
        return part1 + part2 + part3

def to_atoms_dict(molecule):
    # (CH3O)2CO-->{'C':3, 'H':6, 'O':3}
    old_molecule = molecule
    while True:
```

```python
        new_molecule = remove_parentheses(old_molecule)

        if new_molecule==old_molecule:

            break

        else:

            old_molecule = new_molecule

    return analyze_simple_molecule(new_molecule)

def to_molecular_formula(atoms):

    # {'C':1, 'H':4, 'O':1}-->CH3OH

    formula = ''

    for atom in atoms:

        formula += str(atom)

        if atoms[atom]>1:

            formula += str(atoms[atom])

    return formula

def molecular_weight(molecule):

    atoms = to_atoms_dict(molecule)

    MW = 0

    for atom in atoms:

        n = atoms[atom]          # number of atoms

        M = elements[atom]['M']   # atomic weight

        MW += n * M

    return MW

def density(molecule,Z,A):

    # return density in g/cm^3

    MW = molecular_weight(molecule)
```

```
    cell_V = cell_volume(A)

    return Z*MW/0.6022/cell_V

def get_content(molecule,Z):

    atoms = to_atoms_dict(molecule)

    for k in atoms:

        atoms[k] *= Z

    return atoms

def kill_half_hkl(h,k,l,F2):

    hp,kp,lp,F2p = [],[],[],[]

    for i in range(len(h)):

        if l[i]>0:

            hp.append(h[i])

            kp.append(k[i])

            lp.append(l[i])

            F2p.append(F2[i])

        elif l[i]==0:

            if k[i]>0:

                hp.append(h[i])

                kp.append(k[i])

                lp.append(l[i])

                F2p.append(F2[i])

            elif k[i]==0:

                if h[i]>0:

                    hp.append(h[i])
```

```
                kp.append(k[i])

                lp.append(l[i])

                F2p.append(F2[i])

        hp,kp,lp,F2p = np.array(hp),np.array(kp),np.array(lp),np.array(F2p)

        return (hp,kp,lp,F2p)

sct = {

'Ac': (np.array([35.6597 , 23.1032 , 12.5977 ,  4.08655]),

        np.array([ 0.589092,  3.65155 , 18.599  , 117.02   ]),

        13.5266),

 'Ag': (np.array([19.2808, 16.6885,  4.8045,  1.0463]),

        np.array([ 0.6446,  7.4726, 24.6605, 99.8156]),

        5.179),

'Al': (np.array([6.4202, 1.9002, 1.5936, 1.9646]),

        np.array([ 3.0387,  0.7426, 31.5472, 85.0886]),

        1.1151),

 'Am': (np.array([36.6706 , 24.0992 , 17.3415 ,  3.49331]),

        np.array([ 0.483629,  3.20647 , 14.3136 , 102.273   ]),

        13.3592),

 'Ar': (np.array([7.4845, 6.7723, 0.6539, 1.6442]),

        np.array([ 0.9072, 14.8407, 43.8983, 33.3929]),

        1.4445),

'As': (np.array([16.6723,  6.0701,  3.4313,  4.2779]),

        np.array([ 2.6345,  0.2647, 12.9479, 47.7972]),

        2.531),
```

```
'At': (np.array([35.3163 , 19.0211 ,  9.49887,  7.42518]),
       np.array([ 0.68587,  3.97458, 11.3824 , 45.4715 ]),
       13.7108),

'Au': (np.array([16.8819, 18.5913, 25.5582,  5.86  ]),
       np.array([ 0.4611,  8.6216,  1.4826, 36.3956]),
       12.0658),

'B': (np.array([2.0545, 1.3326, 1.0979, 0.7068]),
      np.array([23.2185,  1.021 , 60.3498,  0.1403]),
      -0.1932),

'Ba': (np.array([20.3361, 19.297 , 10.888 ,  2.6959]),
       np.array([ 3.216 ,  0.2756, 20.2073, 167.202 ]),
       2.7731),

'Be': (np.array([1.5919, 1.1278, 0.5391, 0.7029]),
       np.array([ 43.6427,   1.8623, 103.483 ,   0.542 ]),
       0.0385),

'Bi': (np.array([33.3689, 12.951 , 16.5877,  6.4692]),
       np.array([ 0.704 ,  2.9238,  8.7937, 48.0093]),
       13.5782),

'Bk': (np.array([36.7881 , 24.7736 , 17.8919 ,  4.23284]),
       np.array([ 0.451018,  3.04619 , 12.8946  , 86.003   ]),
       13.2754),

'Br': (np.array([17.1789,  5.2358,  5.6377,  3.9851]),
       np.array([ 2.1723, 16.5796,  0.2609, 41.4328]),
       2.9557),

'C': (np.array([2.31  , 1.02  , 1.5886, 0.865 ]),
      np.array([20.8439, 10.2075,  0.5687, 51.6512]),
      0.2156),

'Ca': (np.array([8.6266, 7.3873, 1.5899, 1.0211]),
```

```
        np.array([ 10.4421,   0.6599, 85.7484, 178.437 ]),

        1.3751),

'Cd': (np.array([19.2214, 17.6444,  4.461 ,  1.6029]),

        np.array([ 0.5946,  6.9089, 24.7008, 87.4825]),

        5.0694),

'Ce': (np.array([21.1671 , 19.7695 , 11.8513 ,  3.33049]),

        np.array([ 2.81219 ,  0.226836, 17.6083 , 127.113  ]),

        1.86264),

'Cf': (np.array([36.9185 , 25.1995 , 18.3317 ,  4.24391]),

        np.array([ 0.437533, 3.00775 , 12.4044 , 83.7881 ]),

        13.2674),

'Cl': (np.array([11.4604, 7.1964, 6.2556, 1.6455]),

        np.array([1.04000e-02, 1.16620e+00, 1.85194e+01, 4.77784e+01]),

        -9.5574),

'Cm': (np.array([36.6488 , 24.4096 , 17.399  ,  4.21665]),

        np.array([ 0.465154, 3.08997 , 13.4346 , 88.4834 ]),

        13.2887),

'Co': (np.array([12.2841, 7.3409, 4.0034, 2.3488]),

        np.array([ 4.2791,  0.2784, 13.5359, 71.1692]),

        1.0118),

'Cr': (np.array([10.6406, 7.3537, 3.324 , 1.4922]),

        np.array([ 6.1038,  0.392 , 20.2626, 98.7399]),

        1.1832),

'Cs': (np.array([20.3892, 19.1062, 10.662 , 1.4953]),

        np.array([ 3.569 ,  0.3107, 24.3879, 213.904 ]),

        3.3352),

'Cu': (np.array([13.338 , 7.1676, 5.6158, 1.6735]),

        np.array([ 3.5828,  0.247 , 11.3966, 64.8126]),
```

```
    1.191),
'Dy': (np.array([26.507  , 17.6383 , 14.5596 ,  2.96577]),
       np.array([ 2.1802 ,   0.202172, 12.1899 , 111.874   ]),
       4.29728),
'Er': (np.array([27.6563 , 16.4285 , 14.9779 ,  2.98233]),
       np.array([ 2.07356 ,  0.223545, 11.3604 , 105.703   ]),
       5.92046),
'Eu': (np.array([24.6274, 19.0886, 13.7603,  2.9227]),
       np.array([ 2.3879,   0.1942, 13.7546, 123.174 ]),
       2.5745),
'F': (np.array([3.5392, 2.6412, 1.517 , 1.0243]),
      np.array([10.2825,  4.2944,  0.2615, 26.1476]),
      0.2776),
'Fe': (np.array([11.7695,  7.3573,  3.5222,  2.3045]),
       np.array([ 4.7611,  0.3072, 15.3535, 76.8805]),
       1.0369),
'Fr': (np.array([35.9299 , 23.0547 , 12.1439 ,  2.11253]),
       np.array([ 0.646453,  4.17619 , 23.1052 , 150.645   ]),
       13.7247),
'Ga': (np.array([15.2354,  6.7006,  4.3591,  2.9623]),
       np.array([ 3.0669,  0.2412, 10.7805, 61.4135]),
       1.7189),
'Gd': (np.array([25.0709 , 19.0798 , 13.8518 ,  3.54545]),
       np.array([ 2.25341 ,  0.181951, 12.9331 , 101.398   ]),
       2.4196),
'Ge': (np.array([16.0816,  6.3747,  3.7068,  3.683 ]),
       np.array([ 2.8509,  0.2516, 11.4468, 54.7625]),
       2.1313),
```

'H': (np.array([0.489918, 0.262003, 0.196767, 0.049879]),

np.array([20.6593 , 7.74039, 49.5519 , 2.20159]),

0.001305),

'He': (np.array([0.8734, 0.6309, 0.3112, 0.178 ]),

np.array([ 9.1037, 3.3568, 22.9276, 0.9821]),

0.0064),

'Hf': (np.array([29.144 , 15.1726, 14.7586 , 4.30013]),

np.array([ 1.83262 , 9.5999 , 0.275116, 72.029 ]),

8.58154),

'Hg': (np.array([20.6809, 19.0417, 21.6575, 5.9676]),

np.array([ 0.545 , 8.4484, 1.5729, 38.3246]),

12.6089),

'Ho': (np.array([26.9049 , 17.294 , 14.5583 , 3.63837]),

np.array([ 2.07051, 0.19794, 11.4407 , 92.6566 ]),

4.56796),

'I': (np.array([20.1472, 18.9949, 7.5138, 2.2735]),

np.array([ 4.347 , 0.3814, 27.766 , 66.8776]),

4.0712),

'In': (np.array([19.1624, 18.5596, 4.2948, 2.0396]),

np.array([ 0.5476, 6.3776, 25.8499, 92.8029]),

4.9391),

'Ir': (np.array([27.3049 , 16.7296 , 15.6115 , 5.83377]),

np.array([ 1.59279 , 8.86553 , 0.417916, 45.0011 ]),

11.4722),

'K': (np.array([8.2186, 7.4398, 1.0519, 0.8659]),

np.array([ 12.7949, 0.7748, 213.187 , 41.6841]),

1.4228),

'Kr': (np.array([17.3555, 6.7286, 5.5493, 3.5375]),

```
        np.array([ 1.9384, 16.5623,  0.2261, 39.3972]),
        2.825),
'La': (np.array([20.578 , 19.599  , 11.3727 ,  3.28719]),
        np.array([ 2.94817 ,  0.244475, 18.7726 , 133.124  ]),
        2.14678),
'Li': (np.array([1.1282, 0.7508, 0.6175, 0.4653]),
        np.array([ 3.9546,   1.0524, 85.3905, 168.261 ]),
        0.0377),
'Lu': (np.array([28.9476 , 15.2208 , 15.1    ,  3.71601]),
        np.array([ 1.90182 ,  9.98519 ,  0.261033, 84.3298  ]),
        7.97628),
'Mg': (np.array([5.4204, 2.1735, 1.2269, 2.3073]),
        np.array([ 2.8275, 79.2611,  0.3808,  7.1937]),
        0.8584),
'Mn': (np.array([11.2819,  7.3573,  3.0193,  2.2441]),
        np.array([ 5.3409,  0.3432, 17.8674, 83.7543]),
        1.0896),
'Mo': (np.array([ 3.7025, 17.2356, 12.8876,  3.7429]),
        np.array([ 0.2772,  1.0958, 11.004 , 61.6584]),
        4.3875),
'N': (np.array([12.2126,  3.1322,  2.0125,  1.1663]),
        np.array([5.70000e-03, 9.89330e+00, 2.89975e+01, 5.82600e-01]),
        -11.529),
'Na': (np.array([4.7626, 3.1736, 1.2674, 1.1128]),
        np.array([ 3.285 ,   8.8422,  0.3136, 129.424 ]),
        0.676),
'Nb': (np.array([17.6142 , 12.0144 ,  4.04183,  3.53346]),
        np.array([ 1.18865 , 11.766   ,  0.204785, 69.7957  ]),
```

```
        3.75591),
'Nd': (np.array([22.6845 , 19.6847 , 12.774  ,  2.85137]),
        np.array([ 2.66248 ,  0.210628, 15.885  , 137.903  ]),
        1.98486),
'Ne': (np.array([3.9553, 3.1125, 1.4546, 1.1251]),
        np.array([ 8.4042,  3.4262,  0.2306, 21.7184]),
        0.3515),
'Ni': (np.array([12.8376,  7.292 ,  4.4438,  2.38  ]),
        np.array([ 3.8785,  0.2565, 12.1763, 66.3421]),
        1.0341),
'Np': (np.array([36.1874, 23.5964, 15.6402,  4.1855]),
        np.array([ 0.511929,  3.25396 , 15.3622  , 97.4908  ]),
        13.3573),
'O': (np.array([3.0485, 2.2868, 1.5463, 0.867 ]),
        np.array([13.2771,  5.7011,  0.3239, 32.9089]),
        0.2508),
'Os': (np.array([28.1894, 16.155  , 14.9305,  5.67589]),
        np.array([ 1.62903 ,  8.97948 ,  0.382661, 48.1647  ]),
        11.0005),
'P': (np.array([6.4345, 4.1791, 1.78  , 1.4908]),
        np.array([ 1.9067, 27.157 ,  0.526 , 68.1645]),
        1.1149),
'Pa': (np.array([35.8847, 23.2948, 14.1891,  4.17287]),
        np.array([ 0.547751,  3.41519 , 16.9235  , 105.251   ]),
        13.4287),
'Pb': (np.array([31.0617, 13.0637, 18.442 ,  5.9696]),
        np.array([ 0.6902,  2.3576,  8.618 , 47.2579]),
        13.4118),
```

'Pd': (np.array([19.3319 , 15.5017 , 5.29537 , 0.605844]),

 np.array([ 0.698655, 7.98929 , 25.2052 , 76.8986 ]),

 5.26593),

'Pm': (np.array([23.3405 , 19.6095 , 13.1235 , 2.87516]),

 np.array([ 2.5627 , 0.202088, 15.1009 , 132.721 ]),

 2.02876),

'Po': (np.array([34.6726 , 15.4733 , 13.1138 , 7.02588]),

 np.array([ 0.700999, 3.55078 , 9.55642 , 47.0045 ]),

 13.677),

'Pr': (np.array([22.044 , 19.6697 , 12.3856 , 2.82428]),

 np.array([ 2.77393 , 0.222087, 16.7669 , 143.644 ]),

 2.0583),

'Pt': (np.array([27.0059, 17.7639, 15.7131, 5.7837]),

 np.array([ 1.51293 , 8.81174 , 0.424593, 38.6103 ]),

 11.6883),

'Pu': (np.array([36.5254 , 23.8083 , 16.7707 , 3.47947]),

 np.array([ 0.499384, 3.26371 , 14.9455 , 105.98 ]),

 13.3812),

'Ra': (np.array([35.763 , 22.9064 , 12.4739 , 3.21097]),

 np.array([ 0.616341, 3.87135 , 19.9887 , 142.325 ]),

 13.6211),

'Rb': (np.array([17.1784, 9.6435, 5.1399, 1.5292]),

 np.array([ 1.7888, 17.3151, 0.2748, 164.934 ]),

 3.4873),

'Re': (np.array([28.7621 , 15.7189 , 14.5564 , 5.44174]),

 np.array([ 1.67191, 9.09227, 0.3505, 52.0861 ]),

 10.472),

'Rh': (np.array([19.2957 , 14.3501 , 4.73425 , 1.28918]),

```
        np.array([ 0.751536,  8.21758 , 25.8749  , 98.6062  ]),

        5.328),

'Rn': (np.array([35.5631, 21.2816,  8.0037,  7.4433]),

        np.array([ 0.6631,  4.0691, 14.0422, 44.2473]),

        13.6905),

'Ru': (np.array([19.2674 , 12.9182 ,  4.86337,  1.56756]),

        np.array([ 0.80852,  8.43467, 24.7997 , 94.2928 ]),

        5.37874),

'S': (np.array([6.9053, 5.2034, 1.4379, 1.5863]),

        np.array([ 1.4679, 22.2151,  0.2536, 56.172 ]),

        0.8669),

'Sb': (np.array([19.6418, 19.0455,  5.0371,  2.6827]),

        np.array([ 5.3034,  0.4607, 27.9074, 75.2825]),

        4.5909),

'Sc': (np.array([9.189 , 7.3679, 1.6409, 1.468 ]),

        np.array([ 9.0213,  0.5729, 136.108 ,  51.3531]),

        1.3329),

'Se': (np.array([17.0006,  5.8196,  3.9731,  4.3543]),

        np.array([ 2.4098,  0.2726, 15.2372, 43.8163]),

        2.8409),

'Si': (np.array([6.292,3.035,1.989,1.541]),

        np.array([2.439,32.334,0.678,81.694]),

        1.141),

'Sm': (np.array([24.0042 , 19.4258 , 13.4396 ,  2.89604]),

        np.array([ 2.47274 ,  0.196451, 14.3996 , 128.007  ]),

        2.20963),

'Sn': (np.array([19.1889, 19.1005,  4.4585,  2.4663]),

        np.array([ 5.8303,  0.5031, 26.8909, 83.9571]),
```

4.7821),

'Sr': (np.array([17.5663,  9.8184,  5.422 ,  2.6694]),

np.array([  1.5564,  14.0988,   0.1664, 132.376 ]),

2.5064),

'Ta': (np.array([29.2024 , 15.2293 , 14.5135 ,  4.76492]),

np.array([  1.77333 ,  9.37046,  0.295977, 63.3644 ]),

9.24354),

'Tb': (np.array([25.8976 , 18.2185 , 14.3167 ,  2.95354]),

np.array([  2.24256 ,  0.196143, 12.6648 , 115.362  ]),

3.58324),

'Tc': (np.array([19.1301 , 11.0948 ,  4.64901,  2.71263]),

np.array([  0.864132,  8.14487 , 21.5707 , 86.8472 ]),

5.40428),

'Te': (np.array([19.9644 , 19.0138 ,  6.14487,  2.5239 ]),

np.array([  4.81742 ,  0.420885, 28.5284 , 70.8403 ]),

4.352),

'Th': (np.array([35.5645 , 23.4219 , 12.7473 ,  4.80703]),

np.array([  0.563359,  3.46204 , 17.8309 , 99.1722 ]),

13.4314),

'Ti': (np.array([9.7595, 7.3558, 1.6991, 1.9021]),

np.array([  7.8508,   0.5  ,  35.6338, 116.105 ]),

1.2807),

'Tl': (np.array([27.5446 , 19.1584 , 15.538  ,  5.52593]),

np.array([  0.65515,  8.70751,  1.96347, 45.8149 ]),

13.1746),

'Tm': (np.array([28.1819 , 15.8851 , 15.1542 ,  2.98706]),

np.array([  2.02859 ,  0.238849, 10.9975 , 102.961  ]),

6.75621),

```
'U': (np.array([36.0228, 23.4128, 14.9491,  4.188 ]),

      np.array([  0.5293,   3.3253,  16.0927, 100.613 ]),

      13.3966),

'V': (np.array([10.2971,  7.3511,  2.0703,  2.0571]),

      np.array([  6.8657,   0.4385,  26.8938, 102.478 ]),

      1.2199),

'W': (np.array([29.0818 , 15.43   ,  14.4327 ,  5.11982]),

      np.array([  1.72029 ,  9.2259  ,  0.321703, 57.056   ]),

      9.8875),

'Xe': (np.array([20.2933, 19.0298,  8.9767,  1.99  ]),

       np.array([  3.9282 ,  0.344  , 26.4659, 64.2658]),

      3.7118),

'Y': (np.array([17.776  , 10.2946 ,  5.72629 ,  3.26588]),

      np.array([  1.4029  ,  12.8006  ,   0.125599, 104.354   ]),

      1.91213),

'Yb': (np.array([28.6641 , 15.4345 , 15.3087 ,  2.98963]),

       np.array([  1.9889  ,   0.257119, 10.6647  , 100.417   ]),

      7.56672),

'Zn': (np.array([14.0743,  7.0318,  5.1652,  2.41  ]),

       np.array([  3.2655 ,  0.2333, 10.3163, 58.7097]),

      1.3041),

'Zr': (np.array([17.8765 , 10.948  ,  5.41732 ,  3.65721]),

       np.array([  1.27618 , 11.916   ,  0.117622, 87.6627  ]),

      2.06929)}

def fsc(sl,atom,U=0.05):

    # sl = sin(th)/lambda
```

```python
        B = 8*pi*pi*U*U # by convention, B=8*pi*U should have been used

                        # because of this, when U=0.05 is used,

                        # by convention, effectively U=0.0025 is used

    sa,sb,sc = sct[atom]

    return ((sa*exp(-sb*sl*sl)).sum()+sc)*exp(-B*sl*sl)

def get_sl(h,k,l,A):

    T = matrix_inv(A)

    T11,T22,T33 = T[0][0],T[1][1],T[2][2]

    T12,T13,T23 = T[0][1],T[0][2],T[1][2]

    # sl = sin(th)/lambda

    sl = sqrt(T11*h*h+T22*k*k+T33*l*l+2*T12*h*k+2*T13*h*l+2*T23*l*k)/2

    return sl

def scaling_factor(content,sl,h,k,l,F2,U=0.05):

    # determine scaling factor

    total = 0

    for atom in content:

        fj = np.array([fsc(s,atom,U=0.05) for s in sl])

        F2c_total = (fj**2).sum()

        total += F2c_total*content[atom]

    total_obs = F2.sum()

    scale = total/total_obs

    return scale
```

```python
def atomj_solution(atom_list):

    atomj,atom_labels,solution = [],[],[]

    for atom,label,x,y,z in atom_list:

        x,y,z = float(x),float(y),float(z)

        atomj.append(atom)

        atom_labels.append(label)

        solution.append((x,y,z))

    return (atomj,atom_labels,solution)

def to_atom_list(atomj,atom_labels,solution):

    atom_list = []

    for i,atom in enumerate(atomj):

        x,y,z = solution[i]

        atom_list.append((atom,atom_labels[i],x,y,z))

    return atom_list

def save_solution(heavy_atom,heavy_label,nheavy,solution,light_label='1',

    light_atom='C',do_copy=False):

    if not do_copy: return # not to save solution, but copy to clipboard

    num = 0

    text = ''

    #with open('solution.txt','w') as f:

    if True:

        for x,y,z in solution:
```

```
        num += 1

        if num<nheavy+1:

            q = st3(heavy_atom[num-1]+str(num))+heavy_label[num-1]

        else:

            q = st3(light_atom+str(num))+light_label

        xt = st2(round(x+0,4))

        yt = st2(round(y+0,4))

        zt = st2(round(z+0,4))

        st = '  11.00 0.05  '

        line = q+xt+yt+zt+st

        #print(line)

        #f.write(line+'\n')

        text += line+'\n'

    if do_copy:cp(text[:-1])

    #print('solution peaks are saved')

def direct_solve5(h,k,l,Fo,A,heavy_atom,ntotal,Nheavy,heavy_label,light_label='1',

    light_atom='C',atom_list=None,starttime=None,runs=-1,r1best=1e8):

    # run single atom global min to build model from scratch

    # ntoal = expand to

    # atom_list = already determined part

    # heavy_atom and corresponding Nheavy: prospect atom list

    # note: ntotal could be <, or =, or > Nheavy

    runs='globalmin'

    if starttime is None: starttime = time()

    previoustime = starttime
```

```
a,b,c = abc(A)

sl = get_sl(h,k,l,A)

f2C = np.array([fsc(s,light_atom) for s in sl])

f2a = {}

for atom in heavy_atom:

    if atom not in f2a:

        f2a[atom] = np.array([fsc(s,atom) for s in sl])

f2S = [f2a[heavy_atom[i]]  for i in range(Nheavy)]

f2 = f2S[0]

# calculate correction

for i in range(len(f2S)):

    if i == 0:

        fcorrection = f2S[i]**2

    else:

        fcorrection += f2S[i]**2

# starting point

if atom_list:

    atomj,atom_labels,solution=atomj_solution(atom_list)

    if len(atomj)<Nheavy:

        iheavy = len(atomj)-1
```

```
        else:

            heavy_atom = atomj

            heavy_label = atom_labels

            Nheavy = len(atomj)

            iheavy = Nheavy-1

    else:

        iheavy = 0

        atomj = [heavy_atom[0]]

        atom_labels = [heavy_label[0]]

        solution = [(0.3,0.3,0.3)]

    #save_solution(heavy_atom,heavy_label,Nheavy,solution)

    nheavy = len(atomj)

    nlight = len(solution) - nheavy

    nsofar = len(solution)

    xj = np.array([s[0] for s in solution])

    yj = np.array([s[1] for s in solution])

    zj = np.array([s[2] for s in solution])

    atomj = heavy_atom[:nheavy] + [light_atom for i in range(nheavy,nsofar)]

    fj = np.array([[fsc(s,am) for am in atomj] for s in sl])

    fcorrection -= np.sum(fj**2,axis=-1)
```

```python
Ahj = (fj * sin(tpi*(h[:,np.newaxis]*xj[np.newaxis,:]+k[:,np.newaxis]*yj[np.newaxis,:]
    +l[:,np.newaxis]*zj[np.newaxis,:])) )

Bhj = (fj * cos(tpi*(h[:,np.newaxis]*xj[np.newaxis,:]+k[:,np.newaxis]*yj[np.newaxis,:]
    +l[:,np.newaxis]*zj[np.newaxis,:])) )

Ah1 =np.sum(Ahj ,axis=-1)

Bh1 =np.sum(Bhj ,axis=-1)

iheavy += 1

s_precision =  0.4

print('set up new candidates')

s = s_precision

nx,ny,nz = int(a/s),int(b/s),int(c/s)

candidates = []

for ix in range(nx):

    x = ix/nx

    for iy in range(ny):

        y = iy/ny

        for iz in range(nz):

            z = iz/nz

            p = (x,y,z)

            if notnear3_new(p,solution,atomj,A):

                if True: # not trianglebonding(p,solution,A):

                    candidates.append(p)

dt = int(time()-starttime)

previoustime=time()

print(str(len(candidates))+' new candidates ready at time '+str(dt))
```

```
Ntotal = ntotal

print('expand to '+str(Ntotal)+' atoms')

selected = candidates[:]

nextend = nsofar-1

r1old = 1e300

r1 = 1e300

while len(solution) < Ntotal:

    nextend += 1

    if nextend < Nheavy:

        f2 = f2S[iheavy]

    else:

        f2 = f2C

    fcorrection -= f2**2

    precision = s_precision

    p_found = None

    selected_old, r1s_old, do_it = [], [], False

    while precision > 0.001:

    # coarse search in grids 0.4A to locate the atom

        r1s = []

        for x,y,z in selected:

            try:
```

```
        r1 = r1s_old[selected_old.index((x,y,z))]

    except:

        angle = tpi*(h*x+k*y+l*z)

        Ahj = f2 * sin(angle)

        Bhj = f2 * cos(angle)

        Ah =Ah1 + Ahj

        Bh =Bh1 + Bhj

        F2c = Ah**2+Bh**2

        Dh = sqrt(F2c+fcorrection)-Fo

        r1 = ((Dh)**2).sum()

    r1s.append(r1)

if not do_it:

    do_it = True

else:

    selected_old,r1s_old = selected[:], r1s[:]

r1s = np.array(r1s)

ir = np.argmin(r1s)

p_found = selected[ir]

# improve precision

precision /= 2

lim = precision

x0,y0,z0 = selected[ir]

x = np.arange(x0-lim,x0+1.01*lim,precision)

y = np.arange(y0-lim,y0+1.01*lim,precision)

z = np.arange(z0-lim,z0+1.01*lim,precision)

X,Y,Z = np.meshgrid(x,y,z)
```

```
    selected = []

    for index,x in np.ndenumerate(X):

        y,z = Y[index],Z[index]

        p = (x,y,z)

        if notnear3_new(p,solution,atomj,A):

            if True: # not trianglebonding(p,solution,A):

                selected.append(p) # 1.2

(x,y,z)=p_found

p_found = (put_in_cell(x),put_in_cell(y),put_in_cell(z))

solution.append(p_found)

if nextend < Nheavy:

    atomj.append(heavy_atom[iheavy])

    atom_labels.append(heavy_label[iheavy])

    iheavy += 1

else:

    atomj.append(light_atom)

    atom_labels.append(light_label)

    iheavy += 1

save_history(to_atom_list(atomj,atom_labels,solution),runs=iheavy)

x,y,z = solution[-1]

angle = tpi*(h*x+k*y+l*z)

Ahj = f2 * sin(angle)

Bhj = f2 * cos(angle)
```

```
Ah1 += Ahj

Bh1 += Bhj

F2c = Ah1**2+Bh1**2

r1 = ((sqrt(F2c+fcorrection)-Fo)**2).sum()

#save_solution(heavy_atom,heavy_label,Nheavy,solution,light_label,light_atom)

timenow = time()

timeinterval = timenow-previoustime

totaltime = timenow-starttime

previoustime = timenow

print(runs,iheavy,int(r1/1e3),int(timeinterval),int(totaltime),int(r1best/1e6))

with open('history.txt','a') as f:

    print(",file=f)

    print(runs,iheavy,int(r1/1e3),int(timeinterval),

        int(totaltime),int(r1best/1e3),file=f)

if len(solution)==Ntotal: break

r1old = r1

selected = []

for p in candidates:

    if notnear3_new(p,solution,atomj,A):

        if True: # not trianglebonding(p,solution,A):

            selected.append(p)

candidates = selected[:]

selected_old, r1s_old, do_it = [], [], False

print('finished, solution saved')

with open('history.txt','a') as f:
```

```
    print('total time: ',runs,int(time()-starttime),file=f)

solution = do_arrange(solution,A)

save_history(to_atom_list(atomj,atom_labels,solution),runs='final solution')

save_solution(atomj,atom_labels,len(atomj),solution,light_label='1',

    light_atom='C',do_copy=True)

return (atomj,atom_labels,solution,r1)

def r1_holes(h,k,l,F2,A,heavy_atom,ntotal,Nheavy,heavy_label,light_label='1',

    light_atom='C',atom_list=None,starttime=None,runs=1,s=0.4,n_refine=3,cutlimit=1.5):

    # locate atoms beyond atom_list with single-atom-r1-hole map

a,b,c = abc(A)

sl = get_sl(h,k,l,A)

F2o = F2

f2a = {}

for atom in heavy_atom:

    if atom not in f2a:

        f2a[atom] = np.array([fsc(s,atom) for s in sl])

f2S = [f2a[heavy_atom[i]]  for i in range(Nheavy)]

# starting point

atomj,atom_labels,solution=atomj_solution(atom_list)

while len(solution)<Nheavy:
```

```python
iheavy = len(solution)-1

startfrom = iheavy + 1

f2 = f2S[startfrom]

# the atom and number of it need to be located

atom = heavy_atom[startfrom]

n_atom = 0

for i in range(startfrom, len(heavy_atom)):

    if heavy_atom[i]==atom: n_atom += 1

Ntotal = n_atom+len(solution)

# calculate correction

fcorrection=None

for i in range(startfrom+1,Ntotal):

    if i == startfrom+1:

        fcorrection = f2S[i]**2

    else:

        fcorrection += f2S[i]**2

nheavy = len(solution)

nlight = 0

nsofar = len(solution)

xj = np.array([s[0] for s in solution])

yj = np.array([s[1] for s in solution])

zj = np.array([s[2] for s in solution])
```

```python
atomj = heavy_atom[:nheavy]

fj = np.array([[fsc(s,am) for am in atomj] for s in sl])

Ahj = (fj * sin(tpi*(h[:,np.newaxis]*xj[np.newaxis,:]+k[:,np.newaxis]*yj[np.newaxis,:]
    +l[:,np.newaxis]*zj[np.newaxis,:])) )

Bhj = (fj * cos(tpi*(h[:,np.newaxis]*xj[np.newaxis,:]+k[:,np.newaxis]*yj[np.newaxis,:]
    +l[:,np.newaxis]*zj[np.newaxis,:])) )

Ah1 =np.sum(Ahj ,axis=-1)

Bh1 =np.sum(Bhj ,axis=-1)

if starttime is None: starttime = time()

previoustime = starttime

def rou(x,y,z):

    angle = tpi*(h*x+k*y+l*z)

    Ahj = f2 * sin(angle)

    Bhj = f2 * cos(angle)

    Ah =Ah1 + Ahj

    Bh =Bh1 + Bhj

    F2c = Ah**2+Bh**2

    r1 = ((F2c+fcorrection-F2o)**2).sum()

    return -r1

def rou2(x,y,z):

    angle = tpi*(h*x+k*y+l*z)

    Ahj = f2 * sin(angle)

    Bhj = f2 * cos(angle)
```

```
Ah =Ah1 + Ahj

Bh =Bh1 + Bhj

F2c = Ah**2+Bh**2

r1 = (F2c+fcorrection-F2o)**2

return -r1

#s = 0.4

na,nb,nc = int(a/s),int(b/s),int(c/s)

x = np.array([i/na for i in range(-1,na+1)])

y = np.array([i/nb for i in range(-1,nb+1)])

z = np.array([i/nc for i in range(-1,nc+1)])

X,Y,Z = np.meshgrid(x,y,z)

nb,na,nc = X.shape

peaks = []

for iy in range(1,nb-1):

    if iy%10==0:

        print(runs,iy,int(time()-starttime))

    if iy==1:

        F0 = np.sum(rou2(X[iy-1,:,:,np.newaxis],Y[iy-1,:,:,np.newaxis],Z[iy-
1,:,:,np.newaxis]),axis=-1)

        F = np.sum(rou2(X[iy,:,:,np.newaxis],Y[iy,:,:,np.newaxis],Z[iy,:,:,np.newaxis]),axis=-1)

        F1 =
np.sum(rou2(X[iy+1,:,:,np.newaxis],Y[iy+1,:,:,np.newaxis],Z[iy+1,:,:,np.newaxis]),axis=-1)

    else:

        F0 = F
```

```python
        F = F1

        F1 =
np.sum(rou2(X[iy+1,:,:,np.newaxis],Y[iy+1,:,:,np.newaxis],Z[iy+1,:,:,np.newaxis]),axis=-1)

        for ix in range(1,na-1):

            for iz in range(1,nc-1):

                if F1[ix,iz]<F[ix,iz]>F0[ix,iz]:

                    if F[ix+1,iz]<F[ix,iz]>F[ix-1,iz]:

                    if F[ix,iz+1]<F[ix,iz]>F[ix,iz-1]:

                        peaks.append((X[iy,ix,iz],Y[iy,ix,iz],Z[iy,ix,iz],F[ix,iz]))

    del X

    del Y

    del Z

    del F

    del F0

    del F1

def refine3(x0,y0,z0,sx0,sy0,sz0):

    sx,sy,sz = sx0/2,sy0/2,sz0/2

    grds = {}

    for i in range(-2,3):

        for j in range(-2,3):

            for kk in range(-2,3):

                x,y,z = x0+i*sx,y0+j*sy,z0+kk*sz

                grds[(i,j,kk)] = rou(x,y,z)

    pks = []

    for i in range(-1,2):
```

```
            for j in range(-1,2):

                for kk in range(-1,2):

                    if grds[(i-1,j,kk)]<grds[(i,j,kk)]>grds[(i+1,j,kk)]:

                        if grds[(i,j-1,kk)]<grds[(i,j,kk)]>grds[(i,j+1,kk)]:

                            if grds[(i,j,kk-1)]<grds[(i,j,kk)]>grds[(i,j,kk+1)]:

                                pks.append((x0+i*sx,y0+j*sy,z0+kk*sz,grds[(i,j,kk)]))

        if pks:

            pks.sort(key = lambda s:-s[3])

            return pks[0]

        else:

            return (x0,y0,z0,rou(x0,y0,z0))

    peaks.sort(key = lambda s:-s[3])

    n = len(peaks)

    n_cut = max(100, cutlimit*Ntotal)

    print('refine peaks...')

    for i in range(n):

        if i%50==0:

            print(runs,i,int(time()-starttime))

        if i > n_cut:

            peaks = peaks[:i]

            break

        sx0,sy0,sz0 = 1/na,1/nb,1/nc

        for j in range(n_refine):

            x,y,z,f = peaks[i]

            peaks[i] = refine3(x,y,z,sx0,sy0,sz0)

            sx0,sy0,sz0 = sx0/2,sy0/2,sz0/2
```

```
    peaks.sort(key = lambda s:-s[3])

    atomj = heavy_atom[:]

    atom_labels = heavy_label[:]

    # remove ghost peaks, remove triangle bonding peaks

    pks = peaks[:]

    nn = len(pks)

    for x,y,z,ff in pks:

        n = len(solution)

        if n>=nn: break

        if len(solution)==Nheavy: break

        if notnear3((x,y,z),solution,atomj[:n],A):

            if not trianglebonding((x,y,z),solution,A):

                solution.append((x,y,z))

nn = len(solution)

if len(atomj)>nn:

    atomj =atomj[:nn]

    atom_labels = atom_labels[:nn]

solution = do_arrange(solution,A)

atom_list = to_atom_list(atomj,atom_labels,solution)

save_solution(atomj,atom_labels,len(atomj),solution,light_label='1',

    light_atom='C',do_copy=True)

print('total time: ',runs,int(time()-starttime))

with open('history.txt','a') as f:

    print('total time: ',runs,int(time()-starttime),file=f)

return atom_list
```

```python
# find corrected-peaks - using electron density and FFT

def find_corrected_peaks2(h,k,l,F2,A,atom_list,heavy_atom,heavy_label,Nheavy,
    Ntotal,runs=1,n=1,U=0.05,starttime=None,s=0.4,n_refine=3,cutlimit=1.5,
    do_copy=False,mB=1, patterson=False):
    # dual-space cycling using FFT
    h,k,l,F2 = h.copy(),k.copy(),l.copy(),F2.copy()
    if starttime is None: starttime = time()
    nelectrons = elements[heavy_atom[0]]['Z']
    atom_list.sort(key=lambda a:-elements[a[0]]['Z'])
    atomj,atom_labels,solution = atomj_solution(atom_list)
    n1,n2 = len(atomj),len(heavy_atom)
    N0=n1
    atomj = heavy_atom[:n1]+atomj[n2:]
    atom_labels = heavy_label[:n1]+atom_labels[n2:]
    atomj_input = atomj.copy()

    # final list of atoms
    atomj = atomj[:Ntotal]
    atom_labels = atom_labels[:Ntotal]
    for i in range(len(atomj),Ntotal):
        atomj.append('C')
        atom_labels.append('1')
    nn = min(len(atomj),len(heavy_atom))
    for i in range(nn):
        atomj[i] = heavy_atom[i]
```

```
        atom_labels[i]=heavy_label[i]

a,b,c = abc(A)

if True:

    # get B and E(hkl)

    sl = get_sl(h,k,l,A)

    x = sl**2

    f2a = {}

    for atom in atomj:

        if atom not in f2a:

            f2a[atom] = np.array([fsc(s,atom,U=0) for s in sl])

    fj = np.array([[f2a[atomj[i]]  for i in range(len(atomj))]]).T

    y = F2/np.sum(fj**2,axis=-1)

    # y = C*exp(-2B*x)

    def mismatch(p):

        mis = 0

        yc = p[0]*exp(-2*p[1]*x)

        return ((y-yc)**2).sum()

    res = minimize(mismatch,[1,1],method='BFGS')

    C,B = res.x

    print(C,B)

    z = C*exp(-2*B*x)

    #plt.style.use('seaborn')

    #fig, ax = plt.subplots()

    #ax.plot(x,y,'ro', x,z,'b-')

    #plt.show()
```

```
#input()

#E2 = F2*exp(2*B*sl*sl)

F2 *= exp(mB*B*sl*sl) # this is G2; mB=0 to 2, default 1

hh,kk,ll,FF2 = h.copy(),k.copy(),l.copy(),F2.copy()

h,k,l,F2 = kill_half_hkl(h,k,l,F2)

sl = get_sl(h,k,l,A)

if patterson:

    E2 = F2 *exp(B*sl*sl) # version of E2 after killing half hkl

F2o = F2.copy()

F2o[F2o<0] = 0

Fo = sqrt(F2o)

# real work starts here

xj = np.array([s[0] for s in solution])

yj = np.array([s[1] for s in solution])

zj = np.array([s[2] for s in solution])

fj = np.array([[fsc(s,am) for am in atomj_input] for s in sl])

Ahj = (fj * sin(tpi*(h[:,np.newaxis]*xj[np.newaxis,:]+k[:,np.newaxis]*yj[np.newaxis,:]

    +l[:,np.newaxis]*zj[np.newaxis,:])) )

Bhj = (fj * cos(tpi*(h[:,np.newaxis]*xj[np.newaxis,:]+k[:,np.newaxis]*yj[np.newaxis,:]

    +l[:,np.newaxis]*zj[np.newaxis,:])) )

Ah1 =np.sum(Ahj ,axis=-1)
```

```python
Bh1 =np.sum(Bhj ,axis=-1)

F2c = Ah1**2+Bh1**2

Fc = sqrt(F2c)

Bh1[Bh1==0]=1e-100

phi = atan(Ah1/Bh1)/tpi

phi[Bh1<0] += 0.5

# this is the starting atomic model in the electron density point of view:

def roue(x,y,z):

    return (Fc *cos(tpi * (h*x+k*y+l*z-phi))).sum()*2

hmax = max(hh.max(),-hh.min())

kmax = max(kk.max(),-kk.min())

lmax = max(ll.max(),-ll.min())

Nh,Nk,Nl = 2*hmax,2*kmax,2*lmax

F2o = FF2.copy()

F2min = abs(F2o)[F2o!=0].min()/10000

fc = np.ones((Nh,Nk,Nl))

fo = np.zeros((Nh,Nk,Nl))

dc = np.zeros((Nh,Nk,Nl))

mk = -np.ones((Nh,Nk,Nl))

for i in range(len(hh)):

    ih,ik,il = hh[i],kk[i],ll[i]

    if ih<0: ih += Nh
```

```python
        if ik<0: ik += Nk

        if il<0: il += Nl

        if F2o[i]>0:

            fo[ih,ik,il] = sqrt(F2o[i])

        else:

            fo[ih,ik,il] = sqrt(F2min)

        mk[ih,ik,il] = n

    if patterson:

        Eo = np.zeros(len(E2))

        Eo[E2>=0]=sqrt(E2[E2>=0])

        Eo[E2<0]=0

        # this is patterson function:

        def roup(x,y,z):

            return (Eo *cos(tpi * (h*x+k*y+l*z))).sum()*2

        pts=[(0,0,0),(1,0,0),(0,1,0),(0,0,1),(1,1,0),(1,0,1),(0,1,1),(1,1,1)]

        for ih in range(Nh):

            for ik in range(Nk):

                for il in range(Nl):

                    x,y,z = ih/Nh,ik/Nk,il/Nl

                    if notnear((x,y,z),pts,1.3,A):

                        dc[ih,ik,il] = roup(x,y,z)

                    else:

                        dc[ih,ik,il] = 0

        ih,ik,il = np.unravel_index(np.argmax(dc,axis=None),dc.shape)

        x,y,z = ih/Nh,ik/Nk,il/Nl

        grds = np.zeros((3,3,3))

        sx,sy,sz = 1/Nh,1/Nk,1/Nl
```

```
        for j in range(3):

            sx,sy,sz=sx/2,sy/2,sz/2

            for ih in range(3):

                for ik in range(3):

                    for il in range(3):

                        x0,y0,z0 = x+(ih-1)*sx,y+(ik-1)*sy,z+(il-1)*sz

                        grds[ih,ik,il]=roup(x0,y0,z0)

            ih,ik,il = np.unravel_index(np.argmax(grds,axis=None),grds.shape)

            x,y,z = x+(ih-1)*sx,y+(ik-1)*sy,z+(il-1)*sz

        xm,ym,zm = x,y,z

        # this is patterson superposition min function

        def roupm(x,y,z):

            return min(roup(x,y,z),roup(x+xm,y+ym,z+zm))

        rou0 = roupm

    else:

        rou0 = roue

for ih in range(Nh):

    for ik in range(Nk):

        for il in range(Nl):

            dc[ih,ik,il] = rou0(ih/Nh,ik/Nk,il/Nl)

# print(Nh,Nk,Nl)

# print(int(a/0.4),int(b/0.4),int(c/0.4))

angle_old = np.angle(fc)
```

```python
old_collection, new_collection = [], []

repeats = 0

fosum = (fo**2).sum()

N0-=5

cnt = 0

while True:

    cnt += 1

    dc_new = dc*0

    new_collection = []

    if N0<len(atomj): N0+=5

    solution,nsolution = [],0

    for atom in atomj[:N0]:

        ih,ik,il = np.unravel_index(np.argmax(dc,axis=None),dc.shape)

        new_collection.append((ih,ik,il))

        x,y,z = ih/Nh,ik/Nk,il/Nl

        if atom in the_heavy:

            dd=2.2

        else:

            dd=1.2

        Nh1,Nh2 = int(Nh*(x-dd/a)-1),int(Nh*(x+dd/a)+2)

        Nk1,Nk2 = int(Nk*(y-dd/b)-1),int(Nk*(y+dd/b)+2)

        Nl1,Nl2 = int(Nl*(z-dd/c)-1),int(Nl*(z+dd/c)+2)

        mask1 = np.ones((Nh,Nk,Nl))

        mask2 = np.zeros((Nh,Nk,Nl))

        for ih in range(Nh1,Nh2):

            if ih<0: ih+=Nh

            if ih>=Nh: ih-=Nh
```

```
        for ik in range(Nk1,Nk2):

            if ik<0: ik+=Nk

            if ik>=Nk: ik-=Nk

            for il in range(Nl1,Nl2):

                if il<0: il+=Nl

                if il>=Nl: il-=Nl

                if length((x,y,z),(ih/Nh,ik/Nk,il/Nl),A)<dd:

                    mask1[ih,ik,il] = 0.0

                    dx,dy,dz = (x-ih/Nh),(y-ik/Nk),(z-il/Nl)

                    #mask2[ih,ik,il] += 1/(1+(dx*dx+dy*dy+dz*dz)/0.09)

                    mask2[ih,ik,il] += exp(-(dx*dx+dy*dy+dz*dz)/0.12)

    dc_new += dc*mask2

    dc *= mask1

dc = dc_new

fc = ifftn(dc)

rf = np.absolute(fc)

th2 = (np.angle(fc)).copy()

dth = ((th2-angle_old)**2).sum()/Nh/Nk/Nl

if (dth<0.0005 and old_collection==new_collection):

    repeats+=1

else:

    repeats = 0

if repeats>2 or cnt>20: break

angle_old = th2.copy()

old_collection=new_collection

if cnt%100==0: print(cnt,int(time()-starttime),dth)

scale = sqrt((rf**2).sum()/fosum)

rf = (n+1)*fo-mk*rf/scale
```

```python
    rf[rf<0]=0

    fc = rf*(cos(th2)+1j*sin(th2))

    dc = fftn(fc)

# print('total dual space cycles = ', cnt,int(time()-starttime), 'dth = ', dth)

# with open('history.txt','a') as f:

#    print('total dual space cycles = ',int(time()-starttime), cnt,file=f)

th = np.angle(fc)/tpi

rf = np.absolute(fc)

neg = False

for i in range(len(h)):

    ih,ik,il = h[i],k[i],l[i]

    if ih<0 or ik<0 or il<0:neg=True

    if ih<0: ih += Nh

    if ik<0: ik += Nk

    if il<0: il += Nl

    phi[i] = th[ih,ik,il]

    Fc[i] = rf[ih,ik,il]
# if neg: print('yes, negatives')

scale = sqrt((Fc**2).sum()/(Fo**2).sum())

# recalculate Fc and phi, construct final atomic model

n1 = n+1

dF = n1*Fo-n*Fc/scale
```

```python
def rou(x,y,z):

    return (dF *cos(tpi * (h*x+k*y+l*z-phi))).sum()

def rou2(x,y,z):

    return dF *cos(tpi * (h*x+k*y+l*z-phi))

#s = 0.4

na,nb,nc = int(a/s),int(b/s),int(c/s)

x = np.array([i/na for i in range(-1,na+1)])

y = np.array([i/nb for i in range(-1,nb+1)])

z = np.array([i/nc for i in range(-1,nc+1)])

X,Y,Z = np.meshgrid(x,y,z)

nb,na,nc = X.shape

peaks = []

for iy in range(1,nb-1):

    if iy%10==0:

        print(runs,iy,int(time()-starttime))

    if iy==1:

        F0 = np.sum(rou2(X[iy-1,:,:,np.newaxis],Y[iy-1,:,:,np.newaxis],Z[iy-1,:,:,np.newaxis]),axis=-1)

        F = np.sum(rou2(X[iy,:,:,np.newaxis],Y[iy,:,:,np.newaxis],Z[iy,:,:,np.newaxis]),axis=-1)

        F1 = np.sum(rou2(X[iy+1,:,:,np.newaxis],Y[iy+1,:,:,np.newaxis],Z[iy+1,:,:,np.newaxis]),axis=-1)

    else:

        F0 = F
```

```
        F = F1

        F1 =
np.sum(rou2(X[iy+1,:,:,np.newaxis],Y[iy+1,:,:,np.newaxis],Z[iy+1,:,:,np.newaxis]),axis=-1)

    for ix in range(1,na-1):

        for iz in range(1,nc-1):

            if F1[ix,iz]<F[ix,iz]>F0[ix,iz]:

                if F[ix+1,iz]<F[ix,iz]>F[ix-1,iz]:

                    if F[ix,iz+1]<F[ix,iz]>F[ix,iz-1]:

                        peaks.append((X[iy,ix,iz],Y[iy,ix,iz],Z[iy,ix,iz],F[ix,iz]))

    del X

    del Y

    del Z

    del F

    del F0

    del F1

def refine3(x0,y0,z0,sx0,sy0,sz0):

    sx,sy,sz = sx0/2,sy0/2,sz0/2

    grds = {}

    for i in range(-2,3):

        for j in range(-2,3):

            for kk in range(-2,3):

                x,y,z = x0+i*sx,y0+j*sy,z0+kk*sz

                grds[(i,j,kk)] = rou(x,y,z)

    pks = []

    for i in range(-1,2):
```

```
            for j in range(-1,2):

                for kk in range(-1,2):

                    if grds[(i-1,j,kk)]<grds[(i,j,kk)]>grds[(i+1,j,kk)]:

                        if grds[(i,j-1,kk)]<grds[(i,j,kk)]>grds[(i,j+1,kk)]:

                            if grds[(i,j,kk-1)]<grds[(i,j,kk)]>grds[(i,j,kk+1)]:

                                pks.append((x0+i*sx,y0+j*sy,z0+kk*sz,grds[(i,j,kk)]))

        if pks:

            pks.sort(key = lambda s:-s[3])

            return pks[0]

        else:

            return (x0,y0,z0,rou(x0,y0,z0))

peaks.sort(key = lambda s:-s[3])

n = len(peaks)

print('refine peaks...')

for i in range(n):

    if i%50==0:

        print(runs,i,int(time()-starttime))

    if i > cutlimit*Ntotal:

        peaks = peaks[:i]

        break

    sx0,sy0,sz0 = 1/na,1/nb,1/nc

    for j in range(n_refine):

        x,y,z,f = peaks[i]

        peaks[i] = refine3(x,y,z,sx0,sy0,sz0)

        sx0,sy0,sz0 = sx0/2,sy0/2,sz0/2
```

```
peaks.sort(key = lambda s:-s[3])

x,y,z,f = peaks[0]

x = put_in_cell(x)

y = put_in_cell(y)

z = put_in_cell(z)

peaks1 = peaks[1:]

peaks = [(x,y,z,f)]

for pk1 in peaks1:

    is_new = True

    x1,y1,z1,f1 = pk1

    x1 = put_in_cell(x1)

    y1 = put_in_cell(y1)

    z1 = put_in_cell(z1)

    for pk in peaks:

        x,y,z,f = pk

        dx,dy,dz = x-x1,y-y1,z-z1

        X = [dx,dy,dz]

        if dis_old(X,A)<1.44:  # r < 1.2

            is_new = False

            break

    if is_new : peaks.append((x1,y1,z1,f1))

# save peaks

num = 0

text = ''

ff0 = peaks[0][3]
```

```python
atomj = atomj[:Ntotal]

atom_labels = atom_labels[:Ntotal]

for i in range(len(atomj),Ntotal):

    atomj.append('C')

    atom_labels.append('1')

nn = min(len(atomj),len(heavy_atom))

for i in range(nn):

    atomj[i] = heavy_atom[i]

    atom_labels[i]=heavy_label[i]

# remove ghost peaks, remove triangle bonding peaks

# remove ghost peaks, remove triangle bonding peaks

pks = peaks[1:]

peaks = [peaks[0]]

x,y,z,ff = peaks[0]

solution = [(x,y,z)]

for x,y,z,ff in pks:

    n = len(solution)

    if n>=nn: break

    if notnear3((x,y,z),solution,atomj[:n],A):

        if len(solution)>1:

            trianglebonding = False

            n = len(solution)

            for i in range(n-1):

                xi,yi,zi = solution[i]

                for j in range(i+1,n):

                    xj,yj,zj = solution[j]

                    rij = dis_old((xi-xj,yi-yj,zi-zj),A)
```

```
                if rij < 2.25:

                    ri = dis_old((xi-x,yi-y,zi-z),A)

                    rj = dis_old((x-xj,y-yj,z-zj),A)

                    if ri<2.25 and rj<2.25:

                        trianglebonding = True

                        break

                if trianglebonding: break

            if not trianglebonding:

                solution.append((x,y,z))

                peaks.append((x,y,z,ff))

        else:

            solution.append((x,y,z))

            peaks.append((x,y,z,ff))

#with open('improved.txt','w') as f:

if True:

    for x,y,z,ff in peaks[:Ntotal]:

        num += 1

        q = st3(atomj[num-1]+str(num))+atom_labels[num-1]

        xt = st2(round(x+0,4))

        yt = st2(round(y+0,4))

        zt = st2(round(z+0,4))

        st = '  11.00 10.05  '

        ff *= nelectrons/ff0

        line = q+xt+yt+zt+st+str(round(ff,2))

        #print(line)

        #f.write(line+'\n')

        text += line+'\n'
```

```python
    #print('the corrected-peaks are saved')

    if do_copy:cp(text[:-1])

    solution = []

    for x,y,z,ff in peaks[:Ntotal]:

        solution.append((x,y,z))

    n = len(solution)

    atomj = atomj[:n]

    atom_labels = atom_labels[:n]

    n = len(atomj)

    solution = solution[:n]

    solution = do_arrange(solution,A)

    atom_list = to_atom_list(atomj,atom_labels,solution)

    print('total time: ',runs,int(time()-starttime))

    with open('history.txt','a') as f:

        print('total time: ',runs,int(time()-starttime),file=f)

    return atom_list

# build model from scratch with Patterson technique

# find peaks in sophisticated way, use refine3

def solve_by_patterson3(h,k,l,F2,A,heavy_atom,heavy_label,Nheavy,

    starttime=None,s=0.4,mB=1,cutlimit=30):

    # build model from scratch with Patterson technique

    h,k,l,F2 = h.copy(),k.copy(),l.copy(),F2.copy()

    if starttime is None: starttime = time()
```

```python
# final list of atoms

atomj = heavy_atom[:]

atom_labels = heavy_label[:]

Ntotal=Nheavy

a,b,c = abc(A)

# get B and E(hkl)

sl = get_sl(h,k,l,A)

x = sl**2

f2a = {}

for atom in atomj:

    if atom not in f2a:

        f2a[atom] = np.array([fsc(s,atom,U=0) for s in sl])

fj = np.array([f2a[atomj[i]]  for i in range(len(atomj))]).T

y = F2/np.sum(fj**2,axis=-1)

# y = C*exp(-2B*x)

def mismatch(p):

    mis = 0

    yc = p[0]*exp(-2*p[1]*x)

    return ((y-yc)**2).sum()

res = minimize(mismatch,[1,1],method='BFGS')

C,B = res.x

print(C,B)

z = C*exp(-2*B*x)

#plt.style.use('seaborn')

#fig, ax = plt.subplots()

#ax.plot(x,y,'ro', x,z,'b-')
```

```
#plt.show()

#input()

#E2 = F2*exp(2*B*sl*sl)

F2 *= exp(mB*B*sl*sl) # this is G2; mB=0 to 2, default 1

hh,kk,ll,FF2 = h.copy(),k.copy(),l.copy(),F2.copy()

h,k,l,F2 = kill_half_hkl(h,k,l,F2)

sl = get_sl(h,k,l,A)

E2 = F2 *exp(B*sl*sl) # version of E2 after killing half hkl

Eo = np.zeros(len(E2))

Eo[E2>=0]=sqrt(E2[E2>=0])

Eo[E2<0]=0

Nh,Nk,Nl = int(a/s)+1,int(b/s)+1,int(c/s)+1

dc = np.zeros((Nh,Nk,Nl))

# this is patterson function:

def roup(x,y,z):

    return (Eo *cos(tpi * (h*x+k*y+l*z))).sum()*2

pts=[(0,0,0),(1,0,0),(0,1,0),(0,0,1),(1,1,0),(1,0,1),(0,1,1),(1,1,1)]

for ih in range(Nh):

    for ik in range(Nk):

        for il in range(Nl):

            x,y,z = ih/Nh,ik/Nk,il/Nl

            if notnear((x,y,z),pts,1.3,A):
```

```python
                dc[ih,ik,il] = roup(x,y,z)
            else:
                dc[ih,ik,il] = 0
ih,ik,il = np.unravel_index(np.argmax(dc,axis=None),dc.shape)
x,y,z = ih/Nh,ik/Nk,il/Nl
grds = np.zeros((3,3,3))
sx,sy,sz = 1/Nh,1/Nk,1/Nl
for j in range(3):
    sx,sy,sz=sx/2,sy/2,sz/2
    for ih in range(3):
        for ik in range(3):
            for il in range(3):
                x0,y0,z0 = x+(ih-1)*sx,y+(ik-1)*sy,z+(il-1)*sz
                grds[ih,ik,il]=roup(x0,y0,z0)
    ih,ik,il = np.unravel_index(np.argmax(grds,axis=None),grds.shape)
    x,y,z = x+(ih-1)*sx,y+(ik-1)*sy,z+(il-1)*sz
xm,ym,zm = x,y,z
# this is patterson superposition min function
def roupm(x,y,z):
    return min(roup(x,y,z),roup(x+xm,y+ym,z+zm))

#roupm=roup  # use patterson peaks instead of patterson superposition min peaks

for ih in range(Nh):
    for ik in range(Nk):
        for il in range(Nl):
            x,y,z = ih/Nh,ik/Nk,il/Nl
            dc[ih,ik,il] = roupm(x,y,z)
```

```
candidates=[]

for ih in range(Nh):

    print(ih,Nh)

    for ik in range(Nk):

        for il in range(Nl):

            x,y,z = ih/Nh,ik/Nk,il/Nl

            is_peak=True

            ih1,ih2=ih-1,ih+1

            ik1,ik2=ik-1,ik+1

            il1,il2=il-1,il+1

            if ih1<0:ih1+=Nh

            if ik1<0:ik1+=Nk

            if il1<0:il1+=Nl

            if ih2==Nh:ih2=0

            if ik2==Nk:ik2=0

            if il2==Nl:il2=0

            if dc[ih1,ik,il]>=dc[ih,ik,il]:is_peak=False

            if dc[ih2,ik,il]>=dc[ih,ik,il]:is_peak=False

            if dc[ih,ik1,il]>=dc[ih,ik,il]:is_peak=False

            if dc[ih,ik2,il]>=dc[ih,ik,il]:is_peak=False

            if dc[ih,ik,il1]>=dc[ih,ik,il]:is_peak=False

            if dc[ih,ik,il2]>=dc[ih,ik,il]:is_peak=False

            if is_peak: candidates.append((x,y,z,dc[ih,ik,il]))

rou=roupm

def refine3(x0,y0,z0,sx0,sy0,sz0):
```

```python
    sx,sy,sz = sx0/2,sy0/2,sz0/2

    grds = {}

    for i in range(-2,3):

        for j in range(-2,3):

            for kk in range(-2,3):

                x,y,z = x0+i*sx,y0+j*sy,z0+kk*sz

                grds[(i,j,kk)] = rou(x,y,z)

    pks = []

    for i in range(-1,2):

        for j in range(-1,2):

            for kk in range(-1,2):

                if grds[(i-1,j,kk)]<grds[(i,j,kk)]>grds[(i+1,j,kk)]:

                    if grds[(i,j-1,kk)]<grds[(i,j,kk)]>grds[(i,j+1,kk)]:

                        if grds[(i,j,kk-1)]<grds[(i,j,kk)]>grds[(i,j,kk+1)]:

                            pks.append((x0+i*sx,y0+j*sy,z0+kk*sz,grds[(i,j,kk)]))

    if pks:

        pks.sort(key = lambda s:-s[3])

        return pks[0]

    else:

        return (x0,y0,z0,rou(x0,y0,z0))

peaks=candidates[:]

peaks.sort(key = lambda s:-s[3])

n = len(peaks)

print('refine peaks...')

n_refine=3

for i in range(n):
```

```
    if i%50==0:

        print(n,i,int(time()-starttime))

    if i > cutlimit*Ntotal:

        peaks = peaks[:i]

        break

    sx0,sy0,sz0 = 1/Nh,1/Nk,1/Nl

    for j in range(n_refine):

        x,y,z,f = peaks[i]

        peaks[i] = refine3(x,y,z,sx0,sy0,sz0)

        sx0,sy0,sz0 = sx0/2,sy0/2,sz0/2

peaks.sort(key = lambda s:-s[3])

candidates=[]

for x,y,z,f in peaks:

    x,y,z = put_in_cell(x),put_in_cell(y),put_in_cell(z)

    candidates.append((x,y,z))

h,k,l,F2 = hh.copy(),kk.copy(),ll.copy(),FF2.copy()

Fo=F2.copy()

Fo[Fo<0]=0.0

Fo=sqrt(Fo)

sl = get_sl(h,k,l,A)

f2a = {}
```

```python
for atom in heavy_atom:

    if atom not in f2a:

        f2a[atom] = np.array([fsc(s,atom) for s in sl])

f2S = [f2a[heavy_atom[i]]  for i in range(Nheavy)]

f2 = f2S[0]

# calculate correction

for i in range(len(f2S)):

    if i == 0:

        fcorrection = f2S[i]**2

    else:

        fcorrection += f2S[i]**2

# starting point

iheavy = 0

atomj = [heavy_atom[0]]

atom_labels = [heavy_label[0]]

solution = [candidates[0]]

selected = []

for p in candidates:

    if notnear3(p,solution,atomj[:len(solution)],A):

        selected.append(p)

candidates = selected[:]
```

```python
nheavy = len(atomj)

nlight = 0

nsofar = len(solution)

xj = np.array([s[0] for s in solution])

yj = np.array([s[1] for s in solution])

zj = np.array([s[2] for s in solution])

atomj = heavy_atom[:nheavy]

fj = np.array([[fsc(s,am) for am in atomj] for s in sl])

fcorrection -= np.sum(fj**2,axis=-1)

Ahj = (fj * sin(tpi*(h[:,np.newaxis]*xj[np.newaxis,:]+k[:,np.newaxis]*yj[np.newaxis,:]
    +l[:,np.newaxis]*zj[np.newaxis,:])) )

Bhj = (fj * cos(tpi*(h[:,np.newaxis]*xj[np.newaxis,:]+k[:,np.newaxis]*yj[np.newaxis,:]
    +l[:,np.newaxis]*zj[np.newaxis,:])) )

Ah1 = np.sum(Ahj ,axis=-1)

Bh1 = np.sum(Bhj ,axis=-1)

iheavy += 1

Ntotal = Nheavy

print('expand to '+str(Ntotal)+' atoms')

selected = candidates[:]
```

```python
previoustime=time()

nextend = nsofar-1

r1old = 1e300

r1 = 1e300

while len(solution) < Ntotal:

    print(len(solution),len(candidates))

    nextend += 1

    f2 = f2S[iheavy]

    fcorrection -= f2**2

    p_found = None

    r1s = []

    for x,y,z in selected:

        angle = tpi*(h*x+k*y+l*z)

        Ahj = f2 * sin(angle)

        Bhj = f2 * cos(angle)

        Ah =Ah1 + Ahj

        Bh =Bh1 + Bhj

        F2c = Ah**2+Bh**2

        Dh = sqrt(F2c+fcorrection)-Fo

        r1 = ((Dh)**2).sum()

        r1s.append(r1)

    r1s = np.array(r1s)

    ir = np.argmin(r1s)

    p_found = selected[ir]
```

```
(x,y,z)=p_found

(x,y,z) = (put_in_cell(x),put_in_cell(y),put_in_cell(z))

solution.append((x,y,z))

atomj.append(heavy_atom[iheavy])

atom_labels.append(heavy_label[iheavy])

iheavy += 1

save_history(to_atom_list(atomj,atom_labels,solution),runs=iheavy)

x,y,z = solution[-1]

angle = tpi*(h*x+k*y+l*z)

Ahj = f2 * sin(angle)

Bhj = f2 * cos(angle)

Ah1 += Ahj

Bh1 += Bhj

F2c = Ah1**2+Bh1**2

r1 = ((sqrt(F2c+fcorrection)-Fo)**2).sum()

#save_solution(heavy_atom,heavy_label,Nheavy,solution,light_label,light_atom)

timenow = time()

timeinterval = timenow-previoustime

totaltime = timenow-starttime

previoustime = timenow

runs='patterson_tech'
```

```
        print(runs,iheavy,int(r1/1e3),int(timeinterval),int(totaltime))

        with open('history.txt','a') as f:

            print('',file=f)

            print(runs,iheavy,int(r1/1e3),int(timeinterval),

                int(totaltime),file=f)

        if len(solution)==Ntotal: break

        r1old = r1

        selected = []

        for p in candidates:

            if notnear3(p,solution,atomj[:len(solution)],A):

                selected.append(p)

        candidates = selected[:]

    print('finished, solution saved')

    with open('history.txt','a') as f:

        print('total time: ',runs,int(time()-starttime),file=f)

    solution = do_arrange(solution,A)

    save_history(to_atom_list(atomj,atom_labels,solution),runs='final solution')

    save_solution(atomj,atom_labels,len(atomj),solution,light_label='1',

        light_atom='C',do_copy=True)

    return (atomj,atom_labels,solution,r1)

def save_history(atom_list,runs,do_copy=False):

    from datetime import datetime

    num = 0

    text = ''
```

```python
    with open('history.txt','a') as f:
        f.write('\n\n\n\n\n'+str(datetime.now())+'\n')
        f.write('run number: '+str(runs)+'\n\n')
        for a,l,x,y,z in atom_list:
            num += 1
            if len(a)==2 and num>99:
                str_num = str(num)[1:]
            else:
                str_num = str(num)
            q = st3(a+str_num)+l
            xt = st2(round(x+0,4))
            yt = st2(round(y+0,4))
            zt = st2(round(z+0,4))
            st = '  11.00 0.05  '
            line = q+xt+yt+zt+st+'\n'
            f.write(line)
            text += line
        if do_copy:
            cp(text)

def generate_random_model():
    atom_list = read_atoms('a.res')
    atomj,atom_labels,solution=atomj_solution(atom_list)
    for i in range(len(solution)):
        solution[i] = (random(),random(),random())
    atom_list = to_atom_list(atomj,atom_labels,solution)
    save_history(atom_list,runs='random model')
```

```python
def do_arrange(solution, A):

    solution = solution.copy()

    is1 = list(range(len(solution)))

    is2 = [0]

    is1.remove(0)

    while is1:

        c2 = []

        for i2 in is2:

            p2 = solution[i2]

            c1 = []

            for i1 in is1:

                p1 = solution[i1]

                rmin,p1new = shortest(p2,p1,A)

                c1.append((rmin,p1new,i1))

            c1.sort(key=lambda ss:ss[0])

            c2.append(c1[0])

        c2.sort(key=lambda ss:ss[0])

        rmin,p1new,i1 = c2[0]

        solution[i1]=p1new

        print(len(solution),len(is1),i1)

        is2.append(i1)

        is1.remove(i1)

    X,Y,Z=0,0,0

    N = len(solution)

    for x,y,z in solution:

        X,Y,Z = X+x,Y+y,Z+z

    X,Y,Z=X/N,Y/N,Z/N

    for i in range(N):
```

```python
        x,y,z = solution[i]

        x,y,z = x-X+0.5,y-Y+0.5,z-Z+0.5

        solution[i]=x,y,z

    return solution

def re_arrange():

    atom_list = read_atoms('a.res')

    atomj,atom_labels,solution=atomj_solution(atom_list)

    A = matrix_A('a.res')

    is1 = list(range(len(solution)))

    is2 = [0]

    is1.remove(0)

    while is1:

        c2 = []

        for i2 in is2:

            p2 = solution[i2]

            c1 = []

            for i1 in is1:

                p1 = solution[i1]

                rmin,p1new = shortest(p2,p1,A)

                c1.append((rmin,p1new,i1))

            c1.sort(key=lambda ss:ss[0])

            c2.append(c1[0])

        c2.sort(key=lambda ss:ss[0])

        rmin,p1new,i1 = c2[0]

        solution[i1]=p1new

        print(len(solution),len(is1),i1)
```

```
    is2.append(i1)

    is1.remove(i1)

X,Y,Z=0,0,0

N = len(solution)

for x,y,z in solution:

    X,Y,Z = X+x,Y+y,Z+z

X,Y,Z=X/N,Y/N,Z/N

for i in range(N):

    x,y,z = solution[i]

    x,y,z = x-X+0.5,y-Y+0.5,z-Z+0.5

    solution[i]=x,y,z

atom_list = to_atom_list(atomj,atom_labels,solution)

save_history(atom_list,runs='re-arranged model')

def cacl_r1(atomj,solution,h,k,l,F2,A,U=0.2236):

  F2 = F2.copy()

  sl = get_sl(h,k,l,A)

  contents = {}

  for atom in atomj:

    if atom not in contents:

      contents[atom]=1

    else:

      contents[atom]+=1

  scale = scaling_factor(contents,sl,h,k,l,F2,U)
```

```
F2 *= scale

xj = np.array([s[0] for s in solution])

yj = np.array([s[1] for s in solution])

zj = np.array([s[2] for s in solution])

fj = np.array([[fsc(s,am,U) for am in atomj] for s in sl])

Ahj = (fj * sin(tpi*(h[:,np.newaxis]*xj[np.newaxis,:]+k[:,np.newaxis]*yj[np.newaxis,:]

    +l[:,np.newaxis]*zj[np.newaxis,:])) )

Bhj = (fj * cos(tpi*(h[:,np.newaxis]*xj[np.newaxis,:]+k[:,np.newaxis]*yj[np.newaxis,:]

    +l[:,np.newaxis]*zj[np.newaxis,:])) )

Ah =np.sum(Ahj ,axis=-1)

Bh =np.sum(Bhj ,axis=-1)

F2o=F2.copy()

F2o[F2o<0]=0.0

total = (fj**2).sum()

total_obs = F2o.sum()

scale = total/total_obs

F2o *= scale

# D=Ah**2+Bh**2-F2o

# return (D**2).sum()/(F2o**2).sum()

F2o=F2.copy()

F2o[F2o<0]=0

D=sqrt(Ah**2+Bh**2)-sqrt(F2o)

return (abs(D)).sum()/sqrt(F2o).sum()
```

```python
def get_r1():

    h,k,l,F2,sigF2 = read_hkl3('a.hkl')

    A = matrix_A('a.res')

    atom_list = read_atoms('a.res')

    atomj,atom_labels,solution = atomj_solution(atom_list)

    r1 = cacl_r1(atomj,solution,h,k,l,F2,A,U=0.2236)

    print('r1 = ', r1)

    return r1

def compare_models(correct_res='correct.res',init_res='init.res',

    compare_txt='compare.txt',r0=0.501):

    starttime=time()

    # list heavy_atoms

    heavy_atoms = ['I','Mo','Pd','Se','S','P','Si','Cl','Br','Pt',]

    # read correct model

    atom_list1 = read_atoms(correct_res)

    # read initial model

    atom_list2 = read_atoms(init_res)

    # check if same number of atoms

    if len(atom_list1)!=len(atom_list2):

        print('not same number of atoms, please check!')

        print(len(atom_list1),len(atom_list2))

        #sys.exit()

    A = matrix_A(correct_res)

    goodpairs = None

    smallestnballpark = -1e300
```

```
atom_list2_original=atom_list2[:]

for ii in range(len(atom_list2_original)):

#for ii in range(10):

    atom,label,x00,y00,z00 = atom_list2_original[ii]

    count = 0

    for atom,label,x0,y0,z0 in atom_list1:

        atom_list2=atom_list2_original[:]

        dx,dy,dz = x00-x0,y00-y0,z00-z0

        nballpark=0

        pairs=[]

        for atom,label,x,y,z in atom_list1:

            p1 = (x+dx,y+dy,z+dz)

            for i in range(len(atom_list2)):

                atom2,label2,x2,y2,z2=atom_list2[i]

                p2 = (x2,y2,z2)

                r = bond_length(p1,p2,A)

                if r<r0:

                    nballpark+=1

                    pairs.append((atom,atom2))

                    del atom_list2[i]

                    break

        count+=1

        if nballpark>smallestnballpark:

            smallestnballpark=nballpark

            goodpairs = pairs[:]

        if count%20==0:

            dt = int(time()-starttime)

            print(ii,count,smallestnballpark,dt)
```

```
for i in range(len(atom_list1)):

    atom,label,x,y,z = atom_list1[i]

    atom_list1[i]=(atom,label,-x,-y,-z)

for atom,label,x0,y0,z0 in atom_list1:

    atom_list2=atom_list2_original[:]

    dx,dy,dz = x00-x0,y00-y0,z00-z0

    nballpark=0

    pairs=[]

    for atom,label,x,y,z in atom_list1:

        p1 = (x+dx,y+dy,z+dz)

        for i in range(len(atom_list2)):

            atom2,label2,x2,y2,z2=atom_list2[i]

            p2 = (x2,y2,z2)

            r = bond_length(p1,p2,A)

            if r<r0:

                nballpark+=1

                pairs.append((atom,atom2))

                del atom_list2[i]

                break

    count+=1

    if nballpark>smallestnballpark:

        smallestnballpark=nballpark

        goodpairs = pairs[:]

    if count%20==0:

        dt = int(time()-starttime)

        print(ii,count,smallestnballpark,dt)

for i in range(len(atom_list1)):

    atom,label,x,y,z = atom_list1[i]
```

```python
        atom_list1[i]=(atom,label,-x,-y,-z)
pairs = goodpairs[:]
nheavy = 0
nlight = 0
for atom,label,x,y,z in atom_list1:
    if atom in heavy_atoms:
        nheavy+=1
    else:
        nlight+=1
nheavy_ballpark = 0
nlight_ballpark = 0
nheavy_assign = 0
nlight_assign = 0
for a1,a2 in pairs:
    if a1 in heavy_atoms:
        nheavy_ballpark += 1
        if a2 in heavy_atoms:
            nheavy_assign += 1
    else:
        nlight_ballpark += 1
        if not (a2 in heavy_atoms):
            nlight_assign += 1
with open(compare_txt,'w') as f:
    print('',file=f)
    print(f'{nheavy_ballpark} out of {nheavy} heavy atoms are located within 0.5 A.',file=f)
    print(f'{nheavy_assign} of these {nheavy_ballpark} atoms are correctly assigned as heavy atoms.',file=f)
    print('',file=f)
```

```
        print(f'{nlight_ballpark} out of {nlight} light atoms are located within 0.5 A.',file=f)

        print(f'{nlight_assign} of these {nlight_ballpark} atoms are correctly assigned as light
atoms.',file=f)

    print('all done',int(time()-starttime))

    return smallestnballpark

def matching(solution1,solution2,A,r0=0.501): # solution1 is the correct solution

    return 10

    # check if same number of atoms

    if len(solution1)!=len(solution2):

        print('not same number of atoms, please check!')

        print(len(solution1),len(solution2))

        #sys.exit()

    smallestnballpark = -1e300

    x00,y00,z00 = solution2[0]

    solution2c=solution2[:]

    print('matching...')

    for x0,y0,z0 in solution1:

        solution2=solution2c[:]

        dx,dy,dz = x00-x0,y00-y0,z00-z0

        nballpark=0

        for x,y,z in solution1:

            p1=(x+dx,y+dy,z+dz)
```

```
        for i,p2 in enumerate(solution2):

            r = bond_length(p1,p2,A)

            if r<r0:

                nballpark+=1

                del solution2[i]

                break

    if nballpark>smallestnballpark:

        smallestnballpark=nballpark

for i in range(len(solution1)):

    x,y,z = solution1[i]

    solution1[i]=(-x,-y,-z)

for x0,y0,z0 in solution1:

    solution2=solution2c[:]

    dx,dy,dz = x00-x0,y00-y0,z00-z0

    nballpark=0

    for x,y,z in solution1:

        p1=(x+dx,y+dy,z+dz)

        for i,p2 in enumerate(solution2):

            r = bond_length(p1,p2,A)

            if r<r0:

                nballpark+=1

                del solution2[i]

                break

    if nballpark>smallestnballpark:

        smallestnballpark=nballpark

return smallestnballpark
```

```
def apply_resolution_limit(res_file, hkl_file, dmin):

    A = matrix_A(res_file)

    h,k,l,F2,sigF2 = read_hkl3(hkl_file)

    sl = get_sl(h,k,l,A)

    slmax = 1/2/dmin

    hh,kk,ll,FF2,sigFF2 = [],[],[],[],[]

    for i in range(len(h)):

        if sl[i]<slmax:

            hh.append(h[i])

            kk.append(k[i])

            ll.append(l[i])

            FF2.append(F2[i])

            sigFF2.append(sigF2[i])

    hh,kk,ll,FF2,sigFF2 = np.array(hh),np.array(kk),np.array(ll),np.array(FF2),np.array(sigFF2)

    return (hh,kk,ll,FF2,sigFF2,len(h),len(hh))

def get_structure_solution(res_file,hkl_file,molecule,Z,ntotal_initial,

    starttime,s_dual=0.4,n_refine=3,max_runs=1000,improve_only=False,

    fast=0,n_improve=1,cutlimit=1.5,extension=False,double_first=-1,

    startfrom=1,nextra=0,mB=1,patterson=False):

    # flag fast no longer being used

    starttime=time()

    A = matrix_A(res_file)
```

```python
print('When Z=',Z,', density = ',round(density(molecule,Z,A),2))

content = get_content(molecule,Z)

atoms = list(content.keys())

atoms.sort(key=lambda a:-elements[a]['Z'])

contents,labels = {},{}

lbl = 3

for atom in atoms:

    contents[atom]=content[atom]

    if atom=='C':

        labels[atom]='1'

    else:

        labels[atom]=str(lbl)

        lbl+=1

print(contents)

print(labels)

with open('history.txt','a') as f:

    f.write('\n\n'+molecule+'\n')

    f.write('Z = '+str(Z)+'\n')

    f.write('density = '+str(round(density(molecule,Z,A),2))+'\n')

    for atom in atoms:

        f.write(atom+': '+str(contents[atom])+'\n')

heavy_atom,heavy_label = [],[]

for atom in atoms:
```

```python
        heavy_atom += [atom]*contents[atom]

        heavy_label += [labels[atom]]*contents[atom]

    print(heavy_atom)

    print(heavy_label)

    light_atom = 'C'

    light_label = 'l'

    Nheavy = len(heavy_atom)

    Ntotal = Nheavy+nextra

    if ntotal_initial is None: ntotal_initial = Ntotal

    h,k,l,F2,sigF2 = read_hkl3(hkl_file)

    sl = get_sl(h,k,l,A)

    scale = scaling_factor(contents,sl,h,k,l,F2)

    print('scale = ', scale)

    F2 *= scale

    Fo = F2.copy()

    Fo[Fo<0]=0.0

    Fo = sqrt(Fo)

    if fast==1:
```

```
# use min-cycle to build model from scratch

with open('history.txt','a') as f:

    print('\nStep : use min-cycle to build model from scratch',file=f)

    print('fast = ', fast, file=f)

ntotal = ntotal_initial

atom_list = None

atomj,atom_labels,solution,r1=direct_solve5(h,k,l,Fo,A,heavy_atom,ntotal,

    Nheavy,heavy_label,light_label,light_atom,atom_list=atom_list,

    starttime=starttime)

atom_list = to_atom_list(atomj,atom_labels,solution)

if fast==2:

    # use min-cycle to continue building model

    with open('history.txt','a') as f:

        print('\nStep : use min-cycle to continue building model',file=f)

        print('fast = ', fast, file=f)

    ntotal = ntotal_initial

    atom_list = read_atoms('a.res')

    atomj,atom_labels,solution,r1=direct_solve5(h,k,l,Fo,A,heavy_atom,ntotal,

        Nheavy,heavy_label,light_label,light_atom,atom_list=atom_list,

        starttime=starttime)

    atom_list = to_atom_list(atomj,atom_labels,solution)

if fast==3:

    # use r1 hole to continue building model

    with open('history.txt','a') as f:
```

```
        print('\nStep : use r1 hole to continue building model',file=f)

        print('fast = ', fast, file=f)

        print('startfrom = ',startfrom, file=f)

    if True:

        ntotal = startfrom

        atom_list = None

        atomj,atom_labels,solution,r1=direct_solve5(h,k,l,Fo,A,heavy_atom,ntotal,

            Nheavy,heavy_label,light_label,light_atom,atom_list=atom_list,

            starttime=starttime)

        atom_list = to_atom_list(atomj,atom_labels,solution)

    else:

        atom_list = read_atoms('a.res')

    Ntotal = Nheavy

    ntotal = Ntotal

    s = s_dual

    atom_list = r1_holes(h,k,l,F2,A,heavy_atom,ntotal,Nheavy,

        heavy_label,light_label,light_atom,atom_list=atom_list,

        starttime=starttime,runs='r1 hole map',s=s,n_refine=n_refine,

        cutlimit=cutlimit)

    save_history(atom_list,runs='r1 hole startfrom '+str(startfrom))

    sys.exit()

if fast==4:

    # build model by patterson technique

    with open('history.txt','a') as f:

        print('\nStep : build model by patterson technique',file=f)

        print('fast = ', fast, file=f)
```

```
        print('max_runs = ', max_runs, file=f)

    solve_by_patterson3(h,k,l,F2,A,heavy_atom,heavy_label,Nheavy,
            starttime=starttime,s=0.4,mB=1,cutlimit=30)

    sys.exit()

if fast==5:
    # dual space cycling using FFT
    with open('history.txt','a') as f:
        print('\nStep : dual space cycling using FFT',file=f)
        print('fast = ', fast, file=f)
        print('max_runs = ', max_runs, file=f)
        print('cutlimit = ', cutlimit, file=f)
        print('n_improve = ', n_improve, file=f)
        print('extension = ', extension, file=f)
        print('double_first = ', double_first, file=f)
        print('nextra = ', nextra, file=f)
        print('mB = ', mB, file=f)
        print('patterson = ', patterson, file=f)

    # read correct model
    correct_res='correct.res'
    atom_list1 = read_atoms(correct_res)
    atomj1,atom_labels1,solution1=atomj_solution(atom_list1)

    atom_list = read_atoms('a.res')
```

```python
save_history(atom_list,runs='starting model')

atomj,atom_labels,solution=atomj_solution(atom_list)

nballpark=matching(solution1,solution,A)

r1now=cacl_r1(atomj,solution,h,k,l,F2,A)

print('0: ',nballpark,'/',Nheavy,r1now)

with open('history.txt','a') as f:

    print('0: ',nballpark,'/',Nheavy,r1now,file=f)

with open('matching.cvs','a') as f:

    print(0,',',nballpark,',',r1now,file=f)

# improve solution via dual space cycles

n = n_improve

runs = 1

U=0.1

s = s_dual # 0.25   #0.4

old_list = []

base_list = atom_list[:]

if patterson:

    do_patterson=True

else:

    do_patterson=False

while True:

    atom_list = find_corrected_peaks2(h,k,l,F2,A,atom_list,heavy_atom,

        heavy_label,Nheavy,Ntotal,runs,n,U,starttime,s=s,

        n_refine=n_refine,cutlimit=cutlimit,mB=mB,patterson=do_patterson)

    do_patterson=False

    save_history(atom_list,runs,do_copy=True)
```

```
atomj,atom_labels,solution=atomj_solution(atom_list)

nballpark=matching(solution1,solution,A)

r1now=cacl_r1(atomj,solution,h,k,l,F2,A)

print(runs,':',nballpark,'/',Nheavy,r1now)

with open('history.txt','a') as f:

    print(runs,':',nballpark,'/',Nheavy,r1now,file=f)

with open('matching.cvs','a') as f:

    print(runs,',',nballpark,',',r1now,file=f)

if old_list:

    for j in range(len(old_list)):

        same = True

        for i,(atom,label,x,y,z) in enumerate(old_list[j]):

            try:

                atom,label,x0,y0,z0 = atom_list[i]

                if abs(x-x0)>0.001 or abs(y-y0)>0.001 or abs(z-z0)>0.001:

                    same=False

                    break

            except:

                same = False

        if same:

            print('converged')

            with open('history.txt','a') as f:

                f.write('\nconverged\n')

            sys.exit()

    old_list.append(atom_list[:])

    runs += 1

    if runs > max_runs: break

sys.exit()
```

**S5.4. Module read_data.py**

```python
import os, sys

import numpy as np

from math import radians, degrees

sin = np.sin

cos = np.cos

exp = np.exp

sqrt = np.sqrt

atan = np.arctan

acos = np.arccos

pi = np.pi

tpi = 2*pi

def put_in_cell(x):

    while x < 0:

        x += 1

    while x >= 1:

        x -= 1

    return x

def read_hkl3(choice=None):

        cwd = os.getcwd()

        files = os.listdir(cwd)

        # read hkl
```

```python
if type(choice)!=str:

    print('read hkl into numpy arrays h k l F2 sigF2')

    print('Load hkl file:')

    hkl_files = {}

    i = -1

    for file in files:

        if file.endswith('.hkl'):

            i += 1

            hkl_files[i] = file

            print(i, ": ", file)

if choice is None:

    if i==0:

        choice = 0

    else:

        choice = input('choice 0: ')

        if choice:

            try:

                choice = int(choice)

            except:

                choice = 0

        else:

            choice = 0

if type(choice)==int:

    filename = hkl_files[choice]

else:

    filename = choice

with open(filename,'r') as f:

    fread = f.read()
```

```
lines = fread.split('\n')

ha,ka,la,F2a,sigF2a = [],[],[],[],[]

for line in lines:

        try:

                h,k,l,F2,sigF2 = line.split()

                h,k,l,F2,sigF2 = int(h),int(k),int(l),float(F2),float(sigF2)

                if h or k or l:

                        ha.append(h)

                        ka.append(k)

                        la.append(l)

                        F2a.append(F2)

                        sigF2a.append(sigF2)

        except:

                try:

                        h,k,l,F2,sigF2,n = line.split()

                        h,k,l,F2,sigF2,n = int(h),int(k),int(l),float(F2),float(sigF2),int(n)

                        if h or k or l:

                                ha.append(h)

                                ka.append(k)

                                la.append(l)

                                F2a.append(F2)

                                sigF2a.append(sigF2)

                except:

                        pass

ha = np.array(ha)

ka = np.array(ka)

la = np.array(la)

F2a = np.array(F2a)
```

```python
        sigF2a = np.array(sigF2a)

        return (ha,ka,la,F2a,sigF2a)

def read_hkl2():

        cwd = os.getcwd()

        files = os.listdir(cwd)

        # read hkl

        print('\nLoad hkl file:')

        hkl_files = {}

        i = -1

        for file in files:

                if file.endswith('.hkl'):

                        i += 1

                        hkl_files[i] = file

                        print(i, ": ", file)

        if i==0:

                choice = 0

        else:

                choice = input('choice 0: ')

                if choice:

                        try:

                                choice = int(choice)

                        except:

                                choice = 0

                else:

                        choice = 0

        filename = hkl_files[choice]
```

```python
    with open(filename,'r') as f:

        fread = f.read()

    lines = fread.replace('\r','').split('\n')

    return lines

def read_hkl():

    print('read hkl file to make a dictionary of hkl F2 sigF2')

    lines = read_hkl2()

    hkl = {}

    for line in lines:

        try:

            h,k,l,F2,sigF2 = line.split()

            h,k,l,F2,sigF2 = int(h),int(k),int(l),float(F2),float(sigF2)

            if h or k or l:

                if (h,k,l) not in hkl:

                    hkl[(h,k,l)]=[(F2,sigF2)]

                else:

                    hkl[(h,k,l)].append((F2,sigF2))

        except:

            try:

                h,k,l,F2,sigF2,n = line.split()

                h,k,l,F2,sigF2,n = int(h),int(k),int(l),float(F2),float(sigF2),int(n)

                if h or k or l:

                    if (h,k,l) not in hkl:

                        hkl[(h,k,l)]=[(F2,sigF2,n]

                    else:

                        hkl[(h,k,l)]=append((F2,sigF2,n))

            except:
```

```
                        pass

        #print(hkl)

        return hkl

def read_fcf():

        cwd = os.getcwd()

        files = os.listdir(cwd)

        # read fcf

        print('\nLoad fcf file:')

        fcf_files = {}

        i = -1

        for file in files:

                if file.endswith('.fcf'):

                        i += 1

                        fcf_files[i] = file

                        print(i, ": ", file)

        if i==0:

                choice = 0

        else:

                choice = input('choice 0: ')

                if choice:

                        try:

                                choice = int(choice)

                        except:

                                choice = 0

                else:
```

```python
            choice = 0

    filename = fcf_files[choice]

    with open(filename,'r') as f:

            fread = f.read()

    lines = fread.replace('\r','').split('\n')

    fcf = {}

    for line in lines:

            try:

                    h,k,l,F2,sigF2,F,Ph = line.split()

                    h,k,l,F2,sigF2,F,Ph =
int(h),int(k),int(l),float(F2),float(sigF2),float(F),float(Ph)

                    fcf[(h,k,l)]=(F2,sigF2,F,Ph)

            except:

                    pass

    #print(fcf)

    return fcf

def get_cell(choice=None):

    cwd = os.getcwd()

    files = os.listdir(cwd)

    # read res

    if type(choice)!=str:

            print('get cell parameters from res file')

            print('\nLoad res file:')

            res_files = {}

            i = -1
```

```python
    for file in files:

        if file.endswith('.res'):

            i += 1

            res_files[i] = file

            print(i, ": ", file)

if choice is None:

    if i==0:

        choice = 0

    else:

        choice = input('choice 0: ')

        if choice:

            try:

                choice = int(choice)

            except:

                choice = 0

        else:

            choice = 0

if type(choice)==int:

    filename = res_files[choice]

else:

    filename = choice

with open(filename,'r') as f:

    fread = f.read()

lines = fread.replace('\r','').split('\n')

for line in lines:

    try:

        words = line.split()

        if words[0]=='CELL':
```

```python
                lam = float(words[1])

                a = float(words[2])

                b = float(words[3])

                c = float(words[4])

                alfa = float(words[5])

                beta = float(words[6])

                gamma = float(words[7])

                return (lam,a,b,c,alfa,beta,gamma)

        except:

                pass

def read_res(choice=None):

        cwd = os.getcwd()

        files = os.listdir(cwd)

        # read res

        if type(choice)!=str:

                print('get cell parameters from res file')

                print('\nLoad res file:')

                res_files = {}

                i = -1

                for file in files:

                        if file.endswith('.res'):

                                i += 1

                                res_files[i] = file

                                print(i, ": ", file)

        if choice is None:

                if i==0:
```

```python
                        choice = 0
                else:
                        choice = input('choice 0: ')
                        if choice:
                                try:
                                        choice = int(choice)
                                except:
                                        choice = 0
                        else:
                                choice = 0
        if type(choice)==int:
                filename = res_files[choice]
        else:
                filename = choice
        with open(filename,'r') as f:
                fread = f.read()
        lines = fread.replace('\r','').split('\n')
        return lines

def read_lst():
        cwd = os.getcwd()
        files = os.listdir(cwd)

        # read res
        print('\nLoad lst file:')
        lst_files = {}
        i = -1
        for file in files:
```

```python
        if file.endswith('.lst'):
                i += 1
                lst_files[i] = file
                print(i, ": ", file)
    if i==0:
        choice = 0
    else:
        choice = input('choice 0: ')
        if choice:
                try:
                        choice = int(choice)
                except:
                        choice = 0
        else:
                choice = 0
    filename = lst_files[choice]
    with open(filename,'r') as f:
        fread = f.read()
    lines = fread.replace('\r','').split('\n')
    return lines

def read_atoms(choice=None):
    cwd = os.getcwd()
    files = os.listdir(cwd)

    if type(choice)!=str:
        # read res
```

```python
        print('read res to extract atom list: atom x y z')

        print('\nLoad res file:')

        res_files = {}

        i = -1

        for file in files:

                if file.endswith('.res'):

                        i += 1

                        res_files[i] = file

                        print(i, ": ", file)

if choice is None:

        if i==0:

                choice = 0

        else:

                choice = input('choice 0: ')

                if choice:

                        try:

                                choice = int(choice)

                        except:

                                choice = 0

                else:

                        choice = 0

if type(choice)==int:

        filename = res_files[choice]

else:

        filename = choice

with open(filename,'r') as f:

        fread = f.read()

lines = fread.split('\n')
```

```python
atoms = {}

for line in lines:

    try:

        words = line.split()

        if words[0].upper()=='SFAC':

            for i in range(1,len(words)):

                atoms[str(i)] = words[i]

            break

    except:

        pass

i, n = 0, len(lines)

lines1 = []

while 'LATT' not in lines[i].upper():

    lines1.append(lines[i])

    print(lines[i])

    i += 1

lines2 = [lines[i]]

print(lines[i])

i += 1

while 'SYMM' in lines[i].upper():

    lines2.append(lines[i])

    print(lines[i])

    i += 1

lines3 = []

while 'FVAR' not in lines[i].upper():

    lines3.append(lines[i])

    print(lines[i])
```

```
        i += 1

lines3.append(lines[i])

print(lines[i])

i += 1

lines4 = []

while 'HKLF' not in lines[i].upper():

        lines4.append(lines[i])

        print(lines[i])

        i += 1

lines5 = []

while i < n:

        lines5.append(lines[i])

        print(lines[i])

        i += 1

atomlist = []

for line in lines4:

        try:

                if not line: continue

                if line.startswith(' '): continue

                words = line.split()

                if len(words)<5: continue

                x,y,z = float(words[2]),float(words[3]),float(words[4])

                x,y,z = put_in_cell(x),put_in_cell(y),put_in_cell(z)

                atomlist.append((atoms[words[1]],words[1],x,y,z)) # atom, label, x,y,z

        except:

                pass

return atomlist
```

```python
def read_atoms2():

    cwd = os.getcwd()

    files = os.listdir(cwd)

    # read res

    print('read res to extract atom list: atom_label x y z')

    print('\nLoad res file:')

    res_files = {}

    i = -1

    for file in files:

        if file.endswith('.res'):

            i += 1

            res_files[i] = file

            print(i, ": ", file)

    if i==0:

        choice = 0

    else:

        choice = input('choice 0: ')

        if choice:

            try:

                choice = int(choice)

            except:

                choice = 0

        else:

            choice = 0

    filename = res_files[choice]
```

```python
with open(filename,'r') as f:

        fread = f.read()

lines = fread.split('\n')

atoms = {}

for line in lines:

        try:

                words = line.split()

                if words[0].upper()=='SFAC':

                        for i in range(1,len(words)):

                                atoms[str(i)] = words[i]

                        break

        except:

                pass

i, n = 0, len(lines)

lines1 = []

while 'LATT' not in lines[i].upper():

        lines1.append(lines[i])

        print(lines[i])

        i += 1

lines2 = [lines[i]]

print(lines[i])

i += 1

while 'SYMM' in lines[i].upper():

        lines2.append(lines[i])

        print(lines[i])

        i += 1

lines3 = []
```

```
while 'FVAR' not in lines[i].upper():

        lines3.append(lines[i])

        print(lines[i])

        i += 1

lines3.append(lines[i])

print(lines[i])

i += 1

lines4 = []

while 'HKLF' not in lines[i].upper():

        lines4.append(lines[i])

        print(lines[i])

        i += 1

lines5 = []

while i < n:

        lines5.append(lines[i])

        print(lines[i])

        i += 1

atomlist = []

for line in lines4:

        try:

                if not line: continue

                if line.startswith(' '): continue

                words = line.split()

                if len(words)<5: continue

                x,y,z = float(words[2]),float(words[3]),float(words[4])

                atomlist.append((words[0],x,y,z))

        except:

                pass
```

```python
        return atomlist

def read_analyze_res(choice=None):
    print('read res file and analyze into parts')
    lines = read_res(choice)

    i, n = 0, len(lines)
    lines1 = []
    while 'LATT' not in lines[i].upper():
        lines1.append(lines[i])
        i += 1
    lines2 = [lines[i]]
    i += 1
    while 'SYMM' in lines[i].upper():
        lines2.append(lines[i])
        i += 1
    lines3 = []
    while 'FVAR' not in lines[i].upper():
        lines3.append(lines[i])
        i += 1
    lines3.append(lines[i])
    i += 1
    lines4 = []
    while 'HKLF' not in lines[i].upper():
        lines4.append(lines[i])
        i += 1
    lines5 = []
```

```python
    while i < n:

        lines5.append(lines[i])

        i += 1

    return (lines1,lines2,lines3,lines4,lines5)

def read_analyze_res2(choice=None):

    #print('read res file and analyze into parts')

    lines = read_res(choice)

    i, n = 0, len(lines)

    lines1 = []

    while 'LATT' not in lines[i].upper():

        lines1.append(lines[i])

        i += 1

    lines2 = [lines[i]]

    i += 1

    while 'SYMM' in lines[i].upper():

        lines2.append(lines[i])

        i += 1

    lines3 = []

    while 'FVAR' not in lines[i].upper():

        lines3.append(lines[i])

        i += 1

    lines3.append(lines[i])

    i += 1

    lines4 = []

    while 'HKLF' not in lines[i].upper():
```

```python
        lines4.append(lines[i])

        i += 1

    lines5 = []

    while i < n:

        lines5.append(lines[i])

        i += 1

    for line in lines1:

        try:

            words = line.split()

            if words[0]=='CELL':

                lam = float(words[1])

                a = float(words[2])

                b = float(words[3])

                c = float(words[4])

                alfa = float(words[5])

                beta = float(words[6])

                gamma = float(words[7])

                break

        except:

            pass

    alfa = radians(alfa)

    beta = radians(beta)

    gamma = radians(gamma)

    A = np.array([

    [a*a,a*b*cos(gamma),a*c*cos(beta)],

    [a*b*cos(gamma),b*b,b*c*cos(alfa)],
```

```
    [a*c*cos(beta),b*c*cos(alfa),c*c]

    ])

    res_parts = (lines1,lines2,lines3,lines4,lines5)

    return (res_parts,A)
```

### S5.5. Module elements3.py

```
# density in g/cm^3

# heat in J/g/K

# boiling/melting in C

elements = {'Ac': {'M': '',

        'Z': 89,

        'abundance': '5.5x10-10',

        'block': 'f-block',

        'boiling': '3471',

        'config': '[Rn] 6d1 7s2',

        'density': '10.07',

        'electro_negativity': '1.1',

        'group': 'n/a',

        'heat': '0.12',

        'melting': '1323',

        'name': 'Actinium',

        'name_origin': "Greek aktis, 'ray'",

        'orgin': 'from decay',

        'period': '7',

        'phase': 'solid',

        'rad': '*'},
```

```
'Ag': {'M': 107.87,

        'Z': 47,

        'abundance': '0.075',

        'block': 'd-block',

        'boiling': '2435',

        'config': '[Kr] 4d10 5s1',

        'density': '10.501',

        'electro_negativity': '1.93',

        'group': '11',

        'heat': '0.235',

        'melting': '1234.93',

        'name': 'Silver',

        'name_origin': 'English word Symbol Ag is derived from Latin argentum',

        'orgin': 'primordial',

        'period': '5',

        'phase': 'solid',

        'rad': ''},

'Al': {'M': 26.982,

        'Z': 13,

        'abundance': '82300',

        'block': 'p-block',

        'boiling': '2792',

        'config': '[Ne] 3s2 3p1',

        'density': '2.698',

        'electro_negativity': '1.61',

        'group': '13',

        'heat': '0.897',

        'melting': '933.47',
```

```
          'name': 'Aluminium',

          'name_origin': "Alumina, from Latin alumen (gen. aluminis), 'bitter "

                          "salt, alum'",

          'orgin': 'primordial',

          'period': '3',

          'phase': 'solid',

          'rad': "},

'Am': {'M': '',

          'Z': 95,

          'abundance': '-',

          'block': 'f-block',

          'boiling': '2880',

          'config': '[Rn] 5f7 7s2',

          'density': '13.69',

          'electro_negativity': '1.13',

          'group': 'n/a',

          'heat': '-',

          'melting': '1449',

          'name': 'Americium',

          'name_origin': 'The Americas, where the element was first synthesised, '

                          'by analogy with its homologue europium',

          'orgin': 'synthetic',

          'period': '7',

          'phase': 'solid',

          'rad': '*'},

'Ar': {'M': 39.95,

          'Z': 18,

          'abundance': '3.5',
```

        'block': 'p-block',

        'boiling': '87.30',

        'config': '[Ne] 3s2 3p6',

        'density': '0.0017837',

        'electro_negativity': '-',

        'group': '18',

        'heat': '0.52',

        'melting': '83.80',

        'name': 'Argon',

        'name_origin': "Greek argos, 'idle' (because of its inertness)",

        'orgin': 'primordial',

        'period': '3',

        'phase': 'gas',

        'rad': ''},

'As': {'M': 74.922,

        'Z': 33,

        'abundance': '1.8',

        'block': 'p-block',

        'boiling': '887',

        'config': '[Ar] 3d10 4s2 4p3',

        'density': '5.776',

        'electro_negativity': '2.18',

        'group': '15',

        'heat': '0.329',

        'melting': '1090[l]',

        'name': 'Arsenic',

        'name_origin': "French arsenic, from Greek arsenikon 'yellow arsenic' "

                "(influenced by arsenikos, 'masculine' or 'virile'), "

```
                    'from a West Asian wanderword ultimately from Old '

                    "Iranian *zarniya-ka, 'golden'",

        'orgin': 'primordial',

        'period': '4',

        'phase': 'solid',

        'rad': ''},

'At': {'M': '',

        'Z': 85,

        'abundance': '3x10-20',

        'block': 'p-block',

        'boiling': '610',

        'config': '[Xe] 4f14 5d10 6s2 6p5',

        'density': '7',

        'electro_negativity': '2.2',

        'group': '17',

        'heat': '-',

        'melting': '575',

        'name': 'Astatine',

        'name_origin': "Greek astatos, 'unstable'",

        'orgin': 'from decay',

        'period': '6',

        'phase': 'unknown phase',

        'rad': '*'},

'Au': {'M': 196.97,

        'Z': 79,

        'abundance': '0.004',

        'block': 'd-block',

        'boiling': '3129',
```

      'config': '[Xe] 4f14 5d10 6s1',

      'density': '19.282',

      'electro_negativity': '2.54',

      'group': '11',

      'heat': '0.129',

      'melting': '1337.33',

      'name': 'Gold',

      'name_origin': 'English word Symbol Au is derived from Latin aurum',

      'orgin': 'primordial',

      'period': '6',

      'phase': 'solid',

      'rad': ''},

'B': {'M': 10.81,

      'Z': 5,

      'abundance': '10',

      'block': 'p-block',

      'boiling': '4200',

      'config': '[He] 2s2 2p1',

      'density': '2.34',

      'electro_negativity': '2.04',

      'group': '13',

      'heat': '1.026',

      'melting': '2349',

      'name': 'Boron',

      'name_origin': 'Borax, a mineral (from Arabic bawraq)',

      'orgin': 'primordial',

      'period': '2',

      'phase': 'solid',

```
        'rad': ''},

 'Ba':  {'M': 137.33,

        'Z': 56,

        'abundance': '425',

        'block': 's-block',

        'boiling': '2170',

        'config': '[Xe] 6s2',

        'density': '3.594',

        'electro_negativity': '0.89',

        'group': '2',

        'heat': '0.204',

        'melting': '1000',

        'name': 'Barium',

        'name_origin': "Greek barys, 'heavy'",

        'orgin': 'primordial',

        'period': '6',

        'phase': 'solid',

        'rad': ''},

 'Be':  {'M': 9.0122,

        'Z': 4,

        'abundance': '2.8',

        'block': 's-block',

        'boiling': '2742',

        'config': '[He] 2s2',

        'density': '1.85',

        'electro_negativity': '1.57',

        'group': '2',

        'heat': '1.825',
```

        'melting': '1560',

        'name': 'Beryllium',

        'name_origin': 'Beryl, a mineral (ultimately from the name of Belur in '

                    'southern India)[4]',

        'orgin': 'primordial',

        'period': '2',

        'phase': 'solid',

        'rad': ''},

'Bh': {'M': '',

        'Z': 107,

        'abundance': '-',

        'block': 'd-block',

        'boiling': '-',

        'config': '',

        'density': '(37.1)',

        'electro_negativity': '-',

        'group': '7',

        'heat': '-',

        'melting': '-',

        'name': 'Bohrium',

        'name_origin': 'Niels Bohr, Danish physicist',

        'orgin': 'synthetic',

        'period': '7',

        'phase': 'unknown phase',

        'rad': '*'},

'Bi': {'M': 208.98,

        'Z': 83,

        'abundance': '0.009',

'block': 'p-block',

'boiling': '1837',

'config': '[Xe] 4f14 5d10 6s2 6p3',

'density': '9.807',

'electro_negativity': '2.02',

'group': '15',

'heat': '0.122',

'melting': '544.7',

'name': 'Bismuth',

'name_origin': "German Wismut, from weib Masse 'white mass', unless "

                'from Arabic',

'orgin': 'primordial',

'period': '6',

'phase': 'solid',

'rad': ''},

'Bk': {'M': '',

'Z': 97,

'abundance': '-',

'block': 'f-block',

'boiling': '2900',

'config': '',

'density': '14.79',

'electro_negativity': '1.3',

'group': 'n/a',

'heat': '-',

'melting': '1259',

'name': 'Berkelium',

'name_origin': 'Berkeley, California, where the element was first '

```
            'synthesised',

      'orgin': 'synthetic',

      'period': '7',

      'phase': 'solid',

      'rad': '*'},

'Br': {'M': 79.904,

      'Z': 35,

      'abundance': '2.4',

      'block': 'p-block',

      'boiling': '332.0',

      'config': '[Ar] 3d10 4s2 4p5',

      'density': '3.122',

      'electro_negativity': '2.96',

      'group': '17',

      'heat': '0.474',

      'melting': '265.8',

      'name': 'Bromine',

      'name_origin': "Greek bromos, 'stench'",

      'orgin': 'primordial',

      'period': '4',

      'phase': 'liquid',

      'rad': ''},

'C': {'M': 12.011,

      'Z': 6,

      'abundance': '200',

      'block': 'p-block',

      'boiling': '4300',

      'config': '[He] 2s2 2p2',
```

        'density': '2.267',

        'electro_negativity': '2.55',

        'group': '14',

        'heat': '0.709',

        'melting': '>4000',

        'name': 'Carbon',

        'name_origin': "Latin carbo, 'coal'",

        'orgin': 'primordial',

        'period': '2',

        'phase': 'solid',

        'rad': ''},

    'Ca': {'M': 40.078,

        'Z': 20,

        'abundance': '41500',

        'block': 's-block',

        'boiling': '1757',

        'config': '[Ar] 4s2',

        'density': '1.54',

        'electro_negativity': '1.00',

        'group': '2',

        'heat': '0.647',

        'melting': '1115',

        'name': 'Calcium',

        'name_origin': "Latin calx, 'lime'",

        'orgin': 'primordial',

        'period': '4',

        'phase': 'solid',

        'rad': ''},

'Cd': {'M': 112.41,

        'Z': 48,

        'abundance': '0.159',

        'block': 'd-block',

        'boiling': '1040',

        'config': '[Kr] 4d10 5s2',

        'density': '8.69',

        'electro_negativity': '1.69',

        'group': '12',

        'heat': '0.232',

        'melting': '594.22',

        'name': 'Cadmium',

        'name_origin': 'New Latin cadmia, from King Kadmos',

        'orgin': 'primordial',

        'period': '5',

        'phase': 'solid',

        'rad': ''},

'Ce': {'M': 140.12,

        'Z': 58,

        'abundance': '66.5',

        'block': 'f-block',

        'boiling': '3716',

        'config': '[Xe] 4f1 5d1 6s2',

        'density': '6.77',

        'electro_negativity': '1.12',

        'group': 'n/a',

        'heat': '0.192',

        'melting': '1068',

```
        'name': 'Cerium',

        'name_origin': 'Ceres, a dwarf planet, considered a planet at the time',

        'orgin': 'primordial',

        'period': '6',

        'phase': 'solid',

        'rad': ''},

'Cf': {'M': '',

        'Z': 98,

        'abundance': '-',

        'block': 'f-block',

        'boiling': '(1743)[b]',

        'config': '',

        'density': '15.1',

        'electro_negativity': '1.3',

        'group': 'n/a',

        'heat': '-',

        'melting': '1173',

        'name': 'Californium',

        'name_origin': 'California, where the element was first synthesised in '
                        'the LBNL laboratory',

        'orgin': 'synthetic',

        'period': '7',

        'phase': 'solid',

        'rad': '*'},

'Cl': {'M': 35.45,

        'Z': 17,

        'abundance': '145',

        'block': 'p-block',
```

        'boiling': '239.11',

        'config': '[Ne] 3s2 3p5',

        'density': '0.003214',

        'electro_negativity': '3.16',

        'group': '17',

        'heat': '0.479',

        'melting': '171.6',

        'name': 'Chlorine',

        'name_origin': "Greek chloros, 'greenish yellow'",

        'orgin': 'primordial',

        'period': '3',

        'phase': 'gas',

        'rad': ''},

'Cm': {'M': '',

        'Z': 96,

        'abundance': '-',

        'block': 'f-block',

        'boiling': '3383',

        'config': '',

        'density': '13.51',

        'electro_negativity': '1.28',

        'group': 'n/a',

        'heat': '-',

        'melting': '1613',

        'name': 'Curium',

        'name_origin': 'Pierre Curie and Marie Curie, French physicists and '
                        'chemists',

        'orgin': 'synthetic',

'period': '7',

'phase': 'solid',

'rad': '*'},

'Co': {'M': 58.933,

'Z': 27,

'abundance': '25',

'block': 'd-block',

'boiling': '3200',

'config': '[Ar] 3d7 4s2',

'density': '8.86',

'electro_negativity': '1.88',

'group': '9',

'heat': '0.421',

'melting': '1768',

'name': 'Cobalt',

'name_origin': "German Kobold, 'goblin'",

'orgin': 'primordial',

'period': '4',

'phase': 'solid',

'rad': ''},

'Cr': {'M': 51.996,

'Z': 24,

'abundance': '102',

'block': 'd-block',

'boiling': '2944',

'config': '[Ar] 3d5 4s1',

'density': '7.15',

'electro_negativity': '1.66',

    'group': '6',

    'heat': '0.449',

    'melting': '2180',

    'name': 'Chromium',

    'name_origin': "Greek chroma, 'colour'",

    'orgin': 'primordial',

    'period': '4',

    'phase': 'solid',

    'rad': ''},

'Cs': {'M': 132.91,

    'Z': 55,

    'abundance': '3',

    'block': 's-block',

    'boiling': '944',

    'config': '[Xe] 6s1',

    'density': '1.873',

    'electro_negativity': '0.79',

    'group': '1',

    'heat': '0.242',

    'melting': '301.59',

    'name': 'Caesium',

    'name_origin': "Latin caesius, 'sky-blue'",

    'orgin': 'primordial',

    'period': '6',

    'phase': 'solid',

    'rad': ''},

'Cu': {'M': 63.546,

    'Z': 29,

        'abundance': '60',

        'block': 'd-block',

        'boiling': '2835',

        'config': '[Ar] 3d10 4s1',

        'density': '8.96',

        'electro_negativity': '1.90',

        'group': '11',

        'heat': '0.385',

        'melting': '1357.77',

        'name': 'Copper',

        'name_origin': 'English word, from Latin cuprum, from Ancient Greek '

                    "Kypros 'Cyprus'",

        'orgin': 'primordial',

        'period': '4',

        'phase': 'solid',

        'rad': ''},

'Db': {'M': '',

        'Z': 105,

        'abundance': '-',

        'block': 'd-block',

        'boiling': '-',

        'config': '',

        'density': '(29.3)',

        'electro_negativity': '-',

        'group': '5',

        'heat': '-',

        'melting': '-',

        'name': 'Dubnium',

'name_origin': 'Dubna, Russia, where the element was discovered in the '

                'JINR laboratory',

'orgin': 'synthetic',

'period': '7',

'phase': 'unknown phase',

'rad': '*'},

'Dy': {'M': 162.5,

'Z': 66,

'abundance': '5.2',

'block': 'f-block',

'boiling': '2840',

'config': '[Xe] 4f10 6s2',

'density': '8.55',

'electro_negativity': '1.22',

'group': 'n/a',

'heat': '0.17',

'melting': '1680',

'name': 'Dysprosium',

'name_origin': "Greek dysprositos, 'hard to get'",

'orgin': 'primordial',

'period': '6',

'phase': 'solid',

'rad': ''},

'Er': {'M': 167.26,

'Z': 68,

'abundance': '3.5',

'block': 'f-block',

'boiling': '3141',

'config': '[Xe] 4f12 6s2',

'density': '9.066',

'electro_negativity': '1.24',

'group': 'n/a',

'heat': '0.168',

'melting': '1802',

'name': 'Erbium',

'name_origin': 'Ytterby, Sweden, where it was found; see also yttrium, '

          'terbium, ytterbium',

'orgin': 'primordial',

'period': '6',

'phase': 'solid',

'rad': ''},

'Es': {'M': '',

'Z': 99,

'abundance': '-',

'block': 'f-block',

'boiling': '(1269)',

'config': '',

'density': '8.84',

'electro_negativity': '1.3',

'group': 'n/a',

'heat': '-',

'melting': '1133',

'name': 'Einsteinium',

'name_origin': 'Albert Einstein, German physicist',

'orgin': 'synthetic',

'period': '7',

```
          'phase': 'solid',

          'rad': '*'},

'Eu': {'M': 151.96,

          'Z': 63,

          'abundance': '2',

          'block': 'f-block',

          'boiling': '1802',

          'config': '[Xe] 4f7 6s2',

          'density': '5.243',

          'electro_negativity': '1.2',

          'group': 'n/a',

          'heat': '0.182',

          'melting': '1099',

          'name': 'Europium',

          'name_origin': 'Europe',

          'orgin': 'primordial',

          'period': '6',

          'phase': 'solid',

          'rad': ''},

'F': {'M': 18.998,

          'Z': 9,

          'abundance': '585',

          'block': 'p-block',

          'boiling': '85.03',

          'config': '[He] 2s2 2p5',

          'density': '0.001696',

          'electro_negativity': '3.98',

          'group': '17',
```

```
        'heat': '0.824',

        'melting': '53.53',

        'name': 'Fluorine',

        'name_origin': "Latin fluere, 'to flow'",

        'orgin': 'primordial',

        'period': '2',

        'phase': 'gas',

        'rad': ''},

'Fe': {'M': 55.845,

        'Z': 26,

        'abundance': '56300',

        'block': 'd-block',

        'boiling': '3134',

        'config': '[Ar] 3d6 4s2',

        'density': '7.874',

        'electro_negativity': '1.83',

        'group': '8',

        'heat': '0.449',

        'melting': '1811',

        'name': 'Iron',

        'name_origin': 'English word Symbol Fe is derived from Latin ferrum',

        'orgin': 'primordial',

        'period': '4',

        'phase': 'solid',

        'rad': ''},

'Fm': {'M': '',

        'Z': 100,

        'abundance': '-',
```

    'block': 'f-block',

    'boiling': '-',

    'config': '',

    'density': '(9.7)[b]',

    'electro_negativity': '1.3',

    'group': 'n/a',

    'heat': '-',

    'melting': '(1125)[b]',

    'name': 'Fermium',

    'name_origin': 'Enrico Fermi, Italian physicist',

    'orgin': 'synthetic',

    'period': '7',

    'phase': 'unknown phase',

    'rad': '*'},

'Fr': {'M': '',

    'Z': 87,

    'abundance': '~ 1x10-18',

    'block': 's-block',

    'boiling': '890',

    'config': '[Rn] 7s1',

    'density': '1.87',

    'electro_negativity': '>0.79[6]',

    'group': '1',

    'heat': '-',

    'melting': '281',

    'name': 'Francium',

    'name_origin': 'France, home country of discoverer Marguerite Perey',

    'orgin': 'from decay',

      'period': '7',

      'phase': 'unknown phase',

      'rad': '*'},

'Ga': {'M': 69.723,

      'Z': 31,

      'abundance': '19',

      'block': 'p-block',

      'boiling': '2673',

      'config': '[Ar] 3d10 4s2 4p1',

      'density': '5.907',

      'electro_negativity': '1.81',

      'group': '13',

      'heat': '0.371',

      'melting': '302.9146',

      'name': 'Gallium',

      'name_origin': "Latin Gallia, 'France'",

      'orgin': 'primordial',

      'period': '4',

      'phase': 'solid',

      'rad': ''},

'Gd': {'M': 157.25,

      'Z': 64,

      'abundance': '6.2',

      'block': 'f-block',

      'boiling': '3546',

      'config': '[Xe] 4f7 5d1 6s2',

      'density': '7.895',

      'electro_negativity': '1.2',

```
        'group': 'n/a',

        'heat': '0.236',

        'melting': '1585',

        'name': 'Gadolinium',

        'name_origin': 'Gadolinite, a mineral named after Johan Gadolin, '
                        'Finnish chemist, physicist and mineralogist',

        'orgin': 'primordial',

        'period': '6',

        'phase': 'solid',

        'rad': ''},

'Ge': {'M': 72.63,

        'Z': 32,

        'abundance': '1.5',

        'block': 'p-block',

        'boiling': '3106',

        'config': '[Ar] 3d10 4s2 4p2',

        'density': '5.323',

        'electro_negativity': '2.01',

        'group': '14',

        'heat': '0.32',

        'melting': '1211.40',

        'name': 'Germanium',

        'name_origin': "Latin Germania, 'Germany'",

        'orgin': 'primordial',

        'period': '4',

        'phase': 'solid',

        'rad': ''},

'H': {'M': 1.008,
```

'Z': 1,

'abundance': '1400',

'block': 's-block',

'boiling': '20.28',

'config': '1s1',

'density': '0.00008988',

'electro_negativity': '2.20',

'group': '1',

'heat': '14.304',

'melting': '14.01',

'name': 'Hydrogen',

'name_origin': "Greek elements hydro- and -gen, 'water-forming'",

'orgin': 'primordial',

'period': '1',

'phase': 'gas',

'rad': ''},

'He': {'M': 4.0026,

'Z': 2,

'abundance': '0.008',

'block': 's-block',

'boiling': '4.22',

'config': '1s2',

'density': '0.0001785',

'electro_negativity': '-',

'group': '18',

'heat': '5.193',

'melting': '-[k]',

'name': 'Helium',

'name_origin': "Greek helios, 'sun'",

'orgin': 'primordial',

'period': '1',

'phase': 'gas',

'rad': ''},

'Hf': {'M': 178.49,

'Z': 72,

'abundance': '3',

'block': 'd-block',

'boiling': '4876',

'config': '[Xe] 4f14 5d2 6s2',

'density': '13.31',

'electro_negativity': '1.3',

'group': '4',

'heat': '0.144',

'melting': '2506',

'name': 'Hafnium',

'name_origin': "New Latin Hafnia, 'Copenhagen' (from Danish havn, "

                'harbour)',

'orgin': 'primordial',

'period': '6',

'phase': 'solid',

'rad': ''},

'Hg': {'M': 200.59,

'Z': 80,

'abundance': '0.085',

'block': 'd-block',

'boiling': '629.88',

'config': '[Xe] 4f14 5d10 6s2',

'density': '13.5336',

'electro_negativity': '2.00',

'group': '12',

'heat': '0.14',

'melting': '234.43',

'name': 'Mercury',

'name_origin': 'Mercury, Roman god of commerce, communication, and '

                'luck, known for his speed and mobility Symbol Hg is '

                'derived from its Latin name hydrargyrum, from Greek '

                "hydrargyros, 'water-silver'",

'orgin': 'primordial',

'period': '6',

'phase': 'liquid',

'rad': ''},

'Ho': {'M': 164.93,

        'Z': 67,

        'abundance': '1.3',

        'block': 'f-block',

        'boiling': '2993',

        'config': '[Xe] 4f11 6s2',

        'density': '8.795',

        'electro_negativity': '1.23',

        'group': 'n/a',

        'heat': '0.165',

        'melting': '1734',

        'name': 'Holmium',

        'name_origin': "New Latin Holmia, 'Stockholm'",

        'orgin': 'primordial',

        'period': '6',

        'phase': 'solid',

        'rad': ''},

'Hs': {'M': '',

        'Z': 108,

        'abundance': '-',

        'block': 'd-block',

        'boiling': '-',

        'config': '',

        'density': '(40.7)',

        'electro_negativity': '-',

        'group': '8',

        'heat': '-',

        'melting': '-',

        'name': 'Hassium',

        'name_origin': "New Latin Hassia, 'Hesse', a state in Germany",

        'orgin': 'synthetic',

        'period': '7',

        'phase': 'unknown phase',

        'rad': '*'},

'I': {'M': 126.9,

        'Z': 53,

        'abundance': '0.45',

        'block': 'p-block',

        'boiling': '457.4',

        'config': '[Kr] 4d10 5s2 5p5',

        'density': '4.93',

'electro_negativity': '2.66',

'group': '17',

'heat': '0.214',

'melting': '386.85',

'name': 'Iodine',

'name_origin': "French iode, from Greek ioeides, 'violet'",

'orgin': 'primordial',

'period': '5',

'phase': 'solid',

'rad': ''},

'In': {'M': 114.82,

'Z': 49,

'abundance': '0.25',

'block': 'p-block',

'boiling': '2345',

'config': '[Kr] 4d10 5s2 5p1',

'density': '7.31',

'electro_negativity': '1.78',

'group': '13',

'heat': '0.233',

'melting': '429.75',

'name': 'Indium',

'name_origin': "Latin indicum, 'indigo', the blue colour found in its "

                    'spectrum',

'orgin': 'primordial',

'period': '5',

'phase': 'solid',

'rad': ''},

```
'Ir': {'M': 192.22,

      'Z': 77,

      'abundance': '0.001',

      'block': 'd-block',

      'boiling': '4701',

      'config': '[Xe] 4f14 5d7 6s2',

      'density': '22.56',

      'electro_negativity': '2.20',

      'group': '9',

      'heat': '0.131',

      'melting': '2719',

      'name': 'Iridium',

      'name_origin': 'Iris, the Greek goddess of the rainbow',

      'orgin': 'primordial',

      'period': '6',

      'phase': 'solid',

      'rad': ''},

'K': {'M': 39.098,

      'Z': 19,

      'abundance': '20900',

      'block': 's-block',

      'boiling': '1032',

      'config': '[Ar] 4s1',

      'density': '0.862',

      'electro_negativity': '0.82',

      'group': '1',

      'heat': '0.757',

      'melting': '336.53',
```

'name': 'Potassium',

'name_origin': "New Latin potassa, 'potash', itself from pot and ash "

          'Symbol K is derived from Latin kalium',

'orgin': 'primordial',

'period': '4',

'phase': 'solid',

'rad': ''},

'Kr': {'M': 83.798,

   'Z': 36,

   'abundance': '1x10-4',

   'block': 'p-block',

   'boiling': '119.93',

   'config': '[Ar] 3d10 4s2 4p6',

   'density': '0.003733',

   'electro_negativity': '3.00',

   'group': '18',

   'heat': '0.248',

   'melting': '115.79',

   'name': 'Krypton',

   'name_origin': "Greek kryptos, 'hidden'",

   'orgin': 'primordial',

   'period': '4',

   'phase': 'gas',

   'rad': ''},

'La': {'M': 138.91,

   'Z': 57,

   'abundance': '39',

   'block': 'f-block',

        'boiling': '3737',

        'config': '[Xe] 5d1 6s2',

        'density': '6.145',

        'electro_negativity': '1.1',

        'group': 'n/a',

        'heat': '0.195',

        'melting': '1193',

        'name': 'Lanthanum',

        'name_origin': "Greek lanthanein, 'to lie hidden'",

        'orgin': 'primordial',

        'period': '6',

        'phase': 'solid',

        'rad': ''},

'Li': {'M': 6.94,

        'Z': 3,

        'abundance': '20',

        'block': 's-block',

        'boiling': '1560',

        'config': '[He] 2s1',

        'density': '0.534',

        'electro_negativity': '0.98',

        'group': '1',

        'heat': '3.582',

        'melting': '453.69',

        'name': 'Lithium',

        'name_origin': "Greek lithos, 'stone'",

        'orgin': 'primordial',

        'period': '2',

        'phase': 'solid',

        'rad': ''},

'Lr': {'M': '',

        'Z': 103,

        'abundance': '-',

        'block': 'd-block',

        'boiling': '-',

        'config': '',

        'density': '(15.6)',

        'electro_negativity': '1.3',

        'group': '3',

        'heat': '-',

        'melting': '(1900)',

        'name': 'Lawrencium',

        'name_origin': 'Ernest Lawrence, American physicist',

        'orgin': 'synthetic',

        'period': '7',

        'phase': 'unknown phase',

        'rad': '*'},

'Lu': {'M': 174.97,

        'Z': 71,

        'abundance': '0.8',

        'block': 'd-block',

        'boiling': '3675',

        'config': '[Xe] 4f14 5d1 6s2',

        'density': '9.84',

        'electro_negativity': '1.27',

        'group': '3',

```
        'heat': '0.154',

        'melting': '1925',

        'name': 'Lutetium',

        'name_origin': "Latin Lutetia, 'Paris'",

        'orgin': 'primordial',

        'period': '6',

        'phase': 'solid',

        'rad': ''},

'Md': {'M': '',

        'Z': 101,

        'abundance': '-',

        'block': 'f-block',

        'boiling': '-',

        'config': '',

        'density': '(10.3)',

        'electro_negativity': '1.3',

        'group': 'n/a',

        'heat': '-',

        'melting': '(1100)',

        'name': 'Mendelevium',

        'name_origin': 'Dmitri Mendeleev, Russian chemist who proposed the '
                       'periodic table',

        'orgin': 'synthetic',

        'period': '7',

        'phase': 'unknown phase',

        'rad': '*'},

'Mg': {'M': 24.305,

        'Z': 12,
```

'abundance': '23300',

'block': 's-block',

'boiling': '1363',

'config': '[Ne] 3s2',

'density': '1.738',

'electro_negativity': '1.31',

'group': '2',

'heat': '1.023',

'melting': '923',

'name': 'Magnesium',

'name_origin': 'Magnesia, a district of Eastern Thessaly in Greece',

'orgin': 'primordial',

'period': '3',

'phase': 'solid',

'rad': ''},

'Mn': {'M': 54.938,

'Z': 25,

'abundance': '950',

'block': 'd-block',

'boiling': '2334',

'config': '[Ar] 3d5 4s2',

'density': '7.44',

'electro_negativity': '1.55',

'group': '7',

'heat': '0.479',

'melting': '1519',

'name': 'Manganese',

'name_origin': 'Corrupted from magnesia negra; see magnesium',

```
       'orgin': 'primordial',

       'period': '4',

       'phase': 'solid',

       'rad': "},

'Mo': {'M': 95.95,

       'Z': 42,

       'abundance': '1.2',

       'block': 'd-block',

       'boiling': '4912',

       'config': '[Kr] 4d5 5s1',

       'density': '10.22',

       'electro_negativity': '2.16',

       'group': '6',

       'heat': '0.251',

       'melting': '2896',

       'name': 'Molybdenum',

       'name_origin': "Greek molybdaina, 'piece of lead', from molybdos, "

                     "'lead', due to confusion with lead ore galena (PbS)",

       'orgin': 'primordial',

       'period': '5',

       'phase': 'solid',

       'rad': "},

'Mt': {'M': ",

       'Z': 109,

       'abundance': '-',

       'block': 'd-block',

       'boiling': '-',

       'config': ",
```

```
        'density': '(37.4)',

        'electro_negativity': '-',

        'group': '9',

        'heat': '-',

        'melting': '-',

        'name': 'Meitnerium',

        'name_origin': 'Lise Meitner, Austrian physicist',

        'orgin': 'synthetic',

        'period': '7',

        'phase': 'unknown phase',

        'rad': '*'},

'N': {'M': 14.007,

        'Z': 7,

        'abundance': '19',

        'block': 'p-block',

        'boiling': '77.36',

        'config': '[He] 2s2 2p3',

        'density': '0.0012506',

        'electro_negativity': '3.04',

        'group': '15',

        'heat': '1.04',

        'melting': '63.15',

        'name': 'Nitrogen',

        'name_origin': "Greek nitron and -gen, 'niter-forming'",

        'orgin': 'primordial',

        'period': '2',

        'phase': 'gas',

        'rad': ''},
```

'Na': {'M': 22.99,

    'Z': 11,

    'abundance': '23600',

    'block': 's-block',

    'boiling': '1156',

    'config': '[Ne] 3s1',

    'density': '0.971',

    'electro_negativity': '0.93',

    'group': '1',

    'heat': '1.228',

    'melting': '370.87',

    'name': 'Sodium',

    'name_origin': 'English (from medieval Latin) soda Symbol Na is '

            'derived from New Latin natrium, coined from German '

            "Natron, 'natron'",

    'orgin': 'primordial',

    'period': '3',

    'phase': 'solid',

    'rad': ''},

'Nb': {'M': 92.906,

    'Z': 41,

    'abundance': '20',

    'block': 'd-block',

    'boiling': '5017',

    'config': '[Kr] 4d4 5s1',

    'density': '8.57',

    'electro_negativity': '1.6',

    'group': '5',

```
          'heat': '0.265',

          'melting': '2750',

          'name': 'Niobium',

          'name_origin': 'Niobe, daughter of king Tantalus from Greek mythology; '

                         'see also tantalum',

          'orgin': 'primordial',

          'period': '5',

          'phase': 'solid',

          'rad': ''},

'Nd': {'M': 144.24,

          'Z': 60,

          'abundance': '41.5',

          'block': 'f-block',

          'boiling': '3347',

          'config': '[Xe] 4f4 6s2',

          'density': '7.007',

          'electro_negativity': '1.14',

          'group': 'n/a',

          'heat': '0.19',

          'melting': '1297',

          'name': 'Neodymium',

          'name_origin': "Greek neos didymos, 'new twin'",

          'orgin': 'primordial',

          'period': '6',

          'phase': 'solid',

          'rad': ''},

'Ne': {'M': 20.18,

          'Z': 10,
```

```
        'abundance': '0.005',

        'block': 'p-block',

        'boiling': '27.07',

        'config': '[He] 2s2 2p6',

        'density': '0.0008999',

        'electro_negativity': '-',

        'group': '18',

        'heat': '1.03',

        'melting': '24.56',

        'name': 'Neon',

        'name_origin': "Greek neon, 'new'",

        'orgin': 'primordial',

        'period': '2',

        'phase': 'gas',

        'rad': ''},

'Ni': {'M': 58.693,

        'Z': 28,

        'abundance': '84',

        'block': 'd-block',

        'boiling': '3186',

        'config': '[Ar] 3d8 4s2',

        'density': '8.912',

        'electro_negativity': '1.91',

        'group': '10',

        'heat': '0.444',

        'melting': '1728',

        'name': 'Nickel',

        'name_origin': 'Nickel, a mischievous sprite of German miner mythology',
```

```
      'orgin': 'primordial',

      'period': '4',

      'phase': 'solid',

      'rad': ''},

 'No': {'M': '',

      'Z': 102,

      'abundance': '-',

      'block': 'f-block',

      'boiling': '-',

      'config': '',

      'density': '(9.9)',

      'electro_negativity': '1.3',

      'group': 'n/a',

      'heat': '-',

      'melting': '(1100)',

      'name': 'Nobelium',

      'name_origin': 'Alfred Nobel, Swedish chemist and engineer',

      'orgin': 'synthetic',

      'period': '7',

      'phase': 'unknown phase',

      'rad': '*'},

 'Np': {'M': '',

      'Z': 93,

      'abundance': '<= 3x10-12',

      'block': 'f-block',

      'boiling': '4273',

      'config': '[Rn] 5f4 6d1 7s2',

      'density': '20.45',
```

       'electro_negativity': '1.36',

       'group': 'n/a',

       'heat': '-',

       'melting': '917',

       'name': 'Neptunium',

       'name_origin': 'Neptune, the eighth planet in the Solar System',

       'orgin': 'from decay',

       'period': '7',

       'phase': 'solid',

       'rad': '*'},

'O': {'M': 15.999,

       'Z': 8,

       'abundance': '461000',

       'block': 'p-block',

       'boiling': '90.20',

       'config': '[He] 2s2 2p4',

       'density': '0.001429',

       'electro_negativity': '3.44',

       'group': '16',

       'heat': '0.918',

       'melting': '54.36',

       'name': 'Oxygen',

       'name_origin': "Greek oxy- and -gen, 'acid-forming'",

       'orgin': 'primordial',

       'period': '2',

       'phase': 'gas',

       'rad': ''},

'Os': {'M': 190.23,

'Z': 76,

'abundance': '0.002',

'block': 'd-block',

'boiling': '5285',

'config': '[Xe] 4f14 5d6 6s2',

'density': '22.61',

'electro_negativity': '2.2',

'group': '8',

'heat': '0.13',

'melting': '3306',

'name': 'Osmium',

'name_origin': "Greek osme, 'smell'",

'orgin': 'primordial',

'period': '6',

'phase': 'solid',

'rad': ''},

'P': {'M': 30.974,

'Z': 15,

'abundance': '1050',

'block': 'p-block',

'boiling': '550',

'config': '[Ne] 3s2 3p3',

'density': '1.82',

'electro_negativity': '2.19',

'group': '15',

'heat': '0.769',

'melting': '317.30',

'name': 'Phosphorus',

'name_origin': "Greek phosphoros, 'light-bearing'",

'orgin': 'primordial',

'period': '3',

'phase': 'solid',

'rad': ''},

'Pa': {'M': 231.04,

'Z': 91,

'abundance': '1.4x10-6',

'block': 'f-block',

'boiling': '4300',

'config': '[Rn] 5f2 6d1 7s2',

'density': '15.37',

'electro_negativity': '1.5',

'group': 'n/a',

'heat': '-',

'melting': '1841',

'name': 'Protactinium',

'name_origin': "Proto- (from Greek protos, 'first, before') + "

'actinium, since actinium is produced through the '

'radioactive decay of protactinium',

'orgin': 'from decay',

'period': '7',

'phase': 'solid',

'rad': ''},

'Pb': {'M': 207.2,

'Z': 82,

'abundance': '14',

'block': 'p-block',

```
        'boiling': '2022',

        'config': '[Xe] 4f14 5d10 6s2 6p2',

        'density': '11.342',

        'electro_negativity': '1.87 (2+)2.33 (4+)',

        'group': '14',

        'heat': '0.129',

        'melting': '600.61',

        'name': 'Lead',

        'name_origin': 'English word Symbol Pb is derived from Latin plumbum',

        'orgin': 'primordial',

        'period': '6',

        'phase': 'solid',

        'rad': ''},

'Pd': {'M': 106.42,

        'Z': 46,

        'abundance': '0.015',

        'block': 'd-block',

        'boiling': '3236',

        'config': '[Kr] 4d10',

        'density': '12.02',

        'electro_negativity': '2.20',

        'group': '10',

        'heat': '0.244',

        'melting': '1828.05',

        'name': 'Palladium',

        'name_origin': 'Pallas, an asteroid, considered a planet at the time',

        'orgin': 'primordial',

        'period': '5',
```

```
        'phase': 'solid',

        'rad': ''},

'Pm': {'M': '',

        'Z': 61,

        'abundance': '2x10-19',

        'block': 'f-block',

        'boiling': '3273',

        'config': '[Xe] 4f5 6s2',

        'density': '7.26',

        'electro_negativity': '1.13',

        'group': 'n/a',

        'heat': '-',

        'melting': '1315',

        'name': 'Promethium',

        'name_origin': 'Prometheus, a figure in Greek mythology',

        'orgin': 'from decay',

        'period': '6',

        'phase': 'solid',

        'rad': '*'},

'Po': {'M': '',

        'Z': 84,

        'abundance': '2x10-10',

        'block': 'p-block',

        'boiling': '1235',

        'config': '[Xe] 4f14 5d10 6s2 6p4',

        'density': '9.32',

        'electro_negativity': '2.0',

        'group': '16',
```

```
        'heat': '-',

        'melting': '527',

        'name': 'Polonium',

        'name_origin': "Latin Polonia, 'Poland', home country of Marie Curie",

        'orgin': 'from decay',

        'period': '6',

        'phase': 'solid',

        'rad': '*'},

'Pr': {'M': 140.91,

        'Z': 59,

        'abundance': '9.2',

        'block': 'f-block',

        'boiling': '3793',

        'config': '[Xe] 4f3 6s2',

        'density': '6.773',

        'electro_negativity': '1.13',

        'group': 'n/a',

        'heat': '0.193',

        'melting': '1208',

        'name': 'Praseodymium',

        'name_origin': "Greek prasios didymos, 'green twin'",

        'orgin': 'primordial',

        'period': '6',

        'phase': 'solid',

        'rad': ''},

'Pt': {'M': 195.08,

        'Z': 78,

        'abundance': '0.005',
```

```
        'block': 'd-block',

        'boiling': '4098',

        'config': '[Xe] 4f14 5d9 6s1',

        'density': '21.46',

        'electro_negativity': '2.28',

        'group': '10',

        'heat': '0.133',

        'melting': '2041.4',

        'name': 'Platinum',

        'name_origin': "Spanish platina, 'little silver', from plata 'silver'",

        'orgin': 'primordial',

        'period': '6',

        'phase': 'solid',

        'rad': ''},

'Pu': {'M': '',

        'Z': 94,

        'abundance': '<= 3x10-11',

        'block': 'f-block',

        'boiling': '3501',

        'config': '[Rn] 5f6 7s2',

        'density': '19.84',

        'electro_negativity': '1.28',

        'group': 'n/a',

        'heat': '-',

        'melting': '912.5',

        'name': 'Plutonium',

        'name_origin': 'Pluto, a dwarf planet, considered a planet in the '

                        'Solar System at the time',
```

'orgin': 'from decay',

'period': '7',

'phase': 'solid',

'rad': '*'},

'Ra': {'M': '',

'Z': 88,

'abundance': '9x10-7',

'block': 's-block',

'boiling': '2010',

'config': '[Rn] 7s2',

'density': '5.5',

'electro_negativity': '0.9',

'group': '2',

'heat': '0.094',

'melting': '973',

'name': 'Radium',

'name_origin': "French radium, from Latin radius, 'ray'",

'orgin': 'from decay',

'period': '7',

'phase': 'solid',

'rad': '*'},

'Rb': {'M': 85.468,

'Z': 37,

'abundance': '90',

'block': 's-block',

'boiling': '961',

'config': '[Kr] 5s1',

'density': '1.532',

        'electro_negativity': '0.82',

        'group': '1',

        'heat': '0.363',

        'melting': '312.46',

        'name': 'Rubidium',

        'name_origin': "Latin rubidus, 'deep red'",

        'orgin': 'primordial',

        'period': '5',

        'phase': 'solid',

        'rad': ''},

'Re': {'M': 186.21,

        'Z': 75,

        'abundance': '7x10-4',

        'block': 'd-block',

        'boiling': '5869',

        'config': '[Xe] 4f14 5d5 6s2',

        'density': '21.02',

        'electro_negativity': '1.9',

        'group': '7',

        'heat': '0.137',

        'melting': '3459',

        'name': 'Rhenium',

        'name_origin': "Latin Rhenus, 'the Rhine'",

        'orgin': 'primordial',

        'period': '6',

        'phase': 'solid',

        'rad': ''},

'Rf': {'M': '',

'Z': 104,

'abundance': '-',

'block': 'd-block',

'boiling': '(5800)',

'config': '',

'density': '(23.2)',

'electro_negativity': '-',

'group': '4',

'heat': '-',

'melting': '(2400)',

'name': 'Rutherfordium',

'name_origin': 'Ernest Rutherford, chemist and physicist from New '

                'Zealand',

'orgin': 'synthetic',

'period': '7',

'phase': 'unknown phase',

'rad': '*'},

'Rh': {'M': 102.91,

'Z': 45,

'abundance': '0.001',

'block': 'd-block',

'boiling': '3968',

'config': '[Kr] 4d8 5s1',

'density': '12.41',

'electro_negativity': '2.28',

'group': '9',

'heat': '0.243',

'melting': '2237',

        'name': 'Rhodium',

        'name_origin': "Greek rhodoeis, 'rose-coloured', from rhodon, 'rose'",

        'orgin': 'primordial',

        'period': '5',

        'phase': 'solid',

        'rad': ''},

'Rn': {'M': '',

        'Z': 86,

        'abundance': '4x10-13',

        'block': 'p-block',

        'boiling': '211.3',

        'config': '[Xe] 4f14 5d10 6s2 6p6',

        'density': '0.00973',

        'electro_negativity': '2.2',

        'group': '18',

        'heat': '0.094',

        'melting': '202',

        'name': 'Radon',

        'name_origin': 'Radium emanation, originally the name of the isotope '

                      'Radon-222',

        'orgin': 'from decay',

        'period': '6',

        'phase': 'gas',

        'rad': '*'},

'Ru': {'M': 101.07,

        'Z': 44,

        'abundance': '0.001',

        'block': 'd-block',

```
        'boiling': '4423',

        'config': '[Kr] 4d7 5s1',

        'density': '12.37',

        'electro_negativity': '2.2',

        'group': '8',

        'heat': '0.238',

        'melting': '2607',

        'name': 'Ruthenium',

        'name_origin': "New Latin Ruthenia, 'Russia'",

        'orgin': 'primordial',

        'period': '5',

        'phase': 'solid',

        'rad': ''},

'S': {'M': 32.06,

        'Z': 16,

        'abundance': '350',

        'block': 'p-block',

        'boiling': '717.87',

        'config': '[Ne] 3s2 3p4',

        'density': '2.067',

        'electro_negativity': '2.58',

        'group': '16',

        'heat': '0.71',

        'melting': '388.36',

        'name': 'Sulfur',

        'name_origin': "Latin sulphur, 'brimstone'",

        'orgin': 'primordial',

        'period': '3',
```

        'phase': 'solid',

        'rad': ''},

 'Sb': {'M': 121.76,

        'Z': 51,

        'abundance': '0.2',

        'block': 'p-block',

        'boiling': '1860',

        'config': '[Kr] 4d10 5s2 5p3',

        'density': '6.685',

        'electro_negativity': '2.05',

        'group': '15',

        'heat': '0.207',

        'melting': '903.78',

        'name': 'Antimony',

        'name_origin': 'Latin antimonium, the origin of which is uncertain: '

                       'folk etymologies suggest it is derived from Greek anti '

                       "('against') + monos ('alone'), or Old French "

                       "anti-moine, 'Monk's bane', but it could plausibly be "

                       "from or related to Arabic itmid, 'antimony', "

                       'reformatted as a Latin word Symbol Sb is derived from '

                       "Latin stibium 'stibnite'",

        'orgin': 'primordial',

        'period': '5',

        'phase': 'solid',

        'rad': ''},

 'Sc': {'M': 44.956,

        'Z': 21,

        'abundance': '22',

'block': 'd-block',

'boiling': '3109',

'config': '[Ar] 3d1 4s2',

'density': '2.989',

'electro_negativity': '1.36',

'group': '3',

'heat': '0.568',

'melting': '1814',

'name': 'Scandium',

'name_origin': "Latin Scandia, 'Scandinavia'",

'orgin': 'primordial',

'period': '4',

'phase': 'solid',

'rad': ''},

'Se': {'M': 78.971,

'Z': 34,

'abundance': '0.05',

'block': 'p-block',

'boiling': '958',

'config': '[Ar] 3d10 4s2 4p4',

'density': '4.809',

'electro_negativity': '2.55',

'group': '16',

'heat': '0.321',

'melting': '453',

'name': 'Selenium',

'name_origin': "Greek selene, 'moon'",

'orgin': 'primordial',

        'period': '4',

        'phase': 'solid',

        'rad': ''},

'Sg': {'M': '',

        'Z': 106,

        'abundance': '-',

        'block': 'd-block',

        'boiling': '-',

        'config': '',

        'density': '(35.0)',

        'electro_negativity': '-',

        'group': '6',

        'heat': '-',

        'melting': '-',

        'name': 'Seaborgium',

        'name_origin': 'Glenn T. Seaborg, American chemist',

        'orgin': 'synthetic',

        'period': '7',

        'phase': 'unknown phase',

        'rad': '*'},

'Si': {'M': 28.085,

        'Z': 14,

        'abundance': '282000',

        'block': 'p-block',

        'boiling': '3538',

        'config': '[Ne] 3s2 3p2',

        'density': '2.3296',

        'electro_negativity': '1.9',

```
        'group': '14',

        'heat': '0.705',

        'melting': '1687',

        'name': 'Silicon',

        'name_origin': "Latin silex, 'flint' (originally silicium)",

        'orgin': 'primordial',

        'period': '3',

        'phase': 'solid',

        'rad': ''},

'Sm': {'M': 150.36,

        'Z': 62,

        'abundance': '7.05',

        'block': 'f-block',

        'boiling': '2067',

        'config': '[Xe] 4f6 6s2',

        'density': '7.52',

        'electro_negativity': '1.17',

        'group': 'n/a',

        'heat': '0.197',

        'melting': '1345',

        'name': 'Samarium',

        'name_origin': 'Samarskite, a mineral named after V. '
                        'Samarsky-Bykhovets, Russian mine official',

        'orgin': 'primordial',

        'period': '6',

        'phase': 'solid',

        'rad': ''},

'Sn': {'M': 118.71,
```

'Z': 50,

'abundance': '2.3',

'block': 'p-block',

'boiling': '2875',

'config': '[Kr] 4d10 5s2 5p2',

'density': '7.287',

'electro_negativity': '1.96',

'group': '14',

'heat': '0.228',

'melting': '505.08',

'name': 'Tin',

'name_origin': 'English word Symbol Sn is derived from Latin stannum',

'orgin': 'primordial',

'period': '5',

'phase': 'solid',

'rad': ''},

'Sr': {'M': 87.62,

'Z': 38,

'abundance': '370',

'block': 's-block',

'boiling': '1655',

'config': '[Kr] 5s2',

'density': '2.64',

'electro_negativity': '0.95',

'group': '2',

'heat': '0.301',

'melting': '1050',

'name': 'Strontium',

```
        'name_origin': 'Strontian, a village in Scotland, where it was found',

        'orgin': 'primordial',

        'period': '5',

        'phase': 'solid',

        'rad': ''},

'Ta': {'M': 180.95,

        'Z': 73,

        'abundance': '2',

        'block': 'd-block',

        'boiling': '5731',

        'config': '[Xe] 4f14 5d3 6s2',

        'density': '16.654',

        'electro_negativity': '1.5',

        'group': '5',

        'heat': '0.14',

        'melting': '3290',

        'name': 'Tantalum',

        'name_origin': 'King Tantalus, father of Niobe from Greek mythology; '

                'see also niobium',

        'orgin': 'primordial',

        'period': '6',

        'phase': 'solid',

        'rad': ''},

'Tb': {'M': 158.93,

        'Z': 65,

        'abundance': '1.2',

        'block': 'f-block',

        'boiling': '3503',
```

```
          'config': '[Xe] 4f9 6s2',

          'density': '8.229',

          'electro_negativity': '1.2',

          'group': 'n/a',

          'heat': '0.182',

          'melting': '1629',

          'name': 'Terbium',

          'name_origin': 'Ytterby, Sweden, where it was found; see also yttrium, '
                         'erbium, ytterbium',

          'orgin': 'primordial',

          'period': '6',

          'phase': 'solid',

          'rad': ''},

'Tc': {'M': '',

          'Z': 43,

          'abundance': '~ 3x10-9',

          'block': 'd-block',

          'boiling': '4538',

          'config': '[Kr] 4d5 5s2',

          'density': '11.5',

          'electro_negativity': '1.9',

          'group': '7',

          'heat': '-',

          'melting': '2430',

          'name': 'Technetium',

          'name_origin': "Greek tekhnetos, 'artificial'",

          'orgin': 'from decay',

          'period': '5',
```

        'phase': 'solid',

        'rad': '*'},

'Te': {'M': 127.6,

        'Z': 52,

        'abundance': '0.001',

        'block': 'p-block',

        'boiling': '1261',

        'config': '[Kr] 4d10 5s2 5p4',

        'density': '6.232',

        'electro_negativity': '2.1',

        'group': '16',

        'heat': '0.202',

        'melting': '722.66',

        'name': 'Tellurium',

        'name_origin': "Latin tellus, 'the ground, earth'",

        'orgin': 'primordial',

        'period': '5',

        'phase': 'solid',

        'rad': ''},

'Th': {'M': 232.04,

        'Z': 90,

        'abundance': '9.6',

        'block': 'f-block',

        'boiling': '5061',

        'config': '[Rn] 6d2 7s2',

        'density': '11.72',

        'electro_negativity': '1.3',

        'group': 'n/a',

```
        'heat': '0.113',

        'melting': '2115',

        'name': 'Thorium',

        'name_origin': 'Thor, the Scandinavian god of thunder',

        'orgin': 'primordial',

        'period': '7',

        'phase': 'solid',

        'rad': ''},

'Ti': {'M': 47.867,

        'Z': 22,

        'abundance': '5650',

        'block': 'd-block',

        'boiling': '3560',

        'config': '[Ar] 3d2 4s2',

        'density': '4.54',

        'electro_negativity': '1.54',

        'group': '4',

        'heat': '0.523',

        'melting': '1941',

        'name': 'Titanium',

        'name_origin': 'Titans, the sons of the Earth goddess of Greek '
                        'mythology',

        'orgin': 'primordial',

        'period': '4',

        'phase': 'solid',

        'rad': ''},

'Tl': {'M': 204.38,

        'Z': 81,
```

      'abundance': '0.85',

      'block': 'p-block',

      'boiling': '1746',

      'config': '[Xe] 4f14 5d10 6s2 6p1',

      'density': '11.85',

      'electro_negativity': '1.62',

      'group': '13',

      'heat': '0.129',

      'melting': '577',

      'name': 'Thallium',

      'name_origin': "Greek thallos, 'green shoot or twig'",

      'orgin': 'primordial',

      'period': '6',

      'phase': 'solid',

      'rad': ''},

'Tm': {'M': 168.93,

      'Z': 69,

      'abundance': '0.52',

      'block': 'f-block',

      'boiling': '2223',

      'config': '[Xe] 4f13 6s2',

      'density': '9.321',

      'electro_negativity': '1.25',

      'group': 'n/a',

      'heat': '0.16',

      'melting': '1818',

      'name': 'Thulium',

      'name_origin': 'Thule, the ancient name for an unclear northern '

          'location',

     'orgin': 'primordial',

      'period': '6',

      'phase': 'solid',

      'rad': ''},

'U': {'M': 238.03,

      'Z': 92,

      'abundance': '2.7',

      'block': 'f-block',

      'boiling': '4404',

      'config': '[Rn] 5f3 6d1 7s2',

      'density': '18.95',

      'electro_negativity': '1.38',

      'group': 'n/a',

      'heat': '0.116',

      'melting': '1405.3',

      'name': 'Uranium',

      'name_origin': 'Uranus, the seventh planet in the Solar System',

      'orgin': 'primordial',

      'period': '7',

      'phase': 'solid',

      'rad': ''},

'V': {'M': 50.942,

      'Z': 23,

      'abundance': '120',

      'block': 'd-block',

      'boiling': '3680',

      'config': '[Ar] 3d3 4s2',

    'density': '6.11',

    'electro_negativity': '1.63',

    'group': '5',

    'heat': '0.489',

    'melting': '2183',

    'name': 'Vanadium',

    'name_origin': 'Vanadis, an Old Norse name for the Scandinavian goddess '
                'Freyja',

    'orgin': 'primordial',

    'period': '4',

    'phase': 'solid',

    'rad': ''},

'W': {'M': 183.84,

    'Z': 74,

    'abundance': '1.3',

    'block': 'd-block',

    'boiling': '5828',

    'config': '[Xe] 4f14 5d4 6s2',

    'density': '19.25',

    'electro_negativity': '2.36',

    'group': '6',

    'heat': '0.132',

    'melting': '3695',

    'name': 'Tungsten',

    'name_origin': "Swedish tung sten, 'heavy stone' Symbol W is from "
                'Wolfram, originally from Middle High German wolf-rahm '
                "'wolf's foam' describing the mineral wolframite[5]",

    'orgin': 'primordial',

```
          'period': '6',

          'phase': 'solid',

          'rad': ''},

'Xe': {'M': 131.29,

          'Z': 54,

          'abundance': '3x10-5',

          'block': 'p-block',

          'boiling': '165.03',

          'config': '[Kr] 4d10 5s2 5p6',

          'density': '0.005887',

          'electro_negativity': '2.60',

          'group': '18',

          'heat': '0.158',

          'melting': '161.4',

          'name': 'Xenon',

          'name_origin': "Greek xenon, neuter form of xenos 'strange'",

          'orgin': 'primordial',

          'period': '5',

          'phase': 'gas',

          'rad': ''},

'Y': {'M': 88.906,

          'Z': 39,

          'abundance': '33',

          'block': 'd-block',

          'boiling': '3609',

          'config': '[Kr] 4d1 5s2',

          'density': '4.469',

          'electro_negativity': '1.22',
```

        'group': '3',

        'heat': '0.298',

        'melting': '1799',

        'name': 'Yttrium',

        'name_origin': 'Ytterby, Sweden, where it was found; see also terbium, '
                       'erbium, ytterbium',

        'orgin': 'primordial',

        'period': '5',

        'phase': 'solid',

        'rad': ''},

'Yb': {'M': 173.05,

        'Z': 70,

        'abundance': '3.2',

        'block': 'f-block',

        'boiling': '1469',

        'config': '[Xe] 4f14 6s2',

        'density': '6.965',

        'electro_negativity': '1.1',

        'group': 'n/a',

        'heat': '0.155',

        'melting': '1097',

        'name': 'Ytterbium',

        'name_origin': 'Ytterby, Sweden, where it was found; see also yttrium, '
                       'terbium, erbium',

        'orgin': 'primordial',

        'period': '6',

        'phase': 'solid',

        'rad': ''},

'Zn': {'M': 65.38,

    'Z': 30,

    'abundance': '70',

    'block': 'd-block',

    'boiling': '1180',

    'config': '[Ar] 3d10 4s2',

    'density': '7.134',

    'electro_negativity': '1.65',

    'group': '12',

    'heat': '0.388',

    'melting': '692.88',

    'name': 'Zinc',

    'name_origin': "Most likely from German Zinke, 'prong' or 'tooth', "

            "though some suggest Persian sang, 'stone'",

    'orgin': 'primordial',

    'period': '4',

    'phase': 'solid',

    'rad': ''},

'Zr': {'M': 91.224,

    'Z': 40,

    'abundance': '165',

    'block': 'd-block',

    'boiling': '4682',

    'config': '[Kr] 4d2 5s2',

    'density': '6.506',

    'electro_negativity': '1.33',

    'group': '4',

    'heat': '0.278',

'melting': '2128',

'name': 'Zirconium',

'name_origin': "Zircon, a mineral, from Persian zargun, 'gold-hued'",

'orgin': 'primordial',

'period': '5',

'phase': 'solid',

'rad': ''}}